\documentclass[pra,aps,twocolumn,longbibliography,nobalancelastpage, superscriptaddress]{revtex4-1}

\usepackage[utf8]{inputenc}
\usepackage[T1]{fontenc}
\usepackage{amsmath}
\usepackage{xcolor}
\usepackage{bbold}
\colorlet{myPurple}{blue!40!red}
\colorlet{myPurplee}{blue!10!red}
\colorlet{myCyan}{cyan!60!gray}
\colorlet{myRed}{red!66!black}
\usepackage{tikz}
\usepackage{pgfplots}
\usepackage{float}
\usepackage{geometry}
\pgfplotsset{compat=1.14}
\usepackage[colorlinks=true,citecolor=myRed,urlcolor=myRed,linkcolor=myRed]{hyperref}
\usepackage[normalem]{ulem}
\usepackage{exscale}
\usepackage{bbm}
\usepackage{graphicx}
\usepackage{amsmath}
\usepackage{latexsym}
\usepackage{amsfonts}
\usepackage[separate-uncertainty=true,multi-part-units=single]{siunitx}
\usepackage{amssymb}
\usepackage{times}
\usepackage[T1]{fontenc}
\usepackage{amsthm}
\usepackage{enumerate}
\usepackage{bbold}
\usepackage{color}
\usepackage{nicefrac}
\usepackage{soul}
\newcommand{\sket}[1]{{\ensuremath{\lvert#1\rangle}}}
\newcommand{\lket}[1]{{\ensuremath{\left\lvert#1\right\rangle}}}
\newcommand{\ket}[1]{\if@display\lket{#1}\else\sket{#1}\fi}

\newcommand{\sbra}[1]{{\ensuremath{\langle#1\rvert}}}
\newcommand{\lbra}[1]{{\ensuremath{\left\langle#1\right\rvert}}}
\newcommand{\bra}[1]{\if@display\lbra{#1}\else\sbra{#1}\fi}

\newcommand{\sbraket}[2]{{\ensuremath{\langle#1\rvert#2\rangle}}}
\newcommand{\lbraket}[2]{{\ensuremath{\left\langle#1\!\left\rvert\vphantom{#1}#2\right.\!\right\rangle}}}
\newcommand{\braket}[2]{\if@display\lbraket{#1}{#2}\else\sbraket{#1}{#2}\fi}

\newcommand{\sketbra}[2]{{\ensuremath{\lvert #1\rangle\!\langle #2\rvert}}}
\newcommand{\lketbra}[2]{{\ensuremath{\left\lvert #1\right\rangle\!\!\left\langle #2\right\rvert}}}
\newcommand{\ketbra}[2]{\if@display\lketbra{#1}{#2}\else\sketbra{#1}{#2}\fi}
\usepackage{tcolorbox}
\usepackage{mathtools}

\newcommand{\proj}[1]{\ketbra{#1}{#1}}

\newcommand{\tr}{\textrm{Tr}}

\usepackage{tikz}
\usepackage{lipsum}
\theoremstyle{plain}
\newtheorem{thm}{Theorem}
\newtheorem{lem}{Lemma}

\usepackage[font=small,labelfont=bf,justification=justified,format=plain]{subcaption}

\newtheorem{prop}[thm]{Proposition}
\newtheorem{cor}[thm]{Corollary}
\newtheorem{defi}{Definition}
\usepackage{graphicx}
\usepackage{bm}
\usepackage{dsfont}
\usepackage{tikz}
\usepackage[T1]{fontenc}
\usepackage{amsthm}
\usepackage{array}
\usepackage{amssymb}
\usepackage{amsfonts}
\usepackage{cancel}
\usepackage[toc,page]{appendix}
\usepackage{multirow}
\usepackage{color}
\usepackage{calrsfs}
\usetikzlibrary{backgrounds,decorations.pathreplacing,calc}
\newcommand{\tra}{\textrm{tr}}
\usepackage{tkz-euclide}
\usepackage{tcolorbox}
\usepackage{ragged2e}
\DeclareCaptionJustification{justified}{\justifying}
\newcommand{\laura}[1]{{\color{cyan} #1}}
\DeclareMathAlphabet{\mathcal}{OMS}{cmsy}{m}{n}
\usepackage{thmtools}
\usepackage{thm-restate}

\begin{document}

\title{Experimentally Certified Transmission of a Quantum Message through an Untrusted and Lossy Quantum Channel via Bell's Theorem}
\author{Simon Neves}
\affiliation{Sorbonne Universit\'{e}, CNRS, LIP6, 4 Place Jussieu, Paris F-75005, France}

\author{Laura dos Santos Martins}
\affiliation{Sorbonne Universit\'{e}, CNRS, LIP6, 4 Place Jussieu, Paris F-75005, France}

\author{Verena Yacoub}
\affiliation{Sorbonne Universit\'{e}, CNRS, LIP6, 4 Place Jussieu, Paris F-75005, France}

\author{Pascal Lefebvre}
\affiliation{Sorbonne Universit\'{e}, CNRS, LIP6, 4 Place Jussieu, Paris F-75005, France}

\author{Ivan \v{S}upi\'c}
\affiliation{Sorbonne Universit\'{e}, CNRS, LIP6, 4 Place Jussieu, Paris F-75005, France}

\author{Damian Markham}
\affiliation{Sorbonne Universit\'{e}, CNRS, LIP6, 4 Place Jussieu, Paris F-75005, France}

\author{Eleni Diamanti}
\affiliation{Sorbonne Universit\'{e}, CNRS, LIP6, 4 Place Jussieu, Paris F-75005, France}

\date{\today}


\begin{abstract}
Quantum transmission links are central elements in essentially all protocols involving the exchange of quantum messages. Emerging progress in quantum technologies involving such links needs to be accompanied by appropriate certification tools. In adversarial scenarios, a certification method can be vulnerable to attacks if too much trust is placed on the underlying system. Here, we propose a protocol in a device independent framework, which allows for the certification of practical quantum transmission links in scenarios where minimal assumptions are made about the functioning of the certification setup. In particular, we take unavoidable transmission losses into account by modeling the link as a completely-positive trace-decreasing map. We also, crucially, remove the assumption of independent and identically distributed samples, which is known to be incompatible with adversarial settings. Finally, in view of the use of the certified transmitted states for follow-up applications, our protocol moves beyond certification of the channel to allow us to estimate the quality of the transmitted quantum message itself. To illustrate the practical relevance and the feasibility of our protocol with currently available technology we provide an experimental implementation based on a state-of-the-art polarization entangled photon pair source in a Sagnac configuration and analyze its robustness for realistic losses and errors.
\end{abstract}

\maketitle

\large
\noindent\textbf{Introduction}
\normalsize

\noindent The ability to send and receive quantum information is at the heart of the rapidly developing quantum technologies. Transmitting quantum information over quantum networks promises unparalleled efficiency and security~\cite{QIavision}, as well as new functionalities such as the delegation of quantum computation~\cite{reviewBQC} and quantum sensing~\cite{SM22}. Within quantum computers themselves we will need to input, share and distribute quantum information to different parts, particularly important for architectures relying on multiple quantum processors~\cite{Awschalom21,Saleem21}. The reliable transmission of quantum information is thus an essential building block for future quantum technologies, and, as such, we must be very sure of its working. 
When the physical devices used to test and use these quantum channels are trusted, this question can be answered by standard quantum channel authentication~\cite{barnum2002authentication}, and there are various approaches to this end, from those requiring incredibly expensive entangled resources~\cite{barnum2002authentication,dupuis2012actively,broadbent2013quantum}, to those more achievable, but at cost to security scaling~\cite{markham2015practical,markham2020simple,zhu2019general,takeuchi2019resource}. In this work, we consider a much stronger requirement, where some or all devices used are not trusted, in a so-called device independent setting. This will be a crucial step for testing the transmission through quantum channels for future applications.

Device independence uses Bell-like correlations to imply correct behaviour of quantum hardware, without the need to understand or trust their inner workings~\cite{colbeck_quantum_2011,acin2007}, that is, independently of the physical device used. It is motivated by the inevitable situation where the user of a quantum technology is not necessarily the one who built all the hardware and does not necessarily want to trust it to behave as specified. It has first been applied in quantum information to prove security in quantum key distribution devices, thus making them secure against potential hardware hacks. It has then expanded in many directions, including random number generation~\cite{pironio2010}, verification of quantum computation~\cite{reichardt2013}, and more~\cite{Baccari_2020,SupicBrunner}. The application to quantum channels is relatively recent~\cite{Sekatski2018} (but see also~\cite{magniez2005}), however there are some important missing elements in order to obtain useful certification. 

Here, we address the main remaining obstacles to certify the transmission of quantum information in the device independent framework. 
First, in our approach we explicitly take into account loss. This is particularly important in optical implementations (which is the most natural choice for quantum channels). It is not addressed in current schemes~\cite{Sekatski2018,magniez2005}, which effectively assume that any loss is innocent; this is somewhat against the goals of device independence and opens a security loophole if the loss is controlled by malicious parties. Second, we remove the assumption that each time a channel is used, it is done so in an independent, uncorrelated way, known as identical independent distribution (IID). This assumption similarly makes us vulnerable in terms of security so should be avoided in general. Third, we certify the transmission of quantum information itself. Previous works assume IID, that loss is not malicious, and they certify that the channel that was used during the test was good but without a statement on actual transmitted quantum information~\cite{Sekatski2018}. 
We develop the treatment of loss as a non trace preserving channel, bounding the diamond fidelity between an untrusted channel and an ideal one. We use this to build protocols certifying a transmitted quantum message using this channel. Our protocols are secure in the one-sided device independent setting (where the sender's devices are fully trusted, but not the receiver's), and also in the fully device independent setting when  IID is assumed on the source; in both cases no IID needs to be assumed on the uses of the channel. 

We also demonstrate the feasibility of our protocol and experimentally validate the main elements of one-sided device independent certified transmission with an implementation exploiting a high-quality entangled photon source with polarization encoding obtained in a Sagnac configuration. This allows us to explore the behavior of the minimum fidelity that we can certify for realistic losses in honest channels and confirm the robustness of the protocol against simulated errors introduced by dishonest channels.\\

\large
\noindent\textbf{Results}
\normalsize

\noindent\textbf{Certification protocol.} In our framework, a player Alice wishes to send a qubit state from Hilbert space $\mathcal{H}_i$ to Bob, through a local unitary quantum channel $\mathcal{E}_0$. This \textit{quantum message} is possibly entangled with another system of Hilbert space $\mathcal{S}$ of arbitrary dimension, so the global state reads ${\rho_i\in\mathcal{L}(\mathcal{H}_i\otimes\mathcal{S})}$. The channel takes any qubit from $\mathcal{L}(\mathcal{H}_i)$ to another qubit from $\mathcal{L}(\mathcal{H}_o)$, with output global state ${\rho_o = (\mathcal{E}_0\otimes\mathds{I})[\rho_i]=(U\otimes\mathds{I})\rho_i(U^\dagger\otimes\mathds{I})}$, where $U$ is a local unitary and $\mathds{I}$ is the identity. This model describes a perfect unitary gate in a quantum computer, quantum transmission link (carried on through quantum teleportation or a simple optical fiber) or quantum memory. Without loss of generality, we take $U = \mathds{I}$ and $(\mathcal{E}_0\otimes \mathds{I})[\rho_i]=\rho_i$, as this case encompasses all unitaries in a device independent scenario \cite{Sekatski2018}. This channel is called the \textit{reference channel}.

In real world situations, the channel would be lossy, noisy, or even operated by a malicious party Eve. Also, Alice and Bob normally do not have access to isolated qubit spaces, but operate with physical systems such as photons or atoms, displaying other degrees of freedom. This way, without further assumptions, Alice and Bob have access to a completely positive trace-decreasing (CPTD) map $\mathcal{E}$, \textit{i.e.} a probabilistic channel, that sends density operators from an input Hilbert space $\mathcal{H}_{\mathcal{A}_1}$ to positive operators of trace smaller than 1 on an output Hilbert space $\mathcal{H}_{\mathcal{B}}$. This channel is called the \emph{physical channel}. Alice also possesses a source of bipartite states $\Phi_i$ shared between $\mathcal{H}_{\mathcal{A}_1}$ and a secondary Hilbert space $\mathcal{H}_{\mathcal{A}_2}$, that we call the \textit{probe} input state. She can send one part of $\Phi_i$ through the channel $\mathcal{E}$, resulting in the probe output state $\Phi_o$, shared with~Bob: 
\begin{equation}\label{eq:DefOutputState}
\Phi_o = (\mathcal{E}\otimes \mathds{I})[\Phi_i]/t(\mathcal{E}\vert\Phi_i),    
\end{equation}
where $t(\mathcal{E}\vert\Phi_i)= \tr(\mathcal{E}\otimes \mathds{I})[\Phi_i]$ is the \textit{transmissivity} of $\mathcal{E}$ which \textit{a priori} depends on the input state, as it does in polarizing channels for instance. For more details on this relatively new notion, the reader can refer to SUPP. MAT. \ref{seq:Prel}. Finally, the players can measure states with 2-outcome positive operator-valued measures (POVMs) $\{M^\mathcal{P}_{l|q}\}_{l=0,1}$ where $\mathcal{P} = \mathcal{A}_1,\mathcal{A}_2$ or $\mathcal{B}$ indicating the Hilbert space on which the measurement is acting, and $q$ indicates which POVM is measured, see Eqs.~(\ref{eq:A0}) to (\ref{eq:B1}) below. Fig.~\ref{fig:GeneralScheme} illustrates our setting.

\begin{figure}[htbp]
    \centering
\includegraphics[width=1\linewidth]{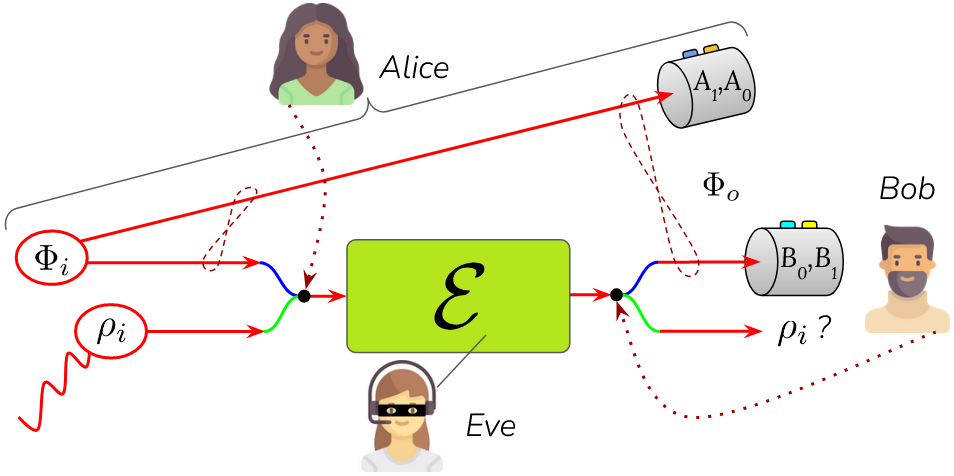}
    \caption{Sketch of the problem. 
    Alice's goal is to send a qubit, potentially part of a larger system, in state $\rho_i$, through an untrusted quantum channel $\mathcal{E}$ (green path). To do so, she sometimes tests the channel by sending half an entangled state (blue path). Alice and Bob can then measure the output state $\Phi_o$, to assess how close the action of the physical channel $\mathcal{E}$ is to an ideal reference channel $\mathcal{E}_0$  on the transmitted state~$\rho_i$.}
    \label{fig:GeneralScheme}
\end{figure}


In an adversarial scenario, Alice and Bob wish to draw device independent conclusions, meaning they make no assumption whatsoever on the states or the measurements. In particular, physical Hilbert spaces are of arbitrarily big dimensions, which include all degrees of freedom of the physical systems and possible entanglement with the rest of the universe. In this way, players can only certify objects up to local isometries, which associate finite-dimension qubit spaces $\mathcal{H}_i$ and $\mathcal{H}_o$, to these infinite-dimension physical spaces $\mathcal{H}_{\mathcal{A}_1}$, $\mathcal{H}_{\mathcal{A}_2}$, $\mathcal{H}_{\mathcal{B}}$. As a device independent procedure, self-testing is actually "blind" to local isometries such that it does not certify a single state, but a whole equivalence class of quantum states mutually related by locally isometric transformations. As shown in \cite{Sekatski2018}, similar conclusions can be drawn in order to device-independently test the equivalence between the physical channel $\mathcal{E}\otimes\mathds{I}$ and the reference operation $\mathcal{E}_0\otimes\mathds{I}$. Note, however, that as a quantum channel is associated to two Hilbert spaces (one in input and the other in output), two isometries are involved in order to extract a qubit-to-qubit channel from a physical channel. This way, the input isometry brings a qubit input state to a physical state that can be fed into the physical channel, while the output isometry extracts a qubit state from the physical channel's output state. However, this formalism, in principle, only applies to completely positive trace-preserving (CPTP) maps. In our case, a trace-decreasing physical channel only returns a state with a certain probability, such that it can only be compared to the reference channel multiplied by a constant $t \leq 1$. Then, one can only make a statement about equivalence between the physical and reference channels, when considering rounds in which the transmission was successful. We capture this intuition with the following definition.

\begin{defi}[Self-testing of a CPTD map]\label{defST}
Let us consider a physical channel ${\mathcal{E}:\mathcal{H}_{\mathcal{A}_1}\longrightarrow\mathcal{H}_{\mathcal{B}}}$. With two local isometries ${\Gamma_i : \mathcal{H}_{\mathcal{A}_1}\otimes\mathcal{H}_i \longrightarrow \mathcal{H}_{\mathcal{A}_1}\otimes \mathcal{H}_i^{ext}}$ (encoding map) and ${\Gamma_o : \mathcal{H}_{\mathcal{B}} \longrightarrow \mathcal{H}_o\otimes \mathcal{H}_o^{ext}}$ (decoding map), and an ancillary state ${\rho_{\mathcal{A}_1}\in\mathcal{L}(\mathcal{H}_{\mathcal{A}_1})}$, we can define an extracted qubit channel $\mathcal{E}_{i,o}$ as:
\begin{equation}\label{eq:TransformedChannel}
    \mathcal{E}_{i,o} : \rho \in \mathcal{L}(\mathcal{H}_i) \longrightarrow \tr_{ext}\bigl((\Gamma_o \circ \mathcal{E} \circ \Gamma_i) [\rho_{\mathcal{A}_1} \otimes \rho \:]\bigr),
\end{equation}
where the trace is taken over $\mathcal{H}_i^{ext}$ and $\mathcal{H}_o^{ext}$ \footnote{The identity channel on $\mathcal{H}_i^{ext}$ is omitted in (\ref{eq:TransformedChannel}) for more clarity.}. The self-testing equivalence between a probabilistic channel $\mathcal{E}$ and the reference channel $\mathcal{E}_0$ is established if there exists $t\in]0;1]$ giving:
\begin{equation}
    \mathcal{E}_{i,o} = t\mathcal{E}_0.
\end{equation}
%
\end{defi}
The reader can refer to SUPP. MAT. \ref{equivalclasses} for more details on the lossy channels' equivalence classes. In experiments, we can never perfectly certify $\mathcal{E}$, therefore we quantify the ability of this probabilistic channel to implement the deterministic channel $\mathcal{E}_0$ by generalizing the diamond fidelity to probabilistic quantum channels:
\begin{equation}\label{eq:diamondFid}
\begin{aligned}
    \mathcal{F}_\diamond^{\Gamma_{i,o}}(\mathcal{E},\mathcal{E}_0) & =\mathcal{F}_\diamond(\mathcal{E}_{i,o},\mathcal{E}_0)\\
    &\hspace{-1cm}= \inf_{\ket{\phi}}  F((\mathcal{E}_{i,o}\otimes\mathds{I})[\phi]/t(\mathcal{E}_{i,o}\vert\phi),(\mathcal{E}_0\otimes\mathds{I})[\phi]),
\end{aligned}
\end{equation}
where $F(\rho,\sigma) = \tr\bigl(\sqrt{\rho^{1/2}\sigma\rho^{1/2}}\bigr)^2$ is the Ulhmann fidelity for quantum states, and the lower bound is taken over all pure states $\ket{\phi}$ from $\mathcal{H}_i^{\otimes 2}$ such that $t(\mathcal{E}_1\vert\phi) \neq 0$ and $t(\mathcal{E}_2\vert\phi) \neq 0$. Note that the left state is normalized by the transmissivity. Consequently, contrary to CPTP maps fidelities, $\mathcal{F}_\diamond(\mathcal{E}_{i,o},\mathcal{E}_0)=1$ does not imply $\mathcal{E}_{i,o}=\mathcal{E}_0$, but only that there exists $t\in]0,1]$ such that $\mathcal{E}_{i,o} = t\mathcal{E}_0$, meaning that the channels are equivalent in the sense of our definition. Physically speaking, these two channels output the same states, under the condition those were not lost. The diamond fidelity is particularly useful here, as it can be interpreted as the minimum probability that $\mathcal{E}\otimes\mathds{I}$ successfully implements the operation $\mathcal{E}_0\otimes\mathds{I}$ on any state, under the condition that a state successfully passes through the channel. The main goal of our protocol is therefore to certify that fidelity.

For that purpose, let us consider the situation where Alice can certify the probe input state $\Phi_i$ up to two local isometries $\Gamma^{A_1/\mathcal{A}_2}: \mathcal{H}_{\mathcal{A}_1/\mathcal{A}_2} \longrightarrow \mathcal{H}_{\mathcal{A}_1/\mathcal{A}_2}\otimes\mathcal{H}_i$ with the following fidelity to a maximally entangled state:
\begin{equation}\label{eq:inputFid}
    F^i = F\bigl((\Lambda^{\mathcal{A}_1}\otimes \Lambda^{\mathcal{A}_2})[\Phi_i],\Phi_+\bigr),
\end{equation}
where $\Phi_+$ is a maximally-entangled state (for instance $\ket{\Phi_+}=\tfrac{\ket{00}+\ket{11}}{\sqrt{2}}$) and $\Lambda^{j}[\cdot] = \tr_{j}(\Gamma^{j}[\cdot])$. We next consider the situation that Alice and Bob are able to certify the probe output state $\Phi_o$ up to local isometries $\Gamma^{\mathcal{A}_2}$ and $\Gamma^{\mathcal{B}}: \mathcal{H}_{\mathcal{B}} \longrightarrow \mathcal{H}_{\mathcal{B}}\otimes\mathcal{H}_o$ with the following fidelity:
\begin{equation}\label{eq:outputFid}
     F^o = F\bigl((\Lambda^{\mathcal{B}}\otimes\Lambda^{\mathcal{A}_2})[(\mathcal{E}\otimes\mathds{I})[\Phi_i]]/t(\mathcal{E}\vert\Phi_i),(\mathcal{E}_0\otimes\mathds{I})[\Phi_+]\bigr).
\end{equation}
Given Eqs.~(\ref{eq:inputFid}) and (\ref{eq:outputFid}), we show in SUPP. MAT.~\ref{sec:certificationBound} that there exist isometries $\Gamma_i,\Gamma_o$ such that Alice and Bob are able to lower bound the diamond fidelity on the corresponding extracted channel $\mathcal{E}_{i,o}$: 
\begin{equation}\label{ineq:MainResult3}
     \mathcal{F}_\diamond(\mathcal{E}_{i,o},\mathcal{E}_0)\geq 1- 4 \sin^2\Bigl(\arcsin\bigl(C^i/t(\mathcal{E}\vert\Phi_i)\bigr) + \arcsin C^o\Bigr),
\end{equation}
where $C^j = \sqrt{1-F^j}$ are sine distances associated to their corresponding fidelities \cite{Rastegin2006}. In this way, checking the input and output fidelities allows us to assess the fidelity of the channel itself. 
This bound generalizes what is shown in \cite{Sekatski2018} to probabilistic channels. It also uses the diamond fidelity, which informs on the behavior of the channel on any state, instead of the Choi-Jamio\l kowski fidelity, which only informs on the behavior of the channel on a maximally entangled state.

This bound gives the direction for estimating the fidelity of a quantum channel. The idea is to evaluate the fidelity $F^i$ of the probe input state to a Bell state, then send one part of that probe state through the channel Alice wishes to send $\rho_i$ through, and finally evaluate the fidelity $F^o$ of the corresponding output state to the same Bell state. Such procedure is possible using recent self-testing results \cite{unnikrishnan2019}, but requires a very large number of experimental rounds in the absence of the IID assumption, as both input and output probe states require certification. We significantly decrease that number by making the IID assumption on the probe state, or by leaving its full characterization to Alice's responsibility. Still, as we make no IID assumption on the channel, optimal security cannot be reached by first testing that channel, and only then using it to send the message state $\rho_i$, as Eve may change the channel's expression in the last moment. Our protocol works around this problem by allowing Alice to hide the message $\rho_i$ among a large number of probe states, at a random position unknown to Eve. In that case, we show in SUPP. MAT. \ref{sec:CertifAvg} that the bound (\ref{ineq:MainResult3}) holds for the average channel $\bar{\mathcal{E}}_{i,o}$ over the whole protocol. Then the \emph{transmission fidelity} between the output quantum message $\bar{\rho}_o = (\bar{\mathcal{E}}_{i,o}\otimes\mathds{I})[\rho_i]/t(\bar{\mathcal{E}}\vert\rho_i)$ and the input quantum message $\rho_i$ is certified:
\begin{equation}
    F(\rho_i,\bar{\rho}_o)\geq \mathcal{F}_\diamond(\bar{\mathcal{E}}_{i,o},\mathds{I}).
\end{equation}
As long as the message's position among the probe states remains hidden, we can use $\bar{\rho}_o$ to describe accurately any statistics that would occur when processing the output state of the protocol, and estimate the quality of an actual transmitted state, instead of a verification of a channel only (see SUPP. MAT. \ref{sec:avgstate} for more details).

\begin{figure}[htbp]
    \centering
\includegraphics[width=1\linewidth]{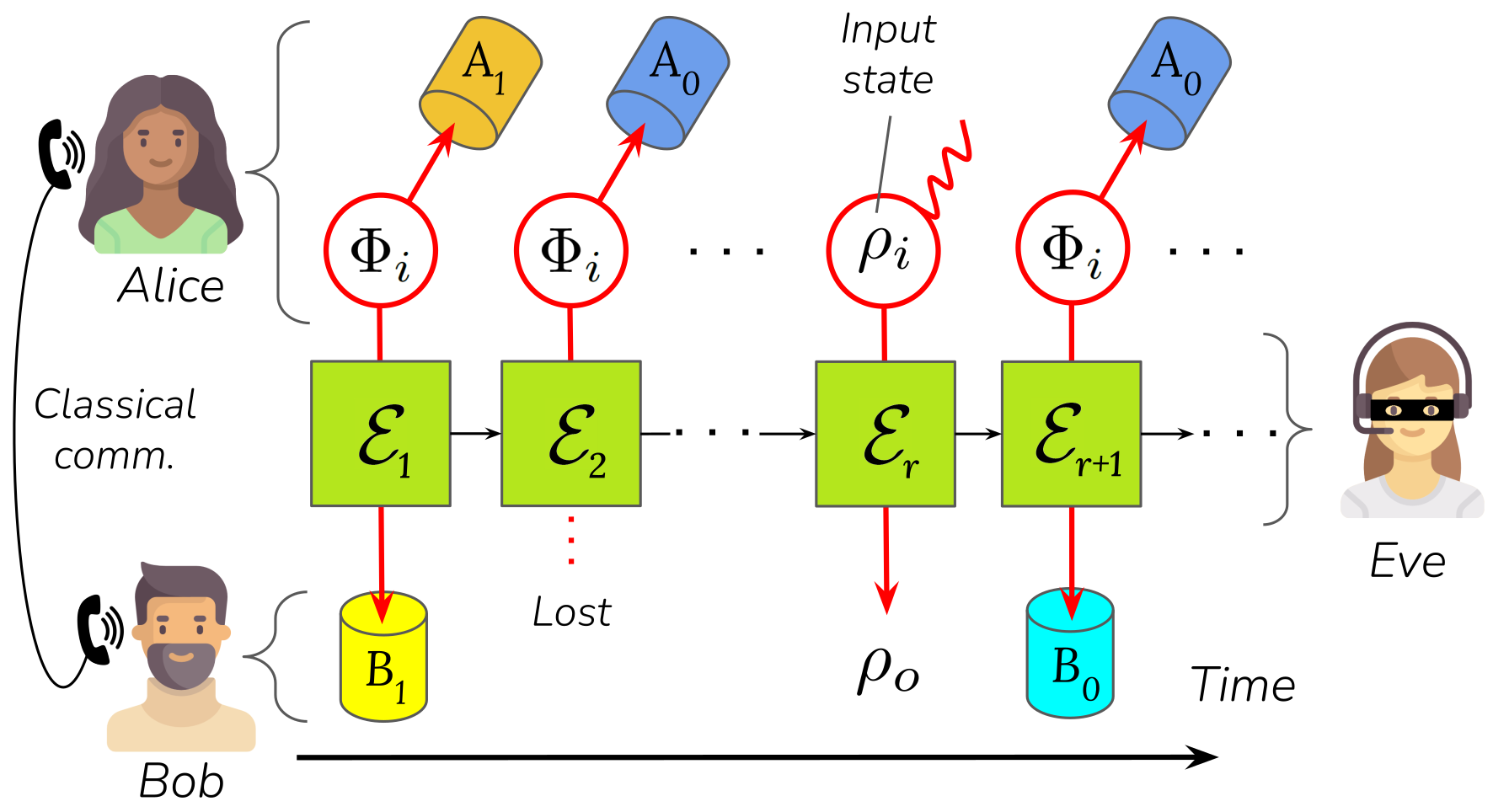}
    \caption{Protocol sketch in a one-sided device independent scenario: Alice prepares $N$ copies of the probe state $\Phi_i$, and sends them through the untrusted channel $\mathcal{E}$ that varies with time, as well as $\rho_i$ at a random secret position $r$. Some states are lost such that Bob only receives a fraction of them. Alice tells Bob the value of $r$. If $\rho_i$ was lost, then the protocol aborts. Otherwise, Bob stores $\rho_i$ and, together with Alice, tests the violation of the steering inequality with the output probe states. They deduce the average channel's quality over the protocol, which informs on the probability that the message $\rho_i$ was accurately transmitted to Bob, up to isometries.}
    \label{fig:ProtocolScheme}
\end{figure}

In SUPP. MAT.~\ref{app:TheoProts} we give detailed protocols where we apply these ideas to test a transmitted quantum message under the device independent (DI) and one-sided device independent (1sDI) scenarios. For the purpose of our demonstration, we focus on an one-sided device independent scenario. A summary of the protocol in this case is given in Fig.~\ref{fig:ProtocolScheme} (for a detailed recipe, the reader can refer to the Supplementary Material). Here, Alice's measurement setup is trusted, such that her Hilbert spaces are qubit spaces $\mathcal{H}_{\mathcal{A}_1} = \mathcal{H}_{\mathcal{A}_2} = \mathcal{H}_i$, her isometries are trivial $\Gamma_i = \Gamma^{\mathcal{A}_1} = \Gamma^{\mathcal{A}_2} = \mathds{I}$, and she performs measurements in the Pauli $X$ and $Z$ bases:
\begin{align}
       &A_0 = M^{\mathcal{A}_2}_{0|0} -  M^{\mathcal{A}_2}_{1|0} = Z,\label{eq:A0}\\
       &A_1 = M^{\mathcal{A}_2}_{0|1} -  M^{\mathcal{A}_2}_{1|1} = X.\label{eq:A1}
\end{align}

This fits a variety of scenarios where Alice is a powerful server, trying to provide states to a weaker client, Bob, whose measurement apparatus is still untrusted. For that reason, Bob's observables, defined as:
\begin{align}
       &B_0 = M^{\mathcal{B}}_{0|0} -  M^{\mathcal{B}}_{1|0},\label{eq:B0}\\
       &B_1 = M^{\mathcal{B}}_{0|1} -  M^{\mathcal{B}}_{1|1},\label{eq:B1}
\end{align}
are \emph{a priori} unknown. In order to bound $F^o$, Alice and Bob use self-testing through steering~\cite{supic2016}. Namely, the maximal violation of the \textit{steering} inequality~\cite{SteerIneq}:
\begin{equation}\label{ineq:steering}
    \beta = \vert\langle A_0B_0\rangle + \langle A_1B_1\rangle\vert \leq \sqrt{2},
\end{equation}
self-tests the maximally entangled pair of qubits. We then combine recent self-testing results \cite{unnikrishnan2019} with further finite statistics methods in a non-IID setting and with a lossy channel, in order to estimate $F^o$ in bound (\ref{ineq:MainResult3}) with high confidence, when a close-to-maximal violation $\beta = 2 - \epsilon$ is measured:
\begin{equation}\label{eq:STboundAnu}
    F^o \geq 1-\alpha f(\epsilon ,K)\simeq 1-\alpha\epsilon,
\end{equation}
with $f$ a function of $\epsilon$ and the number $K$ of states measured by Alice and Bob during the protocol (see Eq.~(\ref{eq:STfunctionAnu}) in Methods), and $\alpha = 1.26$ \cite{unnikrishnan2019}. This outlines the protocol: by sending $N$ characterized probe states through the channel, Alice and Bob estimate $F_o$ and thus the diamond fidelity between the extracted channel and the identity channel, and therefore the transmission fidelity of an unknown state $\rho_i$, as a function of $N$, $\epsilon$, and the number $K$ of transmitted states.\\

\begin{figure*}[htbp]
    \centering
\includegraphics[width=150mm]{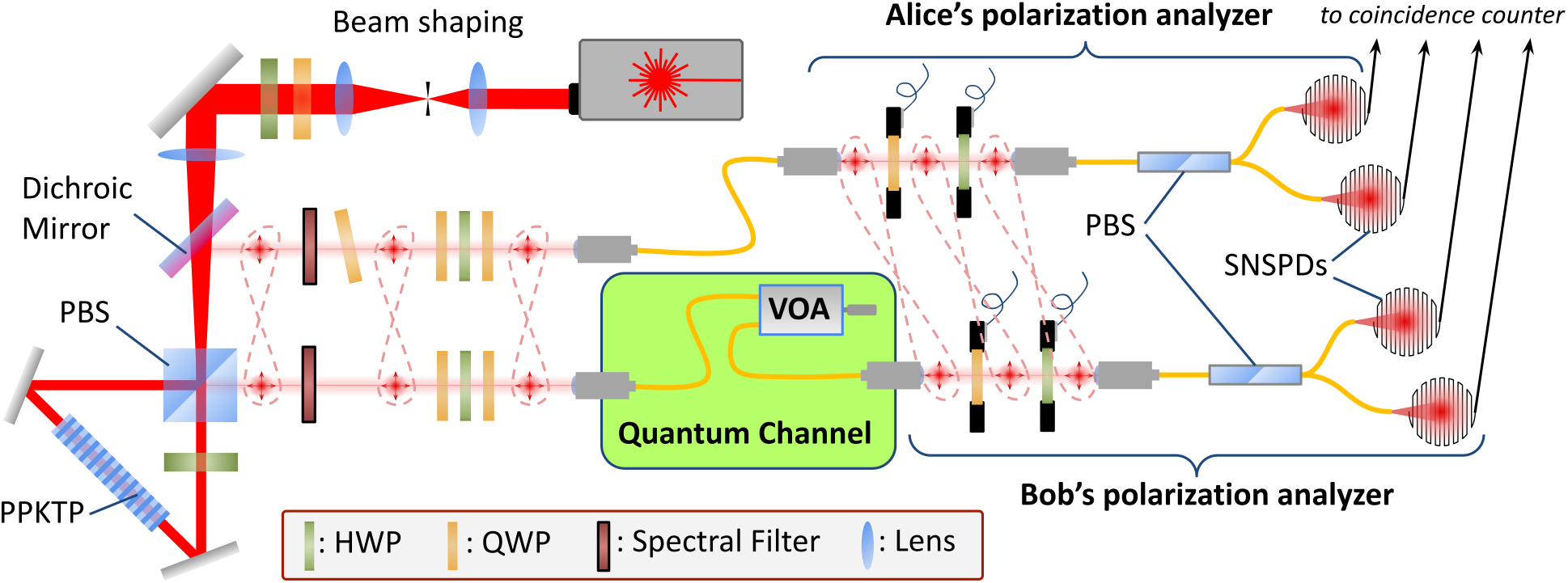}
    \caption{Experimental setup for photonic certified quantum communication through an unstrusted channel. Photon pairs are generated via type-II SPDC, in a ppKTP crystal ($\SI{30}{mm}$-long, $\SI{46.2}{\micro\meter}$ poling period), and entangled in polarization in a Sagnac interferometer. The source is pumped with a $\SI{770}{nm}$ continuous laser. Signal and idler photons are emitted around $\SI{1540}{\nano\meter}$, separated from the pump by a dichroic mirror, and from each other by the polarizing beam splitter (PBS) of the interferometer. They are then coupled into single-mode fibers, and sent to the different players. The idler photon is both used as Alice's part of the maximally-entangled pair and to herald the probe state. The signal photon is sent to Bob through the untrusted lossy channel. A variable optical attenuator (VOA) allows to simulate an honest channel with a tunable amount of loss. The biphoton state is measured with polarization analyzers, each made of two waveplates (WPs), a fibered PBS, and $>80\%$-efficiency Superconducting Nanowire Single-Photon Detectors (SNSPDs). The WPs are mounted on motorized stages, allowing to both regularly randomize the measurement basis and implement dishonest channels. Detection events are then sent to a fast coincidence counter which gathers all the data required in order to evaluate the quantum correlations and channel's transmissivity.}
    \label{fig:XPSetup}
\end{figure*}

\noindent\textbf{Experimental implementation.} In order to test the feasibility of our protocol, we perform a proof-of-principle experiment based on photon pairs, emitted at telecom wavelength via type-II spontaneous parametric down-conversion (SPDC) in a periodically-poled KTP crystal (ppKTP). Photons are entangled in polarization thanks to a Sagnac interferometer~\cite{Fedrizzi2007}, encoding in this way a close-to-maximally entangled pair of qubits. Details of the setup are given in Fig.~\ref{fig:XPSetup}. 

The states emitted by the source are characterized at each iteration of the protocol via quantum state tomography \cite{James2001}, without inserting any untrusted quantum channel (green box in Fig.~\ref{fig:XPSetup}). Polarization analyzers (PA) are trusted for that task, as it is performed by Alice. This way we measured a fidelity of the probe's polarization state to a Bell state of $\overline{F^i} = 99.20\% \pm 0.02\%$ on average over all protocol attempts, with a maximum reached fidelity of $F^i = 99.43\% \pm 0.05\%$. We then send the probe states through an untrusted quantum channel. For this first demonstration we use a variable optical attenuator (VOA) in order to simulate a lossy but honest channel that requires certification. Detecting an idler photon in Alice's PA heralds a signal photon being sent through the quantum channel, which is then detected in Bob's PA. In each protocol attempt, the transmissivity is identified as the probability that Bob detects a state, knowing Alice heralded that state, and is also known as the heralding efficiency $\eta_s$:
\begin{equation}
    t(\mathcal{E}\vert\Phi_i) \simeq \eta_s = R_{si}/R_{i},
\end{equation}
where $R_{si}$ is the pair detection rate and $R_{i}$ the idler detection rate. We measure the pairs in random bases $A_0B_0$ or $A_1B_1$, and evaluate a close-to-maximum violation of steering inequality $\beta = 2-\epsilon$, with an average deviation ${\overline{\epsilon} = 1.42\cdot 10^{-2}}$, and a minimum deviation measured in a protocol $\epsilon_{\min} = 1.32\cdot 10^{-2}$.

\begin{figure}[htbp]
    \centering
\includegraphics[width=0.7\linewidth]{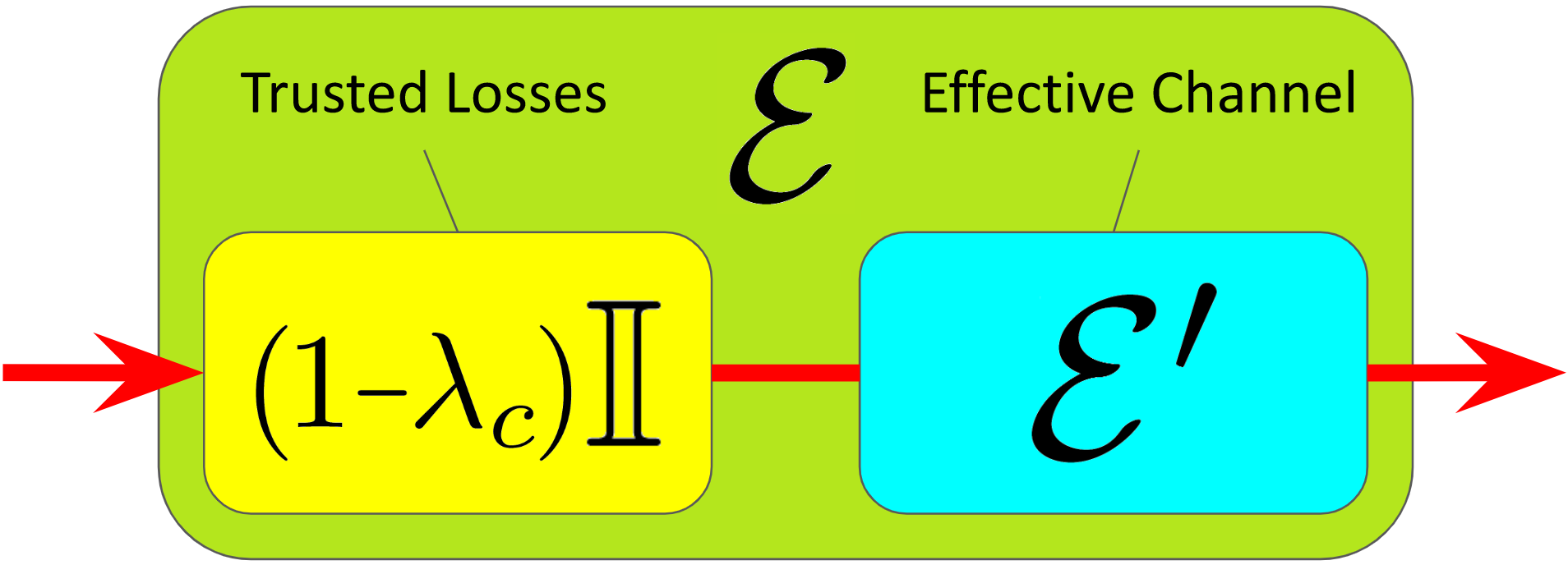}
    \caption{Schematic decomposition of the untrusted channel $\mathcal{E}$, into an equivalent channel $\mathcal{E}'$ that the protocol effectively certifies, and a trusted channel, corresponding to the characterized and homogeneous losses $\lambda_c$ trusted by Alice.}
    \label{fig:trustedLosses}
\end{figure}

For each protocol attempt we set a different transmissivity of the VOA, such that $\eta_s$ ranges from $21.9\%$ to $47.3\%$, the maximum value corresponding to the replacement of the VOA by a simple fiber connector. 
Following the 1sDI setting, Alice trusts her devices, so we are allowed to take losses originating from her equipment as trusted.
However, the experimental set up makes it difficult to distinguish between the source of losses. To allow for all cases we consider that a certain fraction of the losses is not induced by the channel itself, but by other components which are characterized by Alice, as part of the source. 
Such losses are considered homogeneous and trusted, so the channel reads 
\begin{equation}
    \mathcal{E} = (1-\lambda_c) \mathcal{E}',
\end{equation}
with $\lambda_c$ the amount of losses that is trusted and state-independent, and $\mathcal{E}'$ a quantum channel that is strictly equivalent to $\mathcal{E}$ by definition, and therefore returns the same output states; see Fig.~\ref{fig:trustedLosses}. In that case we can certify $\mathcal{E}'$ instead of $\mathcal{E}$, and evaluate the transmissivity in bound (\ref{ineq:MainResult3}) as
\begin{equation}
    t(\mathcal{E}'\vert\Phi_i) = t(\mathcal{E}\vert\Phi_i)/(1-\lambda_c) = \eta_s/(1-\lambda_c).
\end{equation}
This tightens the bound compared to the naive approach where all losses are attributed to the channel. Adopting this interpretation is quite realistic, considering that Alice preforms a full characterization of the probe states, which potentially includes a lower bound on the coupling losses. 
In the most paranoid scenario, we can always set $\lambda_c = 0$ we attribute all loss (including Alice's coupling and detection losses) to the quantum channel.

\begin{figure}[htbp]
    \centering
\includegraphics[width=1\linewidth]{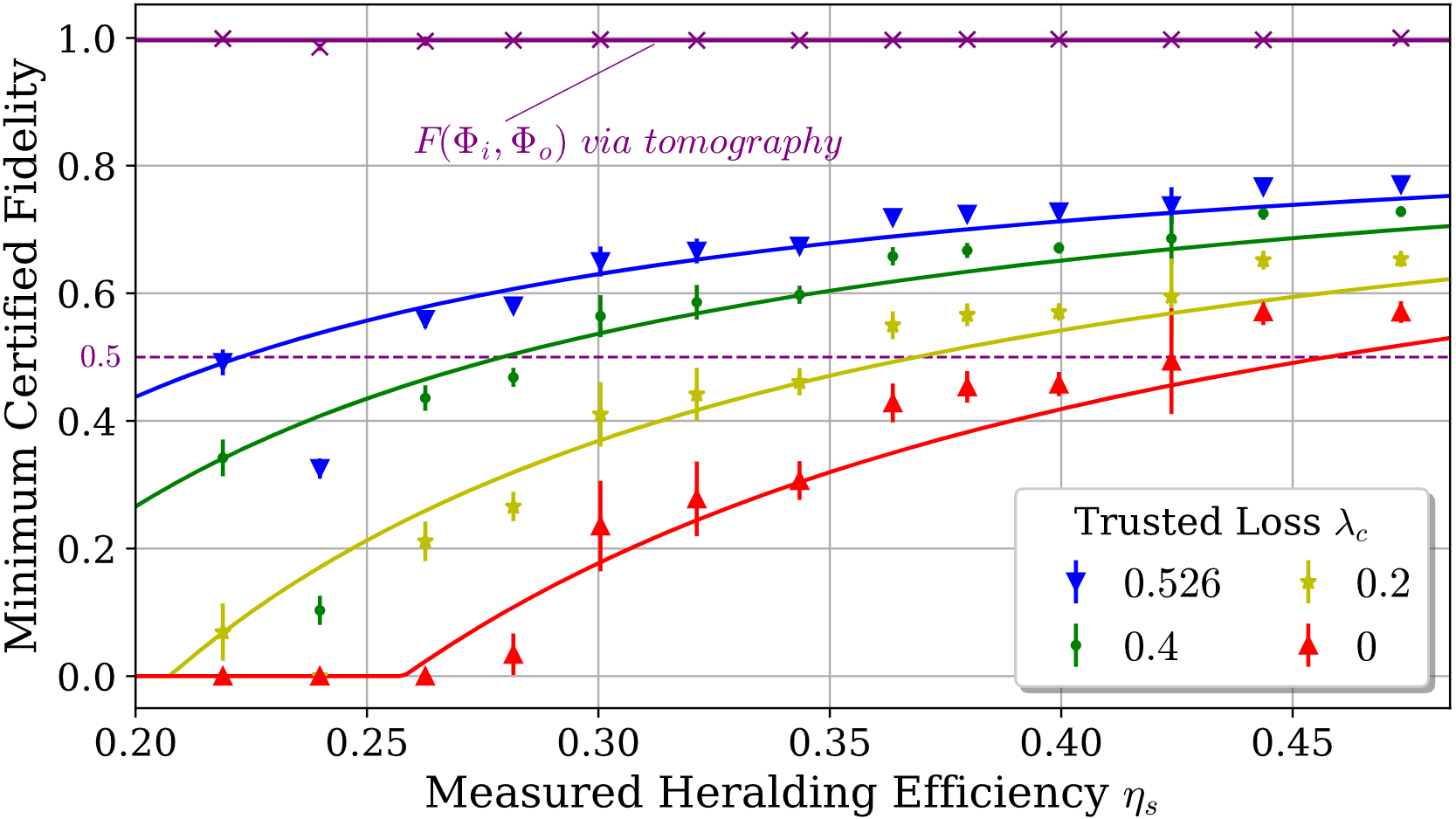}
    \caption{Minimum fidelity $F(\rho_i,\rho_o)$ certified via our protocol as a function of the measured heralding efficiency, tuned with a VOA, and for different trusted losses $\lambda_c$ (colored curves). The curves are plotted by taking the average fidelity of the probe state to a Bell state $\overline{F_i}$, and the average of the deviation from maximum violation $\epsilon$, over all protocol attempts. Experimental results deviate from these curves, as $F^i$ and $\epsilon$ vary between experiments. Errors induced by the finite statistics are directly subtracted from the certified fidelity, as detailed in Methods (see Eqs.~(\ref{eq:RelativeDifferencePassProbaMain}) and (\ref{ineq:FinalResultMain}) in particular). Error bars include effects induced by the unbalance in detectors' efficiency and the propagation of errors on $F^i$. We also display the fidelity $F(\rho_i,\rho_o)$ measured via quantum state tomography, for $\rho_i = \Phi_i$.}
    \label{fig:HonestResults}
\end{figure}

We show the results of our implementations in Fig.~\ref{fig:HonestResults}. Thanks to our close-to-maximum violation of steering inequality and relatively high coupling efficiency, we are able to certify the transmission of an unknown qubit state through the untrusted channel, with a non-trivial transmission fidelity $F(\rho_i,\rho_o)>50\%$. This is true even when Alice attributes all losses to the channel, \emph{i.e.} $\lambda_c = 0$, for channels with the highest transmissivities. The certified fidelity increases as Alice trusts a larger amount of homogeneous losses $\lambda_c$, reaching $F(\rho_i,\rho_o)\geq 77.1\%\pm 0.6\%$ when she assumes a maximum value $\lambda_c = 0.526$ and the channel is close to lossless. In any case, the certified fidelity decreases as the channel gets more lossy, as a direct consequence of bound~(\ref{ineq:MainResult3}), highlighting the difficulties of certifying lossy channels. This gives further motivation to assume that a fraction of the losses is trusted, in order to certify, for example, long-distance quantum communications. In our implementation, assuming maximum trusted losses $\lambda_c = 0.526$, we could certify a non-trivial transmission fidelity $F(\rho_i,\rho_o)>50\%$, for total transmissivities as low as $t(\mathcal{E}\vert\Phi_i) = \eta_s \simeq 0.263$, while such certification was possible only for $\eta_s \gtrsim 0.44$ with no trusted losses $\lambda_c = 0$.

In order to fully demonstrate the protocol, one should send a single quantum message $\rho_i$ through the channel, hidden among the probe states. The value of that state does not matter in our implementation as we do not use it in a later protocol, so we choose $\rho_i = \Phi_i$ and consider that a random copy of the probe state is actually the quantum message. To show the correctness of our protocol, we then perform a tomography of the corresponding transmitted message $\rho_o$ after the channel, and evaluate a transmission fidelity of $F(\rho_i,\rho_o) = 99.79\% \pm 0.02\%$ on average over all protocol attempts, with a minimum value of $F(\rho_i,\rho_o) = 98.7\%\pm 0.5\%$. This is far higher than the values certified by our protocol, as displayed on Fig.~\ref{fig:HonestResults}, which shows the state was indeed properly transmitted. Note that, in this case, the channel and measurement stations are trusted during the tomography of $\rho_o$, as it is performed outside of the protocol. This allows us to measure numerous copies of $\rho_o$, which is necessary for a full characterization of the state. In order to show that the correctness of our certification protocol would hold for other quantum messages $\rho_i$, we perform a full-process tomography of the quantum channel \cite{bongioanni2010experimental}, and lower-bound the fidelity between the physical channel and the identity $\mathcal{F}_\diamond(\mathcal{E},\mathds{I})\geq 94\%\pm 3\%$. We expect this bound to be far from tight, as it is evaluated using the equivalence between diamond and Choi-Jamio\l kowski distances \cite{Choi1975} (see Lemma~\ref{thm:MetricEquivalence} in Methods). Still, the fidelity is greatly above the values certified by our protocol, showing the certification procedure is indeed valid for any quantum message $\rho_i$.

The resilience of the protocol is further shown by experimentally simulating examples of dishonest channels. Let us first recall that the operator of the channel has no information on the position of the quantum message $\rho_i$ before the end of the protocol. This way, a typical attack consists in applying a disruptive transformation with small probability, hoping it will be applied to $\rho_i$ and stay undetected by Alice and Bob. Here we consider such a transformation to be a bit flip and/or a phase flip. For this experimental demonstration, we remove the VOA and consider that all losses are trusted. Note that performing a phase flip is equivalent to turning Bob's first measurement $B_0$ into $-B_0$:
\begin{equation}
    B_0 = M^{\mathcal{B}}_{0\vert 0}- M^{\mathcal{B}}_{1\vert 0} \longrightarrow -B_0 = M^{\mathcal{B}}_{1\vert 0}- M^{\mathcal{B}}_{0\vert 0} .
\end{equation}
Similarly, a bit flip is equivalent to turning Bob's second measurement $B_1$ into $-B_1$. Thus, we perform these flips in practice by randomly changing the waveplate angles in order to get the opposite measurement bases. This simulates dishonest channels of the form:
\begin{equation}\label{eq:MaliciousChannelExample}
\begin{aligned}
    \mathcal{E}_{p,q}[\rho] = (1-p)(1-q)\rho + p(1-q) X\rho X &\\&\hspace{-3cm}+ pq Y\rho Y   + (1-p)q Z\rho Z    ,
\end{aligned}
\end{equation}
with $p$ the bit flip probability and $q$ the phase flip probability.
\begin{figure}[htbp]
    \centering
\includegraphics[width=1\linewidth]{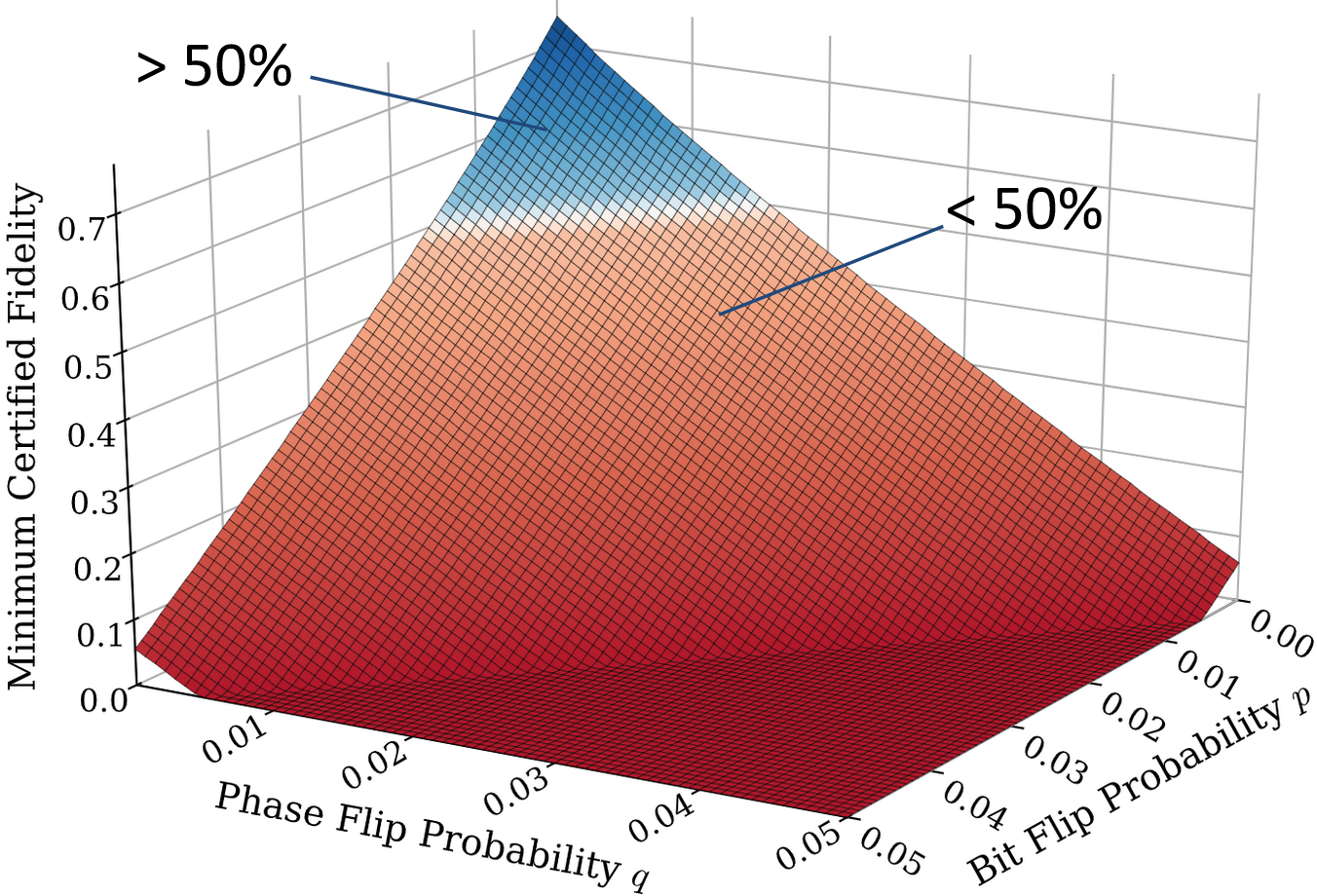}
    \caption{Minimum fidelity $F(\rho_i,\rho_o)$ certified via our protocol, for malicious channels $\mathcal{E}_{p,q}$, where $p$ is the probability of applying gate $X$ and $q$ is the probability of applying gate $Z$. Here we measured a probe state fidelity to a Bell state of $F^i = 99.16\%\pm 0.04\%$, and we trust a maximum amount of losses $\lambda_c = 0.526$.}
    \label{fig:MaliciousResults}
\end{figure}

The certification results are displayed in  Fig.~\ref{fig:MaliciousResults}, for different bit and phase flip probabilities. These show that our implementation is quite sensitive to these attacks, such that a flip probability of $0.01$ induces a collapse of $16\%$ of the certified fidelity, and we only certify $F(\rho_i,\rho_o) \geq 58\%$. The certified fidelity falls below the trivial value $50\%$ for flip probabilities as low as $0.017$. In this way, any attempt of Eve to disrupt the input state $\rho_i$ with such a method can only succeed with very small probabilities $p,q<0.02$, or it will be detected by Alice and Bob. \\ \

\large
\noindent\textbf{Discussion} 
\normalsize

\noindent In this work, we have provided a protocol to certify the transmission of a qubit through an untrusted and lossy quantum channel, by probing the latter with close-to-maximally entangled states and witnessing non-classical correlations at its output. In the DI case these are Bell correlations, in the 1sDI they are steering correlations.
Our theoretical investigations rely only on assumptions made on the probe state's source and the sender's measurement apparatus (in the case of 1sDI), while relaxing assumptions made on the quantum channel and the receiver's measurement apparatus. This setting proves to be an interesting trade-off between realistic experimental conditions and reasonable cryptographic requirements. It also embodies a practical scenario in which a strong server provides a weaker receiver with a quantum bit. 

Compared to previously proposed verification procedures, our protocol not only certifies the probed channels, but also an unmeasured channel through which a single unknown state can be sent. As quantum measurements deteriorate the quantum states, this task can only be performed at the price of measuring a huge amount of probe states, which limits the repeatability of the protocol with current technology. Until further theoretical considerations or technological improvements provide higher repeatability, our protocol can still serve as a practical primitive for other single-shot protocols that require a single quantum state, such as the recently demonstrated quantum weak coin-flipping \cite{Neves2023,bozzio_quantum_2020}.

Our proof-of-principle implementation shows the correctness of this certification procedure, and its feasibility with current technology. This way we could certify non-trivial transmission fidelities for a wide range of losses induced by the channel, by making some mild but realistic assumptions, such as the characterization of a fraction of trusted losses, induced for instance by the coupling of probe states inside optical fibers. By implementing random bit and phase flips, we could show that even a small probability attempt to disrupt the quantum information degrades the certified transmission fidelity, and is therefore detected by the players.

Future developments could demonstrate the feasibility of a fully device independent version of our protocol, in which Alice's measurement or even the probe states' source are not trusted. Such a protocol could be achieved by linking the probe state quality to that of the corresponding output state, or by making the IID assumption on the probe state's source. Also, more investigation on quantum-memory-based attacks could give a sharper idea on the possibilities of deceiving the certification procedure.

Our work opens the way to certification of a wide variety of more sophisticated lossy quantum channels. In particular, the rapid improvements of quantum technologies could soon provide possible applications of this protocol to the authentication of quantum teleportation, memories or repeaters.\\ \

\large
\noindent\textbf{Methods}
\normalsize

\noindent\textbf{Two Useful Lemmas.} The proof of bound (\ref{ineq:MainResult3}) relies on two lemmas, which give fundamental results on lossy quantum channels, and that we provide here.

\begin{restatable}[Extended Processing Inequality]{lem}{ExtProcess}
 For any probabilistic channel $\mathcal{E}$ (CPTD), and any input states $\rho_i$ and $\sigma_i$, the following inequality holds for the sine distance $C(\rho,\sigma) = \sqrt{1-F(\rho,\sigma)}$:
    \begin{equation}\label{eq:extmonotonicityDmain}
        C(\rho_i,\sigma_i) \geq t \cdot C(\rho_o,\sigma_o) ,
    \end{equation}
    where $\rho_o = \mathcal{E}[\rho_i]/t(\mathcal{E}\vert\rho_i)$ and $\sigma_o = \mathcal{E}[\sigma_i]/t(\mathcal{E}\vert\sigma_i)$
    are the output states of the channel, and  $t = t(\mathcal{E}\vert\rho_i)$ or $t = t(\mathcal{E}\vert\sigma_i)$.
    \label{thm:ExtProcess}
\end{restatable}
This first lemma generalizes to CPTD maps the well-known fidelity processing inequality $F(\rho,\sigma) \leq F(\mathcal{E}[\rho],\mathcal{E}[\sigma])$, which holds for any CPTP map $\mathcal{E}$. 

\begin{restatable}[Channel's Metrics Equivalence]{lem}{ChanMetrics}
\label{thm:MetricEquivalence} For any probabilistic channel $\mathcal{E}_1$, and any $\mathcal{E}_2$ that is proportional to a deterministic channel (CPTP map), both acting on $\mathcal{L}(\mathcal{H}_i)$, we have the following inequalities:
   \begin{equation}\label{ineq:equivSineMain}
    \mathcal{C}_J(\mathcal{E}_1,\mathcal{E}_2) \leq \mathcal{C}_\diamond(\mathcal{E}_1,\mathcal{E}_2) \leq \dim\mathcal{H}_i \times \mathcal{C}_J(\mathcal{E}_1,\mathcal{E}_2) ,
    \end{equation}
where the $\mathcal{C}_J$, resp. $\mathcal{C}_\diamond$, are the Choi-Jamio\l kowski, resp. diamond, sine distances of probabilistic quantum channels:
\begin{align}\label{eq:ChannelSineChoi}
 &\mathcal{C}_J(\mathcal{E}_1, \mathcal{E}_2) = C\Bigl(\dfrac{(\mathcal{E}_1\otimes\mathds{I})[\Phi_+]}{t(\mathcal{E}_1\vert\Phi_+)},(\mathcal{E}_2\otimes\mathds{I})[\Phi_+]\Bigr),\\
\label{eq:ChannelTraceDiamond}
&\mathcal{C}_\diamond(\mathcal{E}_1, \mathcal{E}_2)  =\sup_{\ket{\phi}} C\Bigl(\dfrac{(\mathcal{E}_1\otimes\mathds{I})[\phi]}{t(\mathcal{E}_1\vert\phi)},(\mathcal{E}_2\otimes\mathds{I})[\phi]\Bigr).
\end{align}
\end{restatable}
This lemma shows the equivalence between Choi-Jamio\l kowski and diamond distances, which is fundamental when trying to link the behaviour of the channel on a maximally-entangled state, to its behaviour on any quantum state. We also use this lemma in order to bound the diamond fidelity after performing a full process tomography of the channel, by evaluating the more straightforward Choi-Jamio\l kowski fidelity.

Note that both these lemmas also apply to the trace distance $D(\rho,\sigma) = \tfrac{1}{2}\tr\vert\rho -\sigma\vert$, and are proven in SUPP. MAT.~\ref{sec:MetricMonotonicity} and~\ref{sec:MetricEquiv}.

\medskip

\noindent\textbf{Protocol Security.}
In our protocol, the quantum channel is allowed to evolve through time, with some potential memory of the experiment's past history. This way we define the channel $\mathcal{E}_{k\vert [k-1]}$, where $[k-1] = k-1,k-2,...,1$, that operates on the $k$-th state sent by Alice through the protocol. In particular, Alice sends the quantum message $\rho_i$ at a random position $r$ through channel $\mathcal{E}_{r|[r-1]}$. We then define the expected channel over the protocol: 
\begin{equation}\label{eq:AverageChannelDefMain}
\bar{\mathcal{E}} = \frac{1}{N+1}\sum_{k=1}^{N+1}\mathcal{E}_{k|[k-1]} .
\end{equation} 
As $\rho_i$ is sent at a random position that stayed concealed from the channel's operator, the expected transmitted message is $\bar{\rho}_o = (\bar{\mathcal{E}}\otimes\mathds{I})[\rho_i]/t(\bar{\mathcal{E}}|\rho_i)$. As long as $r$ stays hidden and random, any measurement performed on the transmitted message later after the protocol would follow the same statistics as if it was performed on $\bar{\rho}_o$ (see SUPP. MAT. \ref{sec:avgstate} for more details). 
This way, we derive the protocol security by applying bound~(\ref{ineq:MainResult3}) to the average channel $\bar{\mathcal{E}}$, in order to bound the fidelity of $\bar{\rho}_o$ to $\rho_i$, up to isometry. In particular, the output probe state fidelity to a maximally entangled state now reads
\begin{equation}
\begin{aligned}
    F^o = F\bigl((\Lambda^{\mathcal{B}}\otimes\Lambda^{\mathcal{A}_2})[(\mathcal{E}\otimes\mathds{I})[\Phi_i]]/t(\bar{\mathcal{E}}\vert\Phi_i),(\mathcal{E}_0\otimes\mathds{I})[\Phi_+]\bigr).
\end{aligned}
\end{equation}

Using recent self-testing results in a non-IID setting \cite{unnikrishnan2019} applied to the output probe state, we show in SUPP. MAT.~\ref{protocolSecurity} that for any $x > 0$, $C^o = \sqrt{1-F^o}$ can be bounded by two terms, with confidence of at least ${c_x = (1-e^{-x})\cdot (1-2e^{-x})^2}$:
\begin{equation}\label{ineq:boundOutput}
   \arcsin C^o \leq \arcsin\sqrt{\alpha f_x(\epsilon,K)} + \Delta_x(\eta_s,K),
\end{equation}
where $K$ is the number of pairs measured by Alice and Bob, $\eta_s$ is the measured heralding efficiency, $\Delta_x(\eta_s,K)$ is an error function that goes to $0$ for high values of $K$, $\alpha f_x$ gives self-testing bound on the output state, in a non-IID regime, with
\begin{equation}\label{eq:STfunctionAnu}
    f_x(\epsilon,K) = 8\sqrt{\dfrac{x}{K}} + \dfrac{\epsilon}{2} + \dfrac{\epsilon + 8/K}{2 + 1/K} \xrightarrow[K\rightarrow +\infty]{} \epsilon,
\end{equation}
and $\alpha = 1.26$. We choose $x=7$ to get a confidence $c_x > 99.5\%$, and measure $K \simeq 10^9$ copies of the probe state, in order to reach the asymptotic values, which takes from 1 to 3 hours in our experiments depending on the channel transmissivity. Note that the error function is due to both the non-IID regime and the lack of information on channels that do not output any state. A similar error occurs when we evaluate the transmissivity as the measured heralding efficiency:
\begin{equation}\label{eq:RelativeDifferencePassProbaMain}
   t(\bar{\mathcal{E}}\vert\Phi_i) \gtrsim \tau_x(\eta_s,K),
\end{equation}
where $\tau_x(\eta_s,K) \simeq \eta_s$ for high values of $K$. This way, the actual bound on the fidelity between the input and output state reads, with confidence $c_x$,
\begin{equation}\label{ineq:FinalResultMain}
\begin{aligned}
    F(\bar{\rho}_o,\rho_i) \geq 1 - &\:4\cdot\sin^2\biggl( \arcsin\bigl(C^i/\tau_x\bigr) +\\ &\hspace{1cm}\arcsin\sqrt{\alpha f_x(\epsilon,K)} + \Delta_x\biggr) ,
\end{aligned}
\end{equation}
which includes additive error terms compared to bound~(\ref{ineq:MainResult3}). In the analysis of our data, we include these terms that are minimized thanks to the large number $K$ of states measured for each implementation. Note that the expressions for all the mentioned functions are detailed in SUPP. MAT.~\ref{StatError}.

\medskip

\noindent\textbf{Assumptions.} 
For clarity we highlight the assumptions made in our security analysis. 

First, we assume Alice and Bob can communicate via a trusted private classical channel. It allows the players to agree on their measurement settings, Alice to send Bob the position $r$ of the quantum message $\rho_i$, and Bob to tell Alice if the states were properly received. This way, the players can perform measurements on the fly, instead of storing all the states, then deciding of the measurement bases and finally measuring the states, which would require one billion of quantum memories with hours-long storage-time.

Secondly, the fair sampling assumption is required on the measurement apparatus for the self-testing procedure, as we allow a large amount of losses to be induced by the quantum channel. Alice's measurement apparatus is completely trusted and characterized, according to the one-sided device independent scenario. On Bob's side, we assume the efficiency of the measurement apparatus to be independent of the measurement setting $B_0$ or $B_1$. If the efficiency depends on the state measured, then we consider that dependence to be part of the quantum channel. A slight unbalance of efficiency is allowed between the two different measurement outcomes, and we show in the SUPP. MAT. \ref{FairSampling} that the error induced by this unbalance is negligible. 

Finally, in keeping with the 1sDI setting, we make the IID assumption on the probe state source, during each attempt of the protocol. To show the legitimacy of this assumption in our implementation, we performed a series of quantum state tomography measurements, during 8 hours, in order to characterize the fluctuation of the probe state with time. This characterization shows the probe states are stable at the scale of one protocol (see SUPP. MAT.~\ref{sourceDetails} for the detailed results). 


\medskip

\noindent\textbf{Source and Detection.} Probe states are generated via type-II SPDC in a ppKTP crystal combined with a Sagnac interferometer. We maximized the heralding efficiency $\eta_s = R_{si} / R_i$, with $R_i$ the idler photon detection rate and $R_{si}$ the pair detection rate, following the method proposed in \cite{bennink2010optimal,bruno2014pulsed}. For that purpose, the pump's spatial mode and focus as well as the pair's collection modes, were tuned carefully when coupling to single-mode fibers, and losses on the signal photon path were minimized. This way the pump is in a collimated mode at the scale of the crystal, close to a gaussian mode of waist $w_p\simeq\SI{315}{\micro\meter}$, which maximizes the heralding efficiency \cite{guerreiro2013high,bruno2014pulsed}. The signal photon's coupling mode has a waist $w_s \simeq \SI{190}{\micro\meter}$, and the idler photon's is $w_i \simeq \SI{218}{\micro\meter}$. We also used high-efficiency SNSPDs to detect the photons. Losses on the idler photon were not limiting, so we selected the best components and detectors for the signal photon. All detection events were recorded by a time tagger, and dated with picosecond precision. Two detection events were considered simultaneous when measured within the same $\SI{500}{ps}$ coincidence window. In this way, we detect idler photons in Alice's detectors with a rate $R_i = \SI{600\pm 40}{\kilo\hertz}$ (varying from one protocol attempt to another), for a brilliance of $\simeq \SI{670\pm 50}{\kilo\hertz\per\watt\per\nano\meter}$. SNSPDs display dark count rates of $\leq \SI{500}{\hertz}$, such that the probability of falsely heralding a probe state is negligible. Finally, $\SI{1}{\nano\meter}$-bandwidth spectral filters were used to limit the spectrum spread that would otherwise degrade the polarization state because of birefringence and dispersion in optical fibers.

\medskip

\noindent\textbf{Quantum State Tomography.} We perform quantum state tomographies via linear regression estimation \cite{qi2013quantum} and fast maximum likelihood estimation \cite{smolin2012efficient}. Photon counts are corrected by measuring relative efficiencies of the detectors. We use this method in order to reconstruct the probe state $\Phi_i$, and to calculate the probe state fidelity to a maximally entangled state $F^i$. For this calculation, we maximize the fidelity
\begin{equation}
    F^i_{U} = F\bigl((\mathds{I}\otimes U)\Phi_i (\mathds{I}\otimes U^\dagger),\Phi_+\bigr)
\end{equation}
on a local unitary $U$, to evaluate the maximum fidelity up to isometries, as defined in Eq.~(\ref{eq:inputFid}).

The uncertainties on the reconstructed states, induced by the photon counting poissonian statistics as well as by the systematic errors on the measurement bases, are evaluated by using the Monte Carlo method. This way, we simulate 1000 new data samples within the respective uncertainties distributions and reconstruct new density matrices from which we evaluate the average fidelity and standard deviation \cite{altepeter2005photonic}. Slow thermal fluctuation also induce some uncertainty on the fidelity, as our experiment lasts for a relatively long period of time. By continuously performing quantum state tomographies for 8 hours, we are able to evaluate the fluctuations in the quantum state on time spans of the order of a protocol duration. This way, we measure an additional $0.02\%$ error on the quantum state fidelities to Bell states, due to thermal fluctuations. The reader can refer to SUPP. MAT. \ref{sourceDetails} for more details on the evaluation of these thermal fluctuations and the drift of the quantum state through time.

\medskip

\noindent\textbf{Steering measurement.} When testing the violation of steering inequality, players should in principle pick a random measurement basis between $A_0B_0$ and $A_1B_1$ for each new photon pair. However, because of technical limitations of our motorized waveplate stages, we only operate this randomization at a limited rate of $\SI{1}{\hertz}$. A fully secure protocol would therefore require faster electronics and active optical components. 

For the implementation of malicious channels, we perform a 7-hours measurement run. From this single run we generate the data that could be acquired in the certification procedure of a variety of channels $\mathcal{E}_{p,q}$, as defined in Eq.~(\ref{eq:MaliciousChannelExample}). For this run, we randomize the measurement basis, with equal probabilities between $A_0B_0$, $A_1B_1$ (the channel chooses to act honestly), and $-A_0B_0$, $-A_1B_1$ (the channel chooses to disrupt the state). In order to simulate a larger variety of data samples, we perform that randomization at a $\SI{5}{\hertz}$-rate. We then generate the data for the certification of channel $\mathcal{E}_{p,q}$, by picking a random set of samples, with the following proportions:
\begin{itemize}
    \item $q/2$ in basis $-A_0B_0$,
    \item $p/2$ in basis $-A_1B_1$,
    \item $(1-q)/2$ in basis $A_0B_0$,
    \item $(1-p)/2$ in basis $A_1B_1$.
\end{itemize}
The data acquired in basis $-A_0B_0$ and $-A_1B_1$ is treated as if it was acquired in basis $A_0B_0$ and $A_1B_1$, respectively, when calculating the average violation of steering inequality $\beta = \vert\langle A_0B_0\rangle + \langle A_1B_1\rangle\vert$.\\\

\medskip

\noindent\textbf{Note added.} While finishing this manuscript we became aware of a related work by Bock \emph{et al.}~\cite{calibration}.
\medskip

\large
\noindent\textbf{Acknowledgments}
\normalsize
\noindent We acknowledge useful discussions with Anupama Unnikrishnan and Massimiliano Smania on self-testing techniques and assistance from IDQuantique with the single-photon detectors. We also acknowledge financial support from the European Research Council project QUSCO (E.D.), the PEPR integrated projects EPiQ ANR-22-PETQ-0007 and DI-QKD ANR-22-PETQ-0009, which are part of Plan France 2030, and the Marie Sk\l odowska-Curie grant agreement No 956071 (AppQInfo).\\ \

\medskip
\large
\noindent\textbf{Author contributions}\\
\normalsize
S.N.\ developed the theoretical protocols and proofs, together with I.S.,\ D.M.\ and E.D.\ S.N.\ and E.D.\ conceived the experimental setup, S.N.\ developed it, and S.N.,\ L.M.,\ V.Y.\ and P.L.\ performed the protocol implementation. S.N.\ and L.M.\ processed the data. All authors discussed the analysis of the data, and contributed to writing or proofreading the manuscript. D.M.\ and E.D.\ supervised the project.\\ \

\onecolumngrid

\newpage
\appendix

\section*{Supplementary Material}

In addition to the results presented in the main text, we provide the following material in order to prove our different theoretical results and present more experimental details. We also show some interesting theoretical results related to our study, though they are not essential for its understanding. The outline for this material is the following.\\

In appendix \ref{seq:Prel} we give some important definitions, including that of general quantum channels, including lossy channels, equivalence classes of channels, and channels metrics.\\

In appendix \ref{app:Fund} we show some new fundamental results, such as Lemma~\ref{thm:ExtProcess}, \textit{i.e.}, the processing inequality of general lossy channels, Lemma~\ref{thm:MetricEquivalence}, \textit{i.e.}, equivalence inequalities between different metrics of quantum channels, and some useful result on channels' transmissivity. \\

In appendix \ref{app:TheoProts} we provide the detailed theoretical recipes for channel certification protocols, in a one-sided device independent and in a fully device independent. Both these recipes are detailed in the spirit of those provided in \cite{unnikrishnan2019} for authenticated teleportation, and differ slightly from the protocol that we experimentally implement. In particular, the former rely on trusted quantum memories for storing all states sent by Alice, while the latter rely on trusted private classical communications between Alice and Bob.\\

In appendix \ref{protocolSecurity} we use the results of previous paragraphs in order to derive security bounds for our protocols. We first show bound (\ref{ineq:MainResult3}), which relies on the evaluation of the fidelity of a probe state to a maximally entangled state, and the fidelity of the corresponding output state after the channel to the same maximally entangled state. This bounds the fidelity between any state that outputs a quantum channel and the corresponding unknown input state. In the second part of that paragraph, we show how to evaluate the two probe states' fidelities up to isometries, even when no IID assumption is made and the state source might be untrusted. This method relies on self-testing of steering inequalities in a semi-device independent scenario, where Alice's measurement setup is trusted. Still, this method requires the measurement of a large sample of close-to-maximally entangled states, going through a channel that might evolve through time. In particular, the channel might not have the same action on the probe states than on the transmitted state. Therefore, we give some important statistical development in the next part of the paragraph, in order to bound the errors made on the different evaluated fidelities, due to finite state sample in a non-IID setting, as well as losses in the untrusted channel. Finally, we tie up the security proof, combining the previous parts' results in order to provide a bound on the expected fidelity of the transmitted output state to the input state. We then give some way to generalize that security proof to a fully-device independent setting. \\

In appendix \ref{app:ExpProt}, we give additional details on our experimental implementation. In particular, we provide some developments on the probe state source, such as the density matrix of a state emitted by that source, and a characterization of the stability of our source, motivating the IID assumption. We also detail the results of measurements performed during our implementations of the protocol, from which we deduce the bound on the transmission fidelity. We also formalize the fair-sampling assumptions made on the players' measurement apparatus, and discuss the influence of a slight unbalance in the detectors efficiency, which we observe in our experiments.\\
\newpage

\section{Preliminary Definitions}\label{seq:Prel}

In this study, we use the quantum operations formalism \cite{bible}, in order to describe as generally as possible the transformations undergone by quantum states. Such a formalism allows us to include a variety of processes, such as unitary transformations, quantum measurements and ancillary inclusion. Although most studies consider only trace-preserving quantum channels, \textit{i.e.} lossless channels, our study requires the consideration of trace-decreasing channels that account for potentially lossy devices. This section is meant to clarify some important definitions and properties linked to these channels, as well as discuss the physical reality embodied in these mathematical objects. \\

\subsection{Quantum Channels}

A general \textit{quantum channel} $\mathcal{E}$ is a convex, linear and completely-positive non-trace-increasing (CPnTI) map, from operators on space $\mathcal{H}_i$ to operators on space $\mathcal{H}_o$ \textit{i.e.}: 
    \begin{enumerate}
        \item For any sets of probabilities $\{p_i\}$ and density operators $\{\rho_i\}$, the following equality holds:
        
        \begin{equation}\label{eq:operationLinearity}
            \mathcal{E}\bigg[\sum_i p_i \rho_i\bigg] = \sum_i p_i\mathcal{E}[\rho_i]
        \end{equation} 
        
        \item For any secondary system of Hilbert space $\mathcal{S}$, $(\mathcal{E}\otimes\mathds{I}_\mathcal{S})[K]$ is positive for any positive operator $K$ taken in $\mathcal{L}(\mathcal{H}_i\otimes\mathcal{S})$. In particular, $\mathcal{E}$ is completely positive.
        
        \item For any operator $K$ acting on $\mathcal{H}_i$, we have $\tr \mathcal{E}[K]\leq \tr K$.
    \end{enumerate}
    When $\mathcal{E}$ is also \textit{trace-preserving} (CPTP map), in particular $\tr \mathcal{E}[\rho] =1$ for any density operator $\rho$, then we call $\mathcal{E}$ a \textit{deterministic} or \textit{lossless} quantum channel. Otherwise, if the map is \textit{trace-decreasing}, then there exists a state $\rho$ such that $\tr \mathcal{E}[\rho]<1$, we call it a \textit{probabilistic} or \textit{lossy} quantum channel. From this definition can be derived the well known Kraus' theorem, that gives a complete characterization of quantum channels:
\begin{thm}[Kraus' Theorem]
    The map $\mathcal{E}$ from $\mathcal{L}(\mathcal{H}_i)$ to $\mathcal{L}(\mathcal{H}_o)$ is a quantum channel if and only if there exist a set of operators $\{K_j\}_j$ that map $\mathcal{H}_i$ to $\mathcal{H}_o$, such that: 
    \begin{equation}\label{eq:KrausTheorem}
        \mathcal{E}[\rho_i] = \sum_jK_j\rho_iK_j^\dagger
    \end{equation}
    and $\sum_jK_j^\dagger K_j \leq \mathds{I}$. $\mathcal{E}$ is a deterministic quantum channel when this condition holds and $\sum_jK_j^\dagger K_j = \mathds{I}$. When $\sum_jK_j^\dagger K_j < \mathds{I}$, the channel is probabilistic.
\end{thm}

This theorem gives us an operator-sum representation for quantum channels, which will be most useful in the following. The operators $\{K_j\}$ are refered to as \textit{Kraus' operators} of the channel $\mathcal{E}$. \\

The previous axioms and properties imply that for any density operator $\rho \in \mathcal{L}(\mathcal{H}_i\otimes\mathcal{S})$, with $\mathcal{S}$ an arbitrary Hilbert space, we have $0 \leq \tr((\mathcal{E}\otimes\mathds{I})[\rho]) \leq 1$. This means that in the most general case, $(\mathcal{E}\otimes\mathds{I})[\rho]$ is not a density operator. This way, our channel does not operate with absolute certainty, but returns a state only with a certain probability $t(\mathcal{E}\vert\rho) =\tr(\mathcal{E}\otimes\mathds{I})[\rho]$. We call $t(\mathcal{E}\vert\rho)$ the \textit{transmissivity} of channel $\mathcal{E}$. Then for $t(\mathcal{E}\vert\rho) \neq 0$ we define the output state:
    \begin{equation}\label{eq:outputStateDef}
        \rho_o = (\mathcal{E}\otimes\mathds{I})[\rho]/t(\mathcal{E}\vert \rho)
    \end{equation}
and when $t(\mathcal{E}\vert \rho)=0$, \textit{i.e.} no state ever outputs the channel, we set by convention $\rho_o = \mathds{I}/\dim(\mathcal{H}_o\otimes\mathcal{S})$.\\

Quantum channels are fundamental objects that describe any transformation undergone by a quantum state. Still, most studies focus on lossless quantum channels \textit{i.e.} CPTP maps, such that any state passes the channel with absolute certainty. In theory, any situation involving a lossy channel can be described by considering a CPTP map $\mathcal{E}[\bullet] = \mathcal{E}_s[\bullet]\otimes\proj{s} + \mathcal{E}_f[\bullet]\otimes\proj{f}$, with $\mathcal{E}_s$ the \textit{successful} branch and $\mathcal{E}_f$ the \textit{failure} branch, where the state might be considered as lost. However in most experimental situations, we generally have no access to the state when it goes through the failure branch, such that we are only interested in states sent through the success branch. This means we \textit{post-select} states on the success branch, and we only consider the probabilistic channel $\mathcal{E}_s[\rho] = \bra{s}\mathcal{E}[\rho]\ket{s}$. The transmissivity is then the probability that the channel successfully outputs the input state, so that $t(\mathcal{E}_s\vert\rho) = \tr\mathcal{E}_s[\rho] = \tr(\mathcal{E}[\rho]\mathds{I}\otimes\proj{s})$. This way, losses are included in the expression of the channel itself.

Finally we give a few common examples of probabilistic quantum channels. A trivial probabilistic quantum channel is $\mathcal{E} = p\cdot \mathds{I}$ with $p\in\:]0;1]$, that models unbiased losses. In that case the state is simply transmitted without transformation with probability $p$, or lost with probability $1-p$. On the contrary, a channel with fully-biased losses would be a polarizing channel $\mathcal{P}$, with  $\mathcal{P}[\rho] = \proj{\phi}\:\rho\:\proj{\phi}$ for any state $\rho$, with $\ket{\phi}$ a pure state. In that case $t(\mathcal{P}\vert\rho) = 1$ if and only if $\rho = \proj{\phi}$. Finally, probabilistic channels allows us to describe an experiment where one wishes to measure a POVM $\{M_k\}_{1\leq k \leq d}$ but only has access to the first $m$ elements, with $m<d$. We can therefore define the following channel:
\begin{equation}\label{eq:partialPOVM}
    \mathcal{E}[\rho] = \sum_{i=1}^m M_k\rho M_k^\dagger\otimes\proj{k}
\end{equation}
This example is of particular use for Bell state measurements using linear optics, where it was shown that one can measure only two elements out of four~\cite{lutkenhaus_bell_1999}.

\subsection{Equivalence Classes of Quantum Channels}\label{equivalclasses}

Let us consider two channels $\mathcal{E}_1$ and $\mathcal{E}_2$ that are proportional to each other, \textit{i.e.} there exists a factor $p\in\: ]0;1]$ such that $\mathcal{E}_1 = p\cdot\mathcal{E}_2$ (or $\mathcal{E}_2 = p\cdot\mathcal{E}_1$ which is a symmetric case). Then their corresponding transmissivities also display the same proportionality $t(\mathcal{E}_1\vert\rho) = p\cdot t(\mathcal{E}_2\vert\rho)$ for any input state $\rho$. The two channels therefore output the same states when fed the same input state:
\begin{equation}\label{eq:outputStateEquiv}
    \frac{\mathcal{E}_1[\rho]}{t(\mathcal{E}_1\vert\rho)} = \frac{p\cdot\mathcal{E}_2[\rho]}{p\cdot t(\mathcal{E}_2\vert\rho)} = \frac{\mathcal{E}_2[\rho]}{t(\mathcal{E}_2\vert\rho)}
\end{equation}
In numerous practical situations, such as those described in this study, we only consider what happens when the states are not lost, such that we post-select on the states being detected. This way, two channels $\mathcal{E}_1$ and $\mathcal{E}_2$ that are proportional to each other actually describe the same physical situation, and we consider them as equivalent $\mathcal{E}_1 \equiv \mathcal{E}_2$. This defines mathematical equivalence classes of channels that outputs the same quantum states. All channels from a same class can be compared, such that if $\mathcal{E}_1 \equiv \mathcal{E}_2$, then either $\mathcal{E}_1 \geq \mathcal{E}_2$ or $\mathcal{E}_2\geq \mathcal{E}_1$. In the first case, for instance, we have $t(\mathcal{E}_1\vert\rho) \geq t(\mathcal{E}_2\vert\rho)$. For any class of channel, we can find a maximal channel of that class $\mathcal{E}_{max}$ such that $\mathcal{E}_{max} \geq \mathcal{E}$ for any channel $\mathcal{E}$ of the same class. That maximal channel is therefore the most transmissive channel, and there always exists a state $\rho$ that passes the channel with absolute certainty, \textit{i.e.} $t(\mathcal{E}_{max}\vert\rho) = 1$. 

These equivalence classes are of particular interest in our study, as the distances we use do not rigorously define metrics for arbitrary quantum channels, but they do for these classes of quantum channels. They also embody the fact that when certifying a channel $\mathcal{E}$, one can always consider a more transmissive but equivalent channel $\mathcal{E}'$, with $\mathcal{E}' \geq \mathcal{E}$ and $\mathcal{E}' \equiv \mathcal{E}$. We can then use this more transmissive channel in order to describe the physical process, which falls down to assuming a certain amount of losses are trusted, as described in Fig.~\ref{fig:trustedLosses} of the main text.

\subsection{Metrics of Quantum Channels}

We first define the diamond and Choi-Jamio\l kowski trace and sine distances between two channels $\mathcal{E}_1$ and $\mathcal{E}_2$ acting on a space $\mathcal{L}(\mathcal{H}_i)$:
\begin{align}\label{eq:DefinitionMetricDiamond}
    \mathcal{M}_\diamond(\mathcal{E}_1,\mathcal{E}_2) &= \sup_{\ket{\phi}}M\bigl((\mathcal{E}_1\otimes\mathds{I})[\phi]/t(\mathcal{E}_1\vert\rho),(\mathcal{E}_2\otimes\mathds{I})[\phi]/t(\mathcal{E}_2\vert\rho)\bigr)\\
\label{eq:DefinitionMetricChoi}
    \mathcal{M}_J(\mathcal{E}_1,\mathcal{E}_2) &= M\bigl((\mathcal{E}_1\otimes\mathds{I})[\Phi_+]/t(\mathcal{E}_1\vert\Phi_+),(\mathcal{E}_2\otimes\mathds{I})[\Phi_+]/t(\mathcal{E}_2\vert\Phi_+)\bigr)
\end{align}
where $M = D$ or $C$ are the trace and sine distances, $\Phi_+$ is a maximally-entangled state, and the upper bound is taken over pure states $\ket{\phi} \in \mathcal{H}_i^{\otimes 2}$ such that $t(\mathcal{E}_1\vert\phi) \neq 0$ and $t(\mathcal{E}_2\vert\phi) \neq 0$ . These quantities are proper distances only when restricted to deterministic quantum channels, \textit{i.e.} CPTP maps, in which case $M_J(\mathcal{E}_1,\mathcal{E}_2) = 0$ when $M_\diamond(\mathcal{E}_1,\mathcal{E}_2) = 0$, or when $\mathcal{E}_1 = \mathcal{E}_2$. Concerning probabilistic channels, we show that $M_J(\mathcal{E}_1,\mathcal{E}_2) = M_\diamond(\mathcal{E}_1,\mathcal{E}_2) = 0$ if and only if $\mathcal{E}_1 \equiv \mathcal{E}_2$ and the channels are equivalent, in the sense we defined in section~\ref{equivalclasses}, meaning they are proportional to each other. \\

\textbf{\textit{Proof.}} If $\mathcal{E}_1 \equiv \mathcal{E}_2$, then there exists $p\in\: ]0;1]$ such that $\mathcal{E}_1 = p\cdot \mathcal{E}_2$ or $\mathcal{E}_2 = p\cdot \mathcal{E}_1$. Then by definition of $\mathcal{M}_\diamond$ and $\mathcal{M}_J$, we trivially have $\mathcal{M}_\diamond(\mathcal{E}_1,\mathcal{E}_2) = \mathcal{M}_J(\mathcal{E}_1,\mathcal{E}_2) = 0$. Now let us assume $\mathcal{E}_1$ and $\mathcal{E}_2$ are non-zero channels such that $\mathcal{M}_J(\mathcal{E}_1,\mathcal{E}_2) = 0$, and let us show that $\mathcal{E}_1\equiv\mathcal{E}_2$. First, we formulate the following lemma, implicitly introduced earlier in \cite{Sekatski2018}:

\setcounter{lem}{2}

\begin{lem}\label{lemma:bellStateToAny}
 Let $\ket{\psi}\in\mathcal{H}^{\otimes 2}$ be a pure 2-qudits state, with $\dim \mathcal{H} = d$. Then there exists an operator $K_\psi = M_\psi U_\psi$ on $\mathcal{H}$, with $0<M_\psi\leq\mathds{I}$ and $U_\psi$ a unitary, such $\mathds{I}\otimes K_\psi$ transforms the maximally-entangled state $\ket{\Phi_+}=\tfrac{1}{\sqrt{d}}\sum_{i=0}^{d-1}\ket{i}\ket{i}$ into $\ket{\psi}$ with probability $1/d$, \textit{i.e.}:
\begin{equation}
    (\mathds{I}\otimes K_\psi)\ket{\Phi_+} = \dfrac{1}{\sqrt{d}}\ket{\psi}
\end{equation}
\end{lem}
We remind the proof of this lemma, which was detailed in \cite{Sekatski2018}. We use the Schmidt decomposition of $\ket{\psi}$:
\begin{equation}\label{eq:SchmidtDecompDistanceProof}
    \ket{\psi} = \sum_{i=0}^{d-1} \psi_i\ket{i}\ket{i'}
\end{equation}
where $\{\ket{i}\}$ and $\{\ket{i'}\}$ are two orthonormal bases of $\mathcal{H}$. There exists a unitary operator $U_\psi$ acting on $\mathcal{H}$ such that:
\begin{equation}
    (\mathds{I}\otimes U_\psi)\ket{\Phi_+} = \dfrac{1}{\sqrt{d}}\sum_{i=0}^{d-1} \ket{i}\ket{i'} 
\end{equation}
with $d = \dim \mathcal{H}$. We can then define the operator $M_\psi$ that probabilistically transforms $(\mathds{I}\otimes U_\psi)\ket{\Phi_+}$ into $\ket{\psi}$:
\begin{equation}
    M_\psi = \sum_{i=0}^{d-1}  \psi_i\proj{i'}
\end{equation}
Now by we defining the operator $K_\psi = M_\psi U_\psi$, we have:
\begin{equation}
  (\mathds{I}\otimes K_\psi)\ket{\Phi_+} = \tfrac{1}{\sqrt{d}}\ket{\psi} 
\end{equation}
which completes the proof of the lemma.\\ \

From here, as we have $\mathcal{M}_J(\mathcal{E}_1,\mathcal{E}_2) = 0$, then $M\bigl((\mathcal{E}_1\otimes\mathds{I})[\Phi_+]/t(\mathcal{E}_1\vert\Phi_+), (\mathcal{E}_2\otimes\mathds{I})[\Phi_+]/t(\mathcal{E}_2\vert\Phi_+)\bigr) = 0$ which implies:
\begin{equation}\label{eq:DistanceProof2}
(\mathcal{E}_1\otimes\mathds{I})[\Phi_+] = \tfrac{t(\mathcal{E}_1\vert\Phi_+)}{t(\mathcal{E}_2\vert\Phi_+)} \cdot (\mathcal{E}_2\otimes\mathds{I})[\Phi_+]
\end{equation}
For any pure state $\ket{\psi} \in \mathcal{H}^{\otimes 2}$ we define the operator $K_\psi$ from Lemma \ref{lemma:bellStateToAny}, such that $(\mathds{I}\otimes K_\psi)\ket{\Phi_+} = \tfrac{1}{\sqrt{d}}\ket{\psi}$. We can apply that operator on both sides of equation (\ref{eq:DistanceProof2}):
\begin{equation}\label{eq:DistanceProof3}
    (\mathds{I}\otimes K_\psi)(\mathcal{E}_1\otimes\mathds{I})[\Phi_+] (\mathds{I}\otimes K^\dagger_\psi)=\tfrac{t(\mathcal{E}_1\vert\Phi_+)}{t(\mathcal{E}_2\vert\Phi_+)} \cdot (\mathds{I}\otimes K_\psi)(\mathcal{E}_2\otimes\mathds{I})[\Phi_+](\mathds{I}\otimes K^\dagger_\psi)
\end{equation}
which, since $\mathds{I}\otimes K_\psi$ commutes with $\mathcal{E}_1\otimes\mathds{I}$ and $\mathcal{E}_2\otimes\mathds{I}$, implies:
\begin{equation}\label{eq:DistanceProof4}
    (\mathcal{E}_1\otimes\mathds{I})\biggl[(\mathds{I}\otimes K_\psi)\Phi_+(\mathds{I}\otimes K^\dagger_\psi)\biggr]=\tfrac{t(\mathcal{E}_1\vert\Phi_+)}{t(\mathcal{E}_2\vert\Phi_+)} \cdot (\mathcal{E}_2\otimes\mathds{I})\biggl[(\mathds{I}\otimes K_\psi)\Phi_+(\mathds{I}\otimes K^\dagger_\psi)\biggr]
\end{equation}
or equivalently:
\begin{equation}\label{eq:DistanceProof5}
(\mathcal{E}_1\otimes\mathds{I})[\psi]= \dfrac{t(\mathcal{E}_1\vert\Phi_+)}{t(\mathcal{E}_2\vert\Phi_+)} \cdot (\mathcal{E}_2\otimes\mathds{I})[\psi]
\end{equation}

This way, by taking either $p = \dfrac{t(\mathcal{E}_1\vert\Phi_+)}{t(\mathcal{E}_2\vert\Phi_+)}$ or $p = \dfrac{t(\mathcal{E}_2\vert\Phi_+)}{t(\mathcal{E}_1\vert\Phi_+)}$ we have $(\mathcal{E}_1\otimes\mathds{I})[\psi]= p \cdot (\mathcal{E}_2\otimes\mathds{I})[\psi]$ or $(\mathcal{E}_2\otimes\mathds{I})[\psi]= p \cdot (\mathcal{E}_1\otimes\mathds{I})[\psi]$ for all state $\ket{\psi} \in \mathcal{H}^{\otimes 2}$, with $p\in\:]0;1]$. This gives either $\mathcal{E}_1 = p \cdot \mathcal{E}_2$ or $\mathcal{E}_2 = p \cdot \mathcal{E}_1$, and therefore $\mathcal{E}_1 \equiv \mathcal{E}_2$ $\blacksquare$.

As $\mathcal{M}_\diamond(\mathcal{E}_1,\mathcal{E}_2) \geq \mathcal{M}_J(\mathcal{E}_1,\mathcal{E}_2)$, we get the same result when $\mathcal{M}_\diamond(\mathcal{E}_1,\mathcal{E}_2) = 0$.\\

The triangular inequality and symmetry of $\mathcal{M}_J$ and $\mathcal{M}_\diamond$ come trivially from the distance properties of $C$ and $D$. Therefore, $\mathcal{M}_J$ and $\mathcal{M}_\diamond$ define proper distances on classes of non-zero probabilistic channels, that we defined in the last paragraph.

\section{Fundamental Properties of Probabilistic Quantum Channels}\label{app:Fund}

In this section, we show some fundamental results regarding the behaviour of probabilistic quantum channels. The most commonly used distance measure for quantum states is the \textit{trace distance} $D(\rho,\sigma) = \tfrac{1}{2}\tr\vert\rho-\sigma\vert^2$. The \textit{Ulhmann's Fidelity} $F(\rho,\sigma) = \bigl(\tr\sqrt{\sqrt{\rho}\sigma\sqrt{\rho}}\bigr)^2$ is not a metric in itself, but is often more relevant in our context as it can be interpreted as the probability that one state is projected on the other, when the states are purified. Moreover, most self-testing results relate the violation of Bell inequalities to the fidelity between physical and reference states. Finally, we can simply define convenient distances from the fidelity, such as the \textit{sine distance} $C(\rho,\sigma) = \sqrt{1-F(\rho,\sigma)}$ \cite{Gilchrist2005,Rastegin2006}, or the \textit{Bures angle} $A(\rho,\sigma) = \arccos \sqrt{F(\rho,\sigma)}$ \cite{bible}. We show results for these different functions.

\subsection{Metrics Monotonicity Under Quantum Channels}\label{sec:MetricMonotonicity}

Here we give the proof of Lemma~\ref{thm:ExtProcess} from the main text that gives a generalization of the processing inequality, or so-called metric monotonicity, to probabilistic quantum channels and the sine distance:

\setcounter{lem}{0}

\begin{restatable}[Extended Processing Inequality]{lem}{ExtProcessSupp}
 For any probabilistic channel $\mathcal{E}$ (CPTD), and any input states $\rho_i$ and $\sigma_i$, the following inequality holds for the sine distance $C(\rho,\sigma) = \sqrt{1-F(\rho,\sigma)}$:
    \begin{equation}\label{eq:extmonotonicityDmainSupp}
        C(\rho_i,\sigma_i) \geq t \cdot C(\rho_o,\sigma_o),
    \end{equation}
    where $\rho_o = \mathcal{E}[\rho_i]/t(\mathcal{E}\vert\rho_i)$ and $\sigma_o = \mathcal{E}[\sigma_i]/t(\mathcal{E}\vert\sigma_i)$
    are the output states of the channel, and  $t = t(\mathcal{E}\vert\rho_i)$ or $t = t(\mathcal{E}\vert\sigma_i)$.
    \label{thm:ExtProcessSupp}
\end{restatable}

Note that this inequality is also true for the trace distance. We first show that result for the latter, and then extend it to the sine distance.\\ \

\textit{\textbf{Proof.}} Let us first prove the inequality for the trace distance $D$. We follow the guidelines of the proof given in \cite{bible} for CPTP maps. As $\rho_i$ and $\sigma_i$ have a symmetric role, let us consider $t(\mathcal{E}\vert\rho_i) \geq t(\mathcal{E}\vert\sigma_i)$, without loss of generality. We can define two Hermitian positive matrices $P$ and $Q$ with orthogonal support such that $\rho_i-\sigma_i = P - Q$. Therefore, we have $\tr(P) - \tr(Q) = \tr(\rho_i) - \tr(\sigma_i) = 0$ so $\tr(P) = \tr(Q)$. Moreover, $\vert \rho_i-\sigma_i \vert = P + Q$. This way,
\begin{equation}\label{eq:Th3Demo1}
\begin{aligned}
    D(\rho_i,\sigma_i) \: &= \dfrac{1}{2}\tr\vert \rho_i-\sigma_i \vert \\
    &= \dfrac{1}{2}\bigl(\tr(P) + \tr(Q)\bigr) = \tr(P)
\end{aligned}
\end{equation}
There also exists a projector $\Pi$ such that $D(\rho_o,\sigma_o)=\tr\bigr(\Pi\cdot(\rho_o-\sigma_o)\bigl)$. Keeping in mind that $\mathcal{E}$ is trace-decreasing, it follows that for any $t\leq t(\mathcal{E}\vert\rho_i)$: 
\begin{align} \nonumber
    D(\rho_i,\sigma_i) &= \tr(P)\\ \nonumber
    &\geq \tr(\mathcal{E}[P])\\ \nonumber
    &\geq \tr(\Pi\cdot\mathcal{E}[P])\\ \nonumber
    &\geq \tr\bigl(\Pi\cdot(\mathcal{E}[P]-\mathcal{E}[Q])\bigr)\\ \nonumber
    &= \tr\bigl(\Pi\cdot(\mathcal{E}[\rho_i]-\mathcal{E}[\sigma_i])\bigr)\\ \nonumber
    &= t(\mathcal{E}\vert\rho_i)\tr(\Pi\rho_o) - t(\mathcal{E}\vert\sigma_i)\tr(\Pi\sigma_o)\\ \nonumber
    &\geq t(\mathcal{E}\vert\rho_i)\tr\bigl(\Pi\cdot(\rho_o - \sigma_o)\bigr)\\ \nonumber
    &= t(\mathcal{E}\vert\rho_i) \cdot D(\rho_o,\sigma_o)\\ \label{eq:Th3Demo2}
    &\geq t \cdot D(\rho_o,\sigma_o)
\end{align}
This way, we have in particular $D(\rho_i,\sigma_i)\geq t \cdot D(\rho_o,\sigma_o)$ for $t=t(\mathcal{E}\vert\rho_i)$ or $t=t(\mathcal{E}\vert\sigma_i)$ $\blacksquare$.\\

In order to prove the same inequality for the sine distance $C$, let us recall that we can express that distance between any density operators $\rho,\sigma$, as a minimization over their purifications $\ket{r}$ and $\ket{s}$ respectively: $C(\rho,\sigma)=\min \sqrt{1-\braket{r}{s}} = \min D(\proj{r},\proj{s})$, where the minimization is taken over all the purifications. This way, we are going to purify the input and output states in order to extend the inequality from $D$ to $C$. Let us choose two pure states $\ket{r_i},\ket{s_i} \in \mathcal{H}_i\otimes\mathcal{P}$ such that $C(\rho_i,\sigma_i)=  D(\proj{r_i},\proj{s_i})$, with $\mathcal{P}$ a purification space for $\rho_i$ and $\sigma_i$. This purifies the input states. Now let us define the operator $E$ on $\mathcal{H}_i\otimes\mathcal{P}$ such that for any pure state $\ket{\psi}$ in that space:
\begin{equation}\label{eq:PurifOperation}
E\ket{\psi} = \sum_j (K_j \otimes \mathds{I}_\mathcal{P} \ket{\psi}) \otimes \ket{e_j}
\end{equation}
where $\{K_j\}$ are Kraus operators for $\mathcal{E}$ and $\{\ket{e_j}\}$ is an orthonormal basis of an ancillary space $\mathcal{A}$. As $\mathcal{E}$ is trace-decreasing, $E\ket{\psi}$ is not necessarily normalized, but is a pure state when renormalized. This way, we can define the quantum operation $\Tilde{\mathcal{E}}$ such that for any density operator $\rho\in\mathcal{L}(\mathcal{H}_i)\otimes\mathcal{L}(\mathcal{P})$, we have $\Tilde{\mathcal{E}}[\rho]=E\rho E^\dagger$. This operation conserves the purity of pure states, and verifies $\tr_\mathcal{A}(\Tilde{\mathcal{E}}[\rho]) = \mathcal{E}[\rho]$ for any density operator $\rho$. This way, $\Tilde{\mathcal{E}}[\proj{r}]/t(\mathcal{E}\vert\rho_i)$, resp. $\Tilde{\mathcal{E}}[\proj{s}]/t(\mathcal{E}\vert\sigma_i)$, is a purification of $\mathcal{E}[\rho_i]/t(\mathcal{E}\vert\rho_i) = \rho_o$, resp. $\mathcal{E}[\sigma_i]/t(\mathcal{E}\vert\sigma_i) = \sigma_o$. This purifies the output states. Now we only have to apply the extended contractivity of $D$ to the purified states under the quantum operation $\Tilde{\mathcal{E}}$, for $t=t(\mathcal{E}\vert\rho_i)$ or $t=t(\mathcal{E}\vert\sigma_i)$:
\begin{align} \nonumber
    C(\rho_i,\sigma_i) &= D(\proj{r_i},\proj{s_i})\\ \nonumber
    &\geq t\cdot D(\Tilde{\mathcal{E}}[\proj{r}]/t(\mathcal{E}\vert\rho_i),\Tilde{\mathcal{E}}[\proj{s}]/t(\mathcal{E}\vert\sigma_i))\\ \label{eq:Th3Demo3}
    &\geq t \cdot \min D(\proj{r_o},\proj{s_o})\\ \nonumber
    &= t \cdot C(\hat{\rho}_o,\hat{\sigma}_o)
\end{align}
where the minimization is taken over all purifications $\ket{r_o}$, resp. $\ket{s_o}$, of $\rho_o$, resp. $\sigma_o$. This shows the inequality for the sine distance $\blacksquare$.\\

Note that for a trace-preserving quantum operation, $t(\mathcal{E}\vert\rho) = 1$ for any state $\rho$, and we get the well known processing inequality $D(\rho,\sigma)\geq D(\mathcal{E}[\rho],\mathcal{E}[\sigma])$ or $F(\rho,\sigma)\leq F(\mathcal{E}[\rho],\mathcal{E}[\sigma])$, indicating this inequality is tight.

\subsection{Comparison between Quantum Channels Metrics}\label{sec:MetricEquiv}

Choi-Jamio\l kowski and diamond metrics underline different properties of quantum channels. As pointed out in~\cite{Gilchrist2005}, the Choi-Jamio\l kowki metrics are linked to average probability of distinguishing two quantum channels when sending unknown states, while the diamond metrics are linked to the maximum probability of distinguishing these channels. The same can be said about our generalized definitions for probabilistic quantum channels, as long as we condition these probabilities to the detection of a state. For our protocol's security, the worst case scenario is more relevant, which is why diamond distances are preferred. Still, bounding the diamond distance between two channels with the sole knowledge of their actions on a maximally-entangled state is of major importance for our study, which is why we wish to bound diamond distances with their Choi-Jamio\l kowski counterparts. An attempt to show such bounds was done in~\cite{Sekatski2018}, linking the diamond trace distance with the Choi-Jamio\l kowski sine distance. However, it does not give a direct bound on the diamond fidelity, which is more suitable in cryptography in order to evaluate a protocol's success probability. In Lemma~\ref{thm:MetricEquivalence}, presented in the Methods, we demonstrate a tight bound of the diamond sine distance using their Choi-Jamio\l kowski sine distance, without extra information about the channel:

\begin{restatable}[Channel's Metrics Equivalence]{lem}{ChanMetricsSupp}
\label{thm:MetricEquivalenceSupp} For any probabilistic channel $\mathcal{E}_1$, and any $\mathcal{E}_2$ that is proportional to a deterministic channel (CPTP map), both acting on $\mathcal{L}(\mathcal{H}_i)$, we have the following inequalities:
   \begin{equation}\label{ineq:equivSineMainSupp}
    \mathcal{C}_J(\mathcal{E}_1,\mathcal{E}_2) \leq \mathcal{C}_\diamond(\mathcal{E}_1,\mathcal{E}_2) \leq \dim\mathcal{H}_i \times \mathcal{C}_J(\mathcal{E}_1,\mathcal{E}_2) \laura{,}
    \end{equation}
where the $\mathcal{C}_J$, resp. $\mathcal{C}_\diamond$, are the Choi-Jamio\l kowski, resp. diamond, sine distances of probabilistic quantum channels:
\begin{align}\label{eq:ChannelSineChoiSupp}
 &\mathcal{C}_J(\mathcal{E}_1, \mathcal{E}_2) = C\Bigl(\dfrac{(\mathcal{E}_1\otimes\mathds{I})[\Phi_+]}{t(\mathcal{E}_1\vert\Phi_+)},(\mathcal{E}_2\otimes\mathds{I})[\Phi_+]\Bigr)\\
\label{eq:ChannelTraceDiamondSupp}
&\mathcal{C}_\diamond(\mathcal{E}_1, \mathcal{E}_2)  =\sup_{\ket{\phi}} C\Bigl(\dfrac{(\mathcal{E}_1\otimes\mathds{I})[\phi]}{t(\mathcal{E}_1\vert\phi)},(\mathcal{E}_2\otimes\mathds{I})[\phi]\Bigr)
\end{align}
\end{restatable}

Note that once again, the result is also true for trace distances of quantum channels. We provide the proof of this lemma for both trace and sine distances.\\ \  

\textit{\textbf{Proof.}} We want to show the two following inequalities, for any probabilistic channel $\mathcal{E}_1$ and any deterministic channel $\mathcal{E}_2$:
\begin{align}\label{ineq:equivTraceAnnex}
    &\mathcal{D}_J(\mathcal{E}_1,\mathcal{E}_2) \leq  \mathcal{D}_\diamond(\mathcal{E}_1,\mathcal{E}_2)  \leq \dim\mathcal{H}_i \times \mathcal{D}_J(\mathcal{E}_1,\mathcal{E}_2) \\
    \label{ineq:equivSineAnnex}
    &\mathcal{C}_J(\mathcal{E}_1,\mathcal{E}_2) \leq  \mathcal{C}_\diamond(\mathcal{E}_1,\mathcal{E}_2)  \leq \dim\mathcal{H}_i \times \mathcal{C}_J(\mathcal{E}_1,\mathcal{E}_2)
\end{align}
The left-side inequalities are straightforwardly following from the definition of the distances. The right-side comes from the following corollary:

\begin{cor} For any pure state $\rho \in \mathcal{L}(\mathcal{H}_i^{\otimes 2})$ and any pair of probabilistic quantum channels $\mathcal{E}_1$ and $\mathcal{E}_2$ from $\mathcal{L}(\mathcal{H}_i)$ to $\mathcal{L}(\mathcal{H}_o)$, we have:
\begin{align}\label{ineq:equivweakTrace}
    &x \cdot D(\rho_{1},\rho_{2}) \leq \dim\mathcal{H}_i \times \mathcal{D}_J(\mathcal{E}_1,\mathcal{E}_2) \\
    \label{ineq:equivweakSine}
    &x \cdot C(\rho_{1},\rho_{2}) \leq \dim\mathcal{H}_i \times  \mathcal{C}_J(\mathcal{E}_1,\mathcal{E}_2)
\end{align}
for any $ x \leq \max\bigl[\tfrac{t(\mathcal{E}_1\vert\rho)}{t(\mathcal{E}_1\vert\Phi_+)},\tfrac{t(\mathcal{E}_2\vert\rho)}{t(\mathcal{E}_2\vert\Phi_+)}\bigr]$, and with $\rho_{k} = (\mathcal{E}_k\otimes\mathds{I})[\rho]/t(\mathcal{E}_k\vert\rho)$.
\end{cor}

 Let us consider a pure state $\rho = \proj{\psi}$ with $\ket{\psi}\in \mathcal{H}_i\otimes\mathcal{H}_i$, and two probabilistic channels $\mathcal{E}_1$ and $\mathcal{E}_2$. We define the corresponding transmissivities $t(\mathcal{E}_k\vert\rho)$ and output states $\rho_{k} = (\mathcal{E}_k\otimes\mathds{I})[\rho]/t(\mathcal{E}_k\vert\rho)$ for $k=1$ and $2$. Using the operator $K_\psi$ defined in Lemma~\ref{lemma:bellStateToAny}, the map $\mathcal{O}$ defined as $\mathcal{O}[\rho] = K_\psi\rho K^\dagger_\psi$ is a valid quantum operation on $\mathcal{L}(\mathcal{H}_i)$. Furthermore, $\mathds{I}\otimes\mathcal{O}$ transforms $\ket{\Phi_+}$ into $\ket{\psi}$ with probability $1/\dim\mathcal{H}_i$, and commutes with the channels $\mathcal{E}_1\otimes\mathds{I}$ and $\mathcal{E}_2\otimes\mathds{I}$, such that for $k=1$ or $2$ and $d =\dim\mathcal{H}_i$:
\begin{align}
        (\mathds{I}\otimes\mathcal{O})[(\mathcal{E}_k\otimes\mathds{I})[\Phi_+]\:/ t(\mathcal{E}_k\vert\Phi_+)] &= \tfrac{1}{d \cdot t(\mathcal{E}_k\vert\Phi_+)} (\mathcal{E}_k\otimes\mathds{I})[\rho] \\&= \tfrac{t(\mathcal{E}_k\vert\rho)}{d\cdot t(\mathcal{E}_k\vert\Phi_+)} \rho_{k}
\end{align}
This way, $\mathds{I}\otimes\mathcal{O}$ transforms the state $(\mathcal{E}_k\otimes\mathds{I})[\Phi_+]/t(\mathcal{E}_k\vert\Phi_+)$ into $\rho_{k}$, with probability $\tfrac{t(\mathcal{E}_k\vert\rho)}{d\cdot t(\mathcal{E}_k\vert\Phi_+)}$. This way, using Lemma~\ref{thm:ExtProcess} for extented metrics monotonicity to the quantum operation $\mathcal{O}\otimes\mathds{I}$, we deduce the following inequality:
\begin{equation}
    M((\mathcal{E}_1\otimes\mathds{I})[\Phi_+]/t(\mathcal{E}_1\vert\Phi_+),(\mathcal{E}_2\otimes\mathds{I})[\Phi_+]/t(\mathcal{E}_2\vert\Phi_+))\geq t\cdot M(\rho_{1},\rho_{2})
\end{equation} 
for any $t\leq \max\bigl[\tfrac{t(\mathcal{E}_1\vert\rho)}{d\cdot t(\mathcal{E}_1\vert\Phi_+)},\tfrac{t(\mathcal{E}_2\vert\rho)}{d\cdot t(\mathcal{E}_2\vert\Phi_+)}\bigr]$, and $M = C,D$. The left term is $\mathcal{M}_J(\mathcal{E}_1,\mathcal{E}_2)$ for $\mathcal{M}=\mathcal{C}$, and we get inequalities \eqref{ineq:equivweakTrace} and \eqref{ineq:equivweakSine} by taking $x = t\cdot d \leq \max\bigl[\tfrac{t(\mathcal{E}_1\vert\rho)}{t(\mathcal{E}_1\vert\Phi_+)},\tfrac{t(\mathcal{E}_2\vert\rho)}{t(\mathcal{E}_2\vert\Phi_+)}\bigr]$, which shows the corollary. If one of the channels, $\mathcal{E}_2$ for instance, is proportional to a trace-preserving channel, then $t(\mathcal{E}_2\vert\rho) = t(\mathcal{E}_2\vert\Phi_+)$ for any $\rho$. This way, we can take $x=1$, so that the following inequality holds for any pure state $\rho \in \mathcal{L}(\mathcal{H}_i\otimes\mathcal{H}_i)$:
\begin{equation}
    M(\rho_{1},\rho_{2}) \leq d \cdot  \mathcal{M}_J(\mathcal{E}_1,\mathcal{E}_2)
\end{equation}
As it holds for any pure state $\rho$, we showed that $\mathcal{M}_\diamond(\mathcal{E}_1,\mathcal{E}_2) \leq d\times \mathcal{M}_J(\mathcal{E}_1,\mathcal{E}_2)$ for $\mathcal{M}=\mathcal{C}$ or $\mathcal{D}$, which is the right-side of inequalities~(\ref{ineq:equivSineAnnex}) and (\ref{ineq:equivTraceAnnex}) $\blacksquare$.\\

The corollary we just showed allows us to bound the deviation of any output states, with the sole knowledge of the operations actions on a maximally entangled state, even if both channels are probabilistic. Yet in a lot of cases, ours in particular, $\mathcal{E}_2$ is a reference quantum channel $\mathcal{E}_0$ that is trace-preserving, and we can use the special case $\mathcal{M}_\diamond(\mathcal{E},\mathcal{E}_0) \leq \dim\mathcal{H}_i \times \mathcal{M}_J(\mathcal{E},\mathcal{E}_0)$ from the lemma, which does not require to evaluate any transmissivity.

\subsection{Bound on Transmissivity}

One can evaluate the channel's transmissivity $t(\mathcal{E}|\rho_i)$ when sending the input state $\rho_i$, by deriving a bound from the parameters of the problem, as shown in the following lemma.

\setcounter{lem}{3}
\begin{lem}[Bound on the transmissivity]
Let $\mathcal{E}$ be a probabilistic quantum channel on $\mathcal{L}(\mathcal{H}_i)$, and let us consider two states $\Phi_i, \rho_i\in \mathcal{L}(\mathcal{H}_i^{\otimes 2})$ with $\Phi_i$ a close-to-maximally-entangled state. Then the following bound holds:
\begin{equation}\label{ineq:boundproba}
       \vert t(\mathcal{E}|\rho_i) - t(\mathcal{E}|\Phi_i)\vert \leq d\cdot D(\Phi_i,\Phi_+) + d \cdot t(\mathcal{E}|\Phi_i) \cdot \min_{\mathcal{E}_0}D(\Phi_o,(\mathcal{E}_0\otimes\mathds{I})[\Phi_+])
\end{equation}
where $d=\dim \mathcal{H}_i$, $\Phi_o = (\mathcal{E}\otimes\mathds{I})[\Phi_i]/t(\mathcal{E}\vert\Phi_i)$ and the minimization is carried out over all trace-preserving channels $\mathcal{E}_0$.
\end{lem}

Using the parameters of our protocols, knowing $D\leq C$, it follows:
\begin{equation}\label{ineq:corollaryBoundTransmissivity}
    \vert t(\mathcal{E}|\rho_i) - t(\mathcal{E}|\Phi_i)\vert \leq 2\: C^i + 2\: t(\mathcal{E}|\Phi_i) \cdot C^o
\end{equation}
This way, Alice and Bob can predict the abort probability of the protocol from the parameters, in particular the minimum acceptable transmissivity $t(\mathcal{E}|\Phi_i)$ of the channel when sending the probe state (see the following paragraphs). If the transmissivity is too low, one can try to avoid aborting the protocol by asking for more copies of $\rho_i$. Here we provide the proof of the lemma.\\ 

\textit{\textbf{Proof.}} Let us first assume $\rho_i = \proj{\psi} =\psi$ is a pure state, with $\ket{\psi}\in\mathcal{H}_i^{\otimes 2}$. This way we can define the operator $K_\psi$ from Lemma~\ref{lemma:bellStateToAny} such that $(\mathds{I}\otimes K_\psi)\ket{\Phi_+} = \tfrac{1}{\sqrt{d}}\ket{\psi}$, with $d=\dim \mathcal{H}_i$. We recall that for any trace-preserving channel $\mathcal{E}_0$ we have $\tr((\mathcal{E}_0\otimes\mathds{I})[\psi])=1$. This way we have:
\begin{align} \nonumber
    \vert t(\mathcal{E}\vert\psi) - t(\mathcal{E}\vert\phi_i)\vert &= \bigl\vert \tr((\mathcal{E}\otimes\mathds{I})[\psi])-t(\mathcal{E}\vert\Phi_i)\tr((\mathcal{E}_0\otimes \mathds{I})[\psi]\bigr\vert\\ \label{eq:proofTransmissivity1}
    &= d\cdot \bigl\vert \tr((\mathcal{E}\otimes K_\psi)[\Phi_+])-t(\mathcal{E}\vert\Phi_i)\tr((\mathcal{E}_0\otimes K_\psi)[\Phi_+])\bigr\vert\\ \nonumber
    &\leq d\cdot\bigl\vert\tr((\mathcal{E}\otimes K_\psi)[\Phi_+])-\tr((\mathcal{E}\otimes K_\psi)[\Phi_i])\bigr\vert+ d\cdot\bigl\vert\tr((\mathcal{E}\otimes K_\psi)[\Phi_i])- t(\mathcal{E}\vert\Phi_i)\tr((\mathcal{E}_0\otimes K_\psi)[\Phi_+])\bigr\vert.
\end{align}
We use the fact that $D(\rho,\sigma) = \max_{0<P\leq\mathds{I}}\tr(P(\rho-\sigma))$ in order to bound the two terms. The second one is straightforward as $0<K_\psi\leq\mathds{I}$:
\begin{equation}\label{eq:proofTransmissivity2}
       d\cdot\bigl\vert\tr((\mathcal{E}\otimes K_\psi)[\Phi_i])- t(\mathcal{E}\vert\Phi_i)\tr((\mathcal{E}_0\otimes K_\psi)[\Phi_+])\bigr\vert \leq d\cdot t(\mathcal{E}\vert\Phi_i)\cdot D(\Phi_o,(\mathcal{E}_0\otimes\mathds{I})[\Phi_+]).
\end{equation}
%
For the first term we use Kraus' theorem on the probabilistic channel $\mathcal{E}\otimes K_\psi$ such that for any $\rho\in\mathcal{L}(\mathcal{H}_i^{\otimes 2})$ we have ${(\mathcal{E}\otimes K_\psi)[\rho] = \sum_j M_j\rho M_j^\dagger}$, with $0 < \sum M_j^\dagger M_j \leq \mathds{I}$.  The first term therefore gives:
\begin{align}\nonumber
    d\cdot\bigl\vert\tr((\mathcal{E}\otimes K_\psi)[\Phi_+])-\tr((\mathcal{E}\otimes K_\psi)[\Phi_i])\bigr\vert &=  d\cdot\bigl\vert\tr\sum_j M_j(\Phi_+ -\Phi_i)M_j^\dagger\bigr\vert\\ \label{eq:proofTransmissivity3}
    &= d\cdot\biggl\vert\tr\biggl(\bigl(\sum_j M_j^\dagger M_j\bigr)(\Phi_+ -\Phi_i)\biggr)\biggr\vert\\ \nonumber
    &\leq d\cdot D(\Phi_i,\Phi_+).
\end{align}
This gives the bound:
\begin{equation}\label{eq:proofTranmissivity4}
      \vert t(\mathcal{E}|\rho_i) - t(\mathcal{E}|\Phi_i)\vert\leq d\cdot D(\Phi_i,\Phi_+) + d\cdot t(\mathcal{E}|\Phi_i) \cdot D(\Phi_o,(\mathcal{E}_0\otimes\mathds{I})[\Phi_+]). 
\end{equation}
As it is true for any CPTP map $\mathcal{E}_0$, we can minimize the bound on this map, which shows the lemma $\blacksquare$.\\


\section{Detailed Theoretical Protocols}\label{app:TheoProts}

In the following we give the details on the theoretical protocol recipes. 
We start with two protocols for 1sDI and DI transmission certification where  we assume Bob can use trusted quantum memories in order to store all the states he receives, before performing the measurements. 
In fact, these memories can be replaced by the more reasonable assumption that Alice and Bob share a common random source (this is indeed a standard trick in trading memory and communication requirements for shared randomness, see e.g. \cite{unnikrishnan_anonymity_2019}).
It is the latter protocol that we implement in experiments, as we perform the measurements on the fly. 
The method we use in experiment seems more practical with current photonic technology, which does not allow the storage of $\simeq 10^9$ states for a time span of the a few hours. In addition, one can consider these quantum memories to be untrusted channels which require certification. In that sense it also seems more secure to assume trusted classical communications than trusted quantum memories. Here we still provide the recipes for theoretical protocols with quantum memories, as they follow the spirit of the protocol provided in \cite{unnikrishnan2019} for authenticated teleportation. This way, when proving the security, we can apply the bounds from this previous study in order to certify the output probe states after our untrusted channel more directly. However the security caries through all protocols.

\subsection{One-Sided Device Independent Protocol}\label{TheoProtocol}

This first recipe details the protocol that we study in our paper, when Alice's measurement apparatus as well as the probe state source are trusted. This specifically applies to a scenario where a powerful server Alice wants to send a quantum message to a weaker receiver Bob through an untrusted quantum channel.\\ \

Note that from step 1.(b) Alice deduces the minimum amount of state she has to prepare in order to properly certify the channel. If $t$ is overstated and the channel has a lower tranmissivity, then Alice will not prepare enough probe states, which will make the protocol abort in step 5. On the contrary if $t$ is understated, then Alice will prepare more probe states than she and Bob require, which will in fact improve the certification confidence.\\

The security of the protocol is in principle ensured by the fact that Alice and Bob only agree on the measurement after Bob receives all the states. The position of the quantum message $\rho_i$ is also broadcasted after all state are sent through the channel. This way, the channel's operator has no way of guessing the position of the message by spying the communications between Alice and Bob, that can even remain public. As mentioned earlier, in experiment we rely on private classical communication to hide the position $r$ of state $\rho_i$. The full security bound is given in later section \ref{protocolSecurity}.\\

\begin{tcolorbox}[title=Protocol 1: Certified Transmission through a Probabilistic Quantum Channel in 1sDI scenario,title filled] 
$1.$ Prior to the protocol: 
\begin{enumerate}
    \item[(a)] Alice characterizes the probe state $\Phi_i$ emitted by her source and evaluates the quantity $F^i$. She also receives or prepares the quantum message $\rho_i$, possibly shared with an outside party.
    \item[(b)] Alice and Bob agree on parameters $\epsilon, K$, and the minimum transmissivity $t$ allowed for the channel $\mathcal{E}$, depending on their requirements and experimental limitations. 
\end{enumerate}

$2.$ Alice prepares $N = \lceil K/t\rceil$ copies of the probe state $\Phi_i$.\vspace{2mm}

$3.$ Alice successively sends each state through $\mathcal{E}$, including $\rho_i$ in a random $r$-th position, with $r\leq N+1$.\vspace{2mm}

$4.$ Bob establishes the set $\mathbb{S}_P$ of states which successfully passed through $\mathcal{E}$, and broadcast it publicly.\vspace{2mm}

$5.$ If $r \notin \mathbb{S}_P$ or $\vert \mathbb{S}/\{r\} \vert < K$, Alice aborts the protocol. Otherwise, Alice sends $r$ to Bob.\vspace{2mm}

$6.$ Alice separates $\mathbb{S}/\{r\}$ into two random sets $\mathbb{S}_0$ and $\mathbb{S}_1$.\vspace{2mm}

$7.$ For each $k \in \mathbb{S}_q$, $q =0,1$:
\begin{enumerate}
\item[(a)] Alice measures observable $A_q$ on her part of the $k$-th state and gets outcome $a_k$.
\item[(b)] She tells Bob to measure observable $B_q$ on his part of the $k$-th state and he gets outcome $b_k$.
\item[(c)] Alice and Bob calculate their correlation for round $k$ as $c_k = a_k b_k$. 
\end{enumerate} 

$8.$ Alice and Bob deduce the average value over all rounds, of $\beta = \vert\langle A_0B_0\rangle + \langle A_1B_1\rangle\vert$.\vspace{2mm}

$9.$ If $\beta \geq 2 - \epsilon$, then Alice successfully sent the state $\rho_o = \mathcal{E}[\rho_i]/t(\mathcal{E}|\rho_i)$ to Bob, with a certified average fidelity to the target quantum message $\rho_i$, up to isometry.
\end{tcolorbox}
\vspace{0.5cm}

\subsection{Fully Device Independent Protocol}

While the protocol described in the previous section has high relevance when devices in one laboratory can be trusted, the completely adversarial scenario would demand the fully device independent protocol. Theoretically, such protocol can be formulated, but in the absence on any assumptions about the functioning of the devices, which should be the case in the fully device-independent protocol, we argue that the certification procedure would be very resource-demanding and difficult to perform with available resources. To make Protocol 1 fully device independent one needs to certify in a device independent manner the fidelity of probe states $F_i$, which in Protocol 1 figures as a parameter.\\

The input fidelity $F_i$ can be estimated by using self-testing methods, in a similar way like it was done in Protocol 1, with an important difference, that self-testing would be done through the violation of the CHSH inequality. However, without any assumptions about the source or the channel, such protocol would require a very big number of experimental rounds. Namely, if in Protocol $1$ one has to measure $N$ copies to verify that the channel was correctly applied to an unknown quantum message $\rho_i$, in the fully DI scenario, to verify $F_i$ of a single probe state passing through the channel, one would have to measure around $N$ additional states. Hence, the number of experimental rounds would need to be squared, which corresponds to a very low sample-efficiency of the certification protocol. \\

One way to simplify the protocol is by assuming that the source is producing independent and identically distributed copies, i.e. that the source functions in the IID scenario. In that case, we schematically double the sample size $N$ instead of squaring it. Here we provide the recipe for that certification protocol, making the IID assumption on the input probe state. In this framework, our fully device independent protocol simply consists in performing a very similar protocol to the one presented in the previous section, with one difference in step 1.(a), related to using the CHSH inequality~\cite{clauser1969} for certification instead of the steering inequality. In that version, Alice measures the observables $A_3,A_4$ on the part of the system she can send through the channel.\\ \

The security of this protocol can be derived from that of protocol 1, with some slight adjustments. First we use another bound for the self-testing of CHSH inequalities, in a fully device independent and non-IID scenario \cite{unnikrishnan}, in order to certify the output probe state. The input probe state is also certified via self-testing of CHSH inequality in a fully device independent scenario, but keeping the IID assumption. We can then plug the two certified fidelities in our bound (\ref{ineq:MainResult3}).\\

\vspace{1cm}


\begin{tcolorbox}[title=Protocol 2: Certified Transmission through a Probabilistic Quantum Channel in DI scenario,title filled] 

$1.$ Prior to the protocol, Alice and Bob agree on parameters $\epsilon$, $\eta$, $K$, $M$, and the minimum transmissivity $t$ allowed for the channel $\mathcal{E}$, depending on their requirements and experimental limitations.\vspace{2mm}

$2.$ Alice prepares $N+M$ copies of the probe state $\Phi_i$, where $N = \lceil K/t\rceil$. She also receives or prepares the quantum message $\rho_i$, possibly shared with an outside party.\vspace{2mm}

$3.$ Alice measures $M$ random copies of $\Phi_i$, and deduce the value of $E_i = \vert\langle A_0\mathcal{A}_2\rangle + \langle A_0A_3\rangle + \langle A_1\mathcal{A}_2\rangle - \langle A_1A_3\rangle\vert$. \vspace{2mm}

$4.$ If $\beta_i < 2\sqrt{2}-\eta$, Alice aborts the protocol.\vspace{2mm}

$5.$ Alice successively sends each state through $\mathcal{E}$, including $\rho_i$ in a random $r$-th position, with $r\leq N+1$.\vspace{2mm}

$6.$ Bob establishes the set $\mathbb{S}_P$ of states which successfully passed through $\mathcal{E}$, and broadcast it publicly.\vspace{2mm}

$7.$ If $r \notin \mathbb{S}_P$ or $\vert \mathbb{S}/\{r\} \vert < K$, Alice aborts the protocol. Otherwise, Alice sends $r$ to Bob.\vspace{2mm}

$8.$ For each $k \in \mathbb{S}_q$, $q =0,1$:
\begin{enumerate}
\item[(a)] Alice measures observable $A_u$ on her part of the $k$-th statem with $u=0$ or $1$ at random. She gets outcome $a_k$.
\item[(b)] Bob measures observable $B_v$ on his part of the $k$-th state, with $v=0$ or $1$ at random. He gets the outcome $b_k$.
\item[(c)] Alice and Bob calculate their correlation for round $k$ as $c_k = a_k b_k$. 
\end{enumerate} 

$9.$ Alice and Bob deduce the average value over all rounds, of $\beta_o = \vert\langle A_0B_0\rangle + \langle A_0B_1\rangle + \langle A_1B_0\rangle - \langle A_1B_1\rangle\vert$.\vspace{2mm}

$10.$ If $\beta \geq 2\sqrt{2} - \epsilon$, then Alice successfully sent the state $\rho_o = \mathcal{E}[\rho_i]/t(\mathcal{E}|\rho_i)$ to Bob, with a certified average fidelity to the target quantum message $\rho_i$, up to isometry.
\label{alg:CertifChannelDI}
\end{tcolorbox}
\vspace{1cm}

\subsection{Practical Protocol}

We mentioned that the protocol we implement in our experiment differ slightly from the theoretical protocols detailed in previous paragraphs, as the latter rely on Bob being able to store all states he receives from the channel, before agreeing with Alice to measure them. This imposes a strong assumption on Bob's power, which is both impractical for experiments, and unrealistic in our one-sided device independent scenario that assumes the receiver possesses as few trusted resources as possible. Thus, although this protocol follows the recipe from \cite{unnikrishnan2019} which allows for the derivation of the security, we implement a more practical protocol in our experiment. That protocol assumes a private and trusted classical communication channel, but does not rely on trusted quantum memories. Here we detail a theoretical version of that protocol, in a one-sided device independent setting, which fits more to our implementation. We assume the security to be the equivalent to that of protocol 1. In addition, as players perform the measurements \textit{on the fly}, more assumptions are required in order to distinguish potentially biased losses from the channel from detection losses. These are detailed in appendix E.3, and mainly consist in considering detection losses are independent of the measurement basis, which is a form of fair-sampling assumption.\\

\begin{tcolorbox}[title=Protocol 3: Practical Certified Transmission through a Probabilistic Quantum Channel in 1sDI scenario,title filled] 
$1.$ Prior to the protocol: 
\begin{enumerate}
    \item[(a)] Alice characterizes the state probe state $\Phi_i$ emitted by her source and evaluates the quantity $F^i$. She also receives or prepares the quantum message $\rho_i$, possibly shared with an outside party.
    \item[(b)]  Alice and Bob agree on parameters $\epsilon, K$, and the minimum transmissivity $t$ allowed for the channel $\mathcal{E}$, depending on their requirements and experimental limitations. They also share a private random key $r\in[\![1,N+1]\!]$, with $N = \lceil K/t\rceil$. 
\end{enumerate}
For $k\in[\![1,N+1]\!]$:
\begin{enumerate}
    \item[$2.$] If $k\neq r$:
    \begin{enumerate}
        \item[(a)] Alice prepares a copy of the probe state $\Phi_i$ and sends half of it through $\mathcal{E}$. 
        \item[(b)] Alice and Bob privately agree on a random $q \in \{0,1\}$ and measure the observable $A_qB_q$, with an outcome $c_k = a_kb_k$ if Bob received a state, or no outcome if the state was lost through the channel.
    \end{enumerate}
    \item[$3.$]  If $k = r$:
    \begin{enumerate}
        \item[(a)] Alice sends the quantum message $\rho_i$ through $\mathcal{E}$. 
        \item[(b)] If Bob does not receive any state, the protocol aborts. Otherwise, Bob sets the state aside.
    \end{enumerate}
\end{enumerate}

$4.$ If the number of "no-outcome" events during step 2.(b) is bigger than $N-K$, then the protocol aborts.\vspace{2mm}

$5.$ From the correlations $\{c_k\}$, Alice and Bob deduce the average value over all rounds, of $\beta = \vert\langle A_0B_0\rangle + \langle A_1B_1\rangle\vert$.\vspace{2mm}

$6.$ If $\beta \geq 2 - \epsilon$, then Alice successfully sent the state $\rho_o = \mathcal{E}[\rho_i]/t(\mathcal{E}|\rho_i)$ to Bob, with a certified average fidelity to the quantum message $\rho_i$, up to isometry.
\end{tcolorbox}

\section{Protocol Security}\label{protocolSecurity}

When the protocol does not abort, Alice and Bob wish to bound the probability that it successfully implements the channel $\mathcal{E}_0\otimes\mathds{I}_i$ on the input state $\rho_i$. With no IID assumption made on the quantum channel, it is a priori impossible to predict the transformation undergone by $\rho_i$, from the sole measurements performed on other quantum states. However, statistical arguments on a large sample of quantum channels allow us to bound the transmission probability with high confidence, on the condition that the position of the input state $\rho_i$ remains confidential. In the following, we show that bound, using only the measurements performed during the protocol when sending the probe states $\Phi_i$ through the channel.

First, we define the average quantum channel and quantum states, and show it describes accurately the result of the protocol. We also deduce relevant quantities, in term of quantum states and quantum channels fidelities.

Then, we show the certification bound \ref{ineq:MainResult3}, by going through similar guidelines as the certification bound shown in \cite{Sekatski2018} for CPTP maps, and using our new fundamental results on probabilistic channels. 

Next we show how we can apply the recent results from \cite{unnikrishnan2019} to our protocol, in order to certify a virtual and unmeasured probe state, thanks to violation of steering inequality in a one-sided device-independent and non-IID setting, measured on all other probe states. 

Then, we show the expressions of error terms on the fidelity of the probe output state, and on the channel's transmissivity, due to finite number of samples in a non-IID setting. 

Finally we tie all these results together in order to give the full bound on the transmission fidelity $F(\rho_o,\rho_i)$. We also give the modification required to that bound in order to certify the transmission fidelity in protocol 2.


\subsection{Average Channel and States}\label{sec:avgstate}

 During the protocol, Alice sends $N+1$ states through the channel, including $N$ states, and one copy of $\rho_i$. On the $k$-th state, the channel takes the expression $\mathcal{E}_{k|[k-1]}$. If $\rho_i$ is sent through $r$-th channel, then a state $\rho_r$ outputs the channel with probability $t(\mathcal{E}_{r|[r-1]}\vert\rho_i)$. Alice sends the state $\rho_i$ at a random position $r$, meaning it has equal probability to be sent through any of the channels $\{\mathcal{E}_{k|[k-1]}\}_{k=1,...,N+1}$. This way, assuming the channel's operator has no way of guessing the position $r$ where $\rho_i$ is sent, then the expected output state is:
 \begin{equation}
     \bar{\rho}_o = \dfrac{\sum_{r=1}^{N+1}t(\mathcal{E}_{r|[r-1]}\vert\rho_i)\cdot\rho_r}{\sum_{r=1}^{N+1}t(\mathcal{E}_{r|[r-1]}\vert\rho_i)}= \dfrac{\sum_{r=1}^{N+1}(\mathcal{E}_{r|[r-1]}\otimes\mathds{I})[\rho_i]}{\tr\bigl(\sum_{r=1}^{N+1}(\mathcal{E}_{r|[r-1]}\otimes\mathds{I})[\rho_i]\bigr)} = \dfrac{\bigl(\tfrac{1}{N+1}(\sum_{r=1}^{N+1}\mathcal{E}_{r|[r-1]})\otimes\mathds{I}\bigr)[\rho_i]}{\tr\Bigl(\bigl(\tfrac{1}{N+1}(\sum_{r=1}^{N+1}\mathcal{E}_{r|[r-1]})\otimes\mathds{I}\bigr)[\rho_i]\Bigr)}
 \end{equation}
where we omitted the isometries for more simplicity, \textit{i.e.} $\mathcal{E}_{r|[r-1]}$ actually stands for $\tr_{ext}\bigl((\Gamma_o \circ \mathcal{E}_{r|[r-1]} \circ \Gamma_i) [\rho_{\mathcal{A}_1} \otimes \bullet \: ]\bigr)$. From here we naturally define the average channel over the protocol:
\begin{equation}\label{eq:avgChannel2}
     \bar{\mathcal{E}} = \dfrac{1}{N+1}\sum_{k=1}^{N+1}\mathcal{E}_{k|[k-1]}
\end{equation}
which is a physical channel that randomly applies any of the channels $\{\mathcal{E}_{k|[k-1]}\}_{k=1,...,N+1}$. This way the expected output state of the protocol reads:
\begin{equation}
    \bar{\rho}_o = (\bar{\mathcal{E}}_{i,o}\otimes  \mathds{I})[{\rho}_i]/t(\bar{\mathcal{E}}_{i,o}\vert\rho_i) 
\end{equation}
where $\bar{\mathcal{E}}_{i,o} = \tr_{ext}\bigl((\Gamma_o \circ \bar{\mathcal{E}} \circ \Gamma_i) [\rho_{\mathcal{A}_1} \otimes \bullet \: ]\bigr)$. Similarly the expected ouput state when sending the probe state $\Phi_i$ reads:
\begin{equation}\label{eq:avgBell}
    \bar{\Phi}_o = (\bar{\mathcal{E}}\otimes  \mathds{I})[{\Phi}_i]/t(\bar{\mathcal{E}}\vert\Phi_i)
\end{equation}
such that states are expected to undergo the operation $\bar{\mathcal{E}}$. Following previous studies lifting the IID assumption \cite{Anu2019,govcanin2022sample}, we aim at certifying the average output state $\bar{\rho}_o$. To support this choice, we can predict the result of a measurement performed after the protocol, following the idea of \cite{govcanin2022sample}. If the state $\rho_i$ was sent through the $r$-th channel, then we can express the probability that it was not lost in the channel ($\checkmark$) \textit{and} we measure the $k$-th outcome of any POVM $\{E_k\}_k$:
\begin{equation}
    \Pr\bigl((\checkmark \cap k)\vert r\bigr) = t(\mathcal{E}_{r\vert[r-1]}\vert\rho_i)\cdot\tr(E_k \cdot \rho_r) = \tr\bigr(E_k \cdot (\mathcal{E}_{r|[r-1]}\otimes\mathds{I})[\rho_i]\bigl)
\end{equation}
The input state $\rho_i$ has equal probability to be sent through any channel $\{\mathcal{E}_{r\vert[r-1]}\}$, so the probability that, after the protocol, the state was not lost in the channel ($\checkmark$) \textit{and} we measure the $k$-th outcome of the POVM $\{E_k\}_k$ reads:
\begin{equation}
    \Pr(\checkmark \cap k) = \dfrac{1}{N+1}\sum_{r=1}^{N+1}\Pr\bigl((\checkmark \cap k)\vert r\bigr) =\tr\bigr(E_k \cdot (\bar{\mathcal{E}}_{i,o}\otimes  \mathds{I})[{\rho}_i]\bigl)
\end{equation}
Similarly, the probability that the state was not lost in the channel reads:
\begin{equation}
    \Pr(\checkmark) = \dfrac{1}{N+1}\sum_{r=1}^{N+1}\Pr(\checkmark \vert r)  = \dfrac{1}{N+1}\sum_{r=1}^{N+1}t(\mathcal{E}_{r\vert[r-1]}\vert\rho_i) =t(\bar{\mathcal{E}}_{i,o}\vert\rho_i) 
\end{equation}
Knowing the protocol succeeded, so the input state was not lost in the channel, the probability that we measure the $k$-th outcome of the POVM reads:
\begin{equation}
    \Pr(k \vert \checkmark) = \dfrac{\Pr(\checkmark \cap k)}{\Pr(\checkmark)} = \dfrac{\tr\bigr(E_k \cdot (\bar{\mathcal{E}}_{i,o}\otimes  \mathds{I})[{\rho}_i]\bigl)}{t(\bar{\mathcal{E}}_{i,o}\vert\rho_i) } =  \tr(E_k\cdot \bar{\rho}_o ) 
\end{equation}
This shows that as long as $r$, the position of $\rho_i$ among probe states, remains hidden and random, any measurement performed on the output state later after the protocol would follow the same statistics as if it was performed on the expected output state $\bar{\rho}_o$. By extension, the fidelity $F\bigl(\bar{\rho}_o,(\mathcal{E}_0\otimes\mathds{I})[\rho_i]\bigr)$ can be interpreted as the average probability of successfully implementing the channel $\mathcal{E}_0$ on $\rho_i$, up to isometry \cite{Anu2019}. In the following, we show how to bound that fidelity using only the measurements performed during the protocol when sending the probe states $\Phi_i$ through the channel.

\subsection{Bounding Channel Fidelity with State Fidelities}\label{sec:certificationBound}

In the following, we prove the key theoretical result of this study (\ref{ineq:MainResult3}), which allows one to bound the quality of a channel with probe states fidelities to a maximally-entangled state, up to isometries. More precisely, we show the following lemma:

\begin{lem}[Probabilistic Channel Certification]
Let us consider a deterministic channel $\mathcal{E}_0$ from $\mathcal{L}(\mathcal{H}_i)$ to $\mathcal{L}(\mathcal{H}_o)$, a probabilistic channel $\mathcal{E}$ from $\mathcal{L}(\mathcal{H}_{\mathcal{A}_1})$ to $\mathcal{L}(\mathcal{H}_{\mathcal{B}})$, and a secondary space $\mathcal{L}(\mathcal{H}_{\mathcal{A}_2})$. For any isometries $\Gamma^{\mathcal{B}}: \mathcal{H}_{\mathcal{B}} \longrightarrow \mathcal{H}_{\mathcal{B}}\otimes\mathcal{H}_o$ and $\Gamma^{\mathcal{A}_1/\mathcal{A}_2}: \mathcal{H}_{\mathcal{A}_1/\mathcal{A}_2} \longrightarrow \mathcal{H}_{\mathcal{A}_1/\mathcal{A}_2}\otimes\mathcal{H}_i$ we define the corresponding fidelities of a state $\Phi_i \in \mathcal{L}(\mathcal{H}_{\mathcal{A}_1}\otimes\mathcal{H}_{\mathcal{A}_2})$ to a maximally-entangled state $\Phi_+\in\mathcal{L}(\mathcal{H}_i^{\otimes 2})$, before and after application of the channels:
\begin{equation}\label{eq:InputOuputFidDI}
    \begin{aligned}
    F^i = &\: F((\Lambda^{\mathcal{A}_1}\otimes \Lambda^{\mathcal{A}_2})[\Phi_i],\Phi_+)\\
    F^o =&\: F((\Lambda^{\mathcal{B}}\otimes\Lambda^{\mathcal{A}_2})[(\mathcal{E}\otimes\mathds{I})[\Phi_i]]/t(\mathcal{E}\vert\Phi_i),(\mathcal{E}_0\otimes\mathds{I})[\Phi_+])
\end{aligned}
\end{equation}
where $\Lambda^\mathcal{P}[\cdot] = \tr_{\mathcal{P}}(\Gamma^{\mathcal{P}}[\cdot])$ for $\mathcal{P}=\mathcal{A}_1,\mathcal{A}_2$ or $\mathcal{B}$. Then there exist two isometries $\Gamma_i$ (encoding map) and $\Gamma_o$ (decoding map), built from $\Gamma^{\mathcal{A}_1}$, $\Gamma^{\mathcal{A}_2}$ and $\Gamma^{\mathcal{B}}$, such that channel fidelities between $\mathcal{E}$ and $\mathcal{E}_0$ are bounded, up to isometries: 
\begin{equation}\label{ineq:MainResult2}
     \sqrt{1-\mathcal{F}_\diamond(\mathcal{E}_{i,o},\mathcal{E}_0)} \leq d \cdot  \sqrt{1-\mathcal{F}_J(\mathcal{E}_{i,o},\mathcal{E}_0)} \leq d \cdot \sin\biggl( \arcsin\bigl(C^i/t(\mathcal{E}\vert\Phi_i)\bigr) + \arcsin C^o\biggr)
\end{equation}
where $d=\dim \mathcal{H}_i$, $\mathcal{E}_{i,o}=\tr_{ext}\bigl((\Gamma_o \circ \mathcal{E} \circ \Gamma_i) [\rho_{\mathcal{A}_1} \otimes \bullet \: ]\bigr)$, $\rho_{\mathcal{A}_1}$ an ancillary state in $\mathcal{L}(\mathcal{H}_{\mathcal{A}_1})$, and $C^i = \sqrt{1-F^i}$ and $C^o = \sqrt{1-F^o}$.
\label{thm:channelCertif}
\end{lem}

\textbf{\textit{Proof:}} This theorem is a generalization of the result from~\cite{Sekatski2018} to trace-decreasing channels. We follow the same guidelines for our proof. First we define $\Phi_i' = (\mathds{I}\otimes\Lambda^{\mathcal{A}_2})[\Phi_i]$, in order to forget about the injection on Alice's second subsystem, that does not have much relevance here as the channel $\mathcal{E}$ leaves it unaffected. Then, we note that according to Proposition 2 from~\cite{Sekatski2018}, if one is given a target pure state $\rho_0 \in \mathcal{L}(\mathcal{H}_{sys})$ and any state $\Gamma[\rho] \in \mathcal{L}(\mathcal{H}_{ext}\otimes\mathcal{H}_{sys})$ with $\Lambda[\rho] = \tr_{ext}(\Gamma[\rho])\in \mathcal{L}(\mathcal{H}_{sys})$, then the following relation holds
\begin{equation}
    F(\Lambda[\rho],\rho_0) = F(\Gamma[\rho],\rho_{ext}\otimes\rho_0)
\end{equation}
with $\rho_{ext} = \dfrac{\tr_{sys}(\Gamma[\rho]\rho_0\otimes\mathds{I})}{\tr(\Gamma[\rho]\rho_0\otimes\mathds{I})}$ . We start by applying this proposition to $F^i$, with $\mathcal{H}_{sys} = \mathcal{H}_i\otimes\mathcal{H}_i$ and $\mathcal{H}_{ext} = \mathcal{H}_{\mathcal{A}_1}$, so we get a new expression of that fidelity:
\begin{equation}
    F^i = F\bigl((\Gamma^{\mathcal{A}_1}\otimes\mathds{I})[\Phi_i'],\rho_{\mathcal{A}_1}\otimes\Phi_+\bigr)
\end{equation}
with $\rho_{\mathcal{A}_1} = \dfrac{\tr_{\mathcal{H}_i\otimes\mathcal{H}_i}\bigl((\Gamma^{\mathcal{A}_1}\otimes\mathds{I})[\Phi'_i]\proj{\Phi_+}\otimes\mathds{I}\bigr)}{\tr\bigl((\Gamma^{\mathcal{A}_1}\otimes\mathds{I})[\Phi'_i]\proj{\Phi_+}\otimes\mathds{I}\bigr)}$. The isometry $\Gamma^{\mathcal{A}_1}$ can be written as a unitary, applied on a Hilbert state of larger dimension, so that $(\Gamma^{\mathcal{A}_1}\otimes\mathds{I})[\Phi'_i] = (U^i\otimes\mathds{I})[\sigma_{ext} \otimes \Phi'_i]$ with $\sigma_{ext}$ an ancillary pure state and $U^i$ a unitary operation applied on that state and $\mathcal{H}_{\mathcal{A}_1}$. This way we get:
\begin{equation}
    \begin{aligned}
    F^i =&\  F\bigl((U^i\otimes\mathds{I})[\sigma_{ext} \otimes \Phi_i'],\rho_{\mathcal{A}_1}\otimes\Phi_+\bigr)\\
    =&\  F\bigl(\sigma_{ext} \otimes \Phi_i',({U^i}^\dagger\otimes\mathds{I})[\rho_{\mathcal{A}_1}\otimes\Phi_+]\bigr)\\
    \leq&\  F\bigl(\Phi_i',\tr_{ext,i}({U^i}^\dagger\otimes\mathds{I})[\rho_{\mathcal{A}_1}\otimes\Phi_+]\bigr)
    \end{aligned}
\end{equation}
where we use the fidelity invariance under unitary operation, and the fact that it can only increase upon tracing out, here of the Hilbert space of $\sigma_{ext}$. This allows us to define the encoding map $\Gamma^i = ({U^i}^\dagger\otimes\mathds{I})[\: \bullet \:]$ so we have:
\begin{equation}
    F^i \leq F\bigl(\Phi_i',\tr_{ext,i}(\Gamma^i[\rho_{\mathcal{A}_1}\otimes\Phi_+])\bigr)
\end{equation}
Now by defining the decoding map $\Gamma^o = \Gamma^{\mathcal{B}}$, we can apply the map $\Gamma^o \circ \bar{\mathcal{E}} \otimes \mathds{I}$ to both states on the right-hand side of the inequality, and use Lemma~\ref{thm:ExtProcess} for extended metric monotonicity, and once again fidelity monotonicity when tracing out subsystems:
\begin{equation}
    \begin{aligned}
    C^i &= \sqrt{1- F^i} \\&\geq C\bigl(\Phi_i',\tr_{ext,i}(\Gamma^i[\rho_{\mathcal{A}_1}\otimes\Phi_+])\bigr)\\&
    \geq t(\bar{\mathcal{E}}\vert\Phi_i')\cdot C\bigl((\Gamma^o\circ\bar{\mathcal{E}}\otimes\mathds{I})[\Phi_i']/t(\bar{\mathcal{E}}\vert\Phi_i'),\tr_{ext,i}((\Gamma^o\circ\bar{\mathcal{E}}\circ\Gamma^i\otimes \mathds{I})[\rho_{\mathcal{A}_1}\otimes\Phi_+])/\Tilde{t}\: \bigr)\\&
    \geq t(\bar{\mathcal{E}}\vert\Phi_i')\cdot C\bigl((\Lambda^{\mathcal{B}}\circ\bar{\mathcal{E}}\otimes\mathds{I})[\Phi_i']/t(\bar{\mathcal{E}}\vert\Phi_i'),\tr_{ext}((\Gamma^o\circ\bar{\mathcal{E}}\circ\Gamma^i\otimes \mathds{I})[\rho_{\mathcal{A}_1}\otimes\Phi_+])/\Tilde{t}\: \bigr)
    \end{aligned}
\end{equation}
Here in order to apply Lemma \ref{thm:ExtProcess}, we noted that $t(\bar{\mathcal{E}}\vert \Phi_i') = \tr((\bar{\mathcal{E}}\otimes\mathds{I})[\Phi_i'])$ is the transmissivity of the first state, which does not vary under application of isometry $\Gamma^o$. Also $\Tilde{t}$ is the transmissivity of the second state, $i.e.$ $\tilde{t} = t(\bar{\mathcal{E}}_{i,o}\vert\Phi_+)$ as we define $\bar{\mathcal{E}}_{i,o} =\tr_{ext}((\Gamma^o\circ\bar{\mathcal{E}}\circ\Gamma^i)[\rho_{\mathcal{A}_1}\otimes\:\bullet\:])$. The last partial trace in the inequality is carried out over all subsystems except $\mathcal{L}(\mathcal{H}_o\otimes\mathcal{H}_i)$, such that the distance can only decrease. Noting that $(\Lambda^{\mathcal{B}}\circ\bar{\mathcal{E}}\otimes\mathds{I})[\Phi_i']/t(\bar{\mathcal{E}}\vert\Phi_i') = (\Lambda^{\mathcal{B}}\otimes\Lambda^{\mathcal{A}_2})\circ(\bar{\mathcal{E}}\otimes\mathds{I})[\Phi_i]/t(\bar{\mathcal{E}}\vert\Phi_i)$ we get:
\begin{equation}\label{ineq:Cinputdemo}
    C^i/ t(\bar{\mathcal{E}}\vert\Phi_i') \geq C\bigl((\Lambda^{\mathcal{B}}\otimes\Lambda^{\mathcal{A}_2})[\bar{\Phi}_o],(\bar{\mathcal{E}}_{i,o}\otimes \mathds{I})[\Phi_+]/t(\bar{\mathcal{E}}_{i,o}\vert\Phi_+)\bigr)
\end{equation}
Finally, we can apply an equivalent of triangular inequality to Ulhmann's fidelity:
\begin{equation}
\begin{aligned}
     \arccos\sqrt{F(\rho_1,\rho_3)} &= \arcsin C(\rho_1,\rho_3)\\
     &\hspace{-2cm}\leq \arccos\sqrt{F(\rho_1,\rho_2)} + \arccos\sqrt{F(\rho_2,\rho_3)}\\ 
     &\hspace{-2cm}= \arcsin C(\rho_1,\rho_2) + \arcsin C(\rho_2,\rho_3)
\end{aligned}
\end{equation}
with the following states
\begin{align}
    \rho_1 &= (\bar{\mathcal{E}}_{i,o}\otimes\mathds{I})[\Phi_+]/t(\bar{\mathcal{E}}_{i,o}\vert\Phi_+)\\
    \rho_2 &= (\Lambda^{\mathcal{B}}\otimes\Lambda^{\mathcal{A}_2})[\bar{\Phi}_o]\\
    \rho_3 &= (\mathcal{E}_0\otimes\mathds{I})[\Phi_+].
\end{align}
$\rho_1$ is the output state of the real channel when sending a perfect maximally entangled state, $\rho_2$ the average output state we effectively measure after application of the real channel on a close-to-maximally-entangled state, and $\rho_3$ the output state of the target channel when sending a perfect maximally entangled state. This way we have $C(\rho_2,\rho_3) = C^o$ and $C(\rho_1,\rho_3)=\arccos\sqrt{\mathcal{F}_J(\bar{\mathcal{E}}_{i,o},\mathcal{E}_0)}$ by definition, and $C(\rho_1,\rho_2)\leq C^i/t(\bar{\mathcal{E}}\vert\Phi_i)$ via inequality (\ref{ineq:Cinputdemo}). This gives the final result: 
\begin{equation}
    \arccos\sqrt{F_J(\bar{\mathcal{E}}_{i,o},\mathcal{E}_0)}=\arcsin \mathcal{C}_J(\bar{\mathcal{E}}_{i,o},\mathcal{E}_0) \leq  \arcsin\bigl(C^i/t(\bar{\mathcal{E}}\vert\Phi_i)\bigr) + \arcsin(C^o)
\end{equation}
From here, one just has to use the comparison between diamond and Choi-Jamio\l kowski distances, as we showed in Lemma~\ref{thm:MetricEquivalence}, in order to get the bound (\ref{ineq:MainResult2}) and lemma~\ref{thm:channelCertif} $\blacksquare$.\\

\vspace{0.3cm}
\begin{figure}[htbp]
    \centering
\includegraphics[height=.18\linewidth]{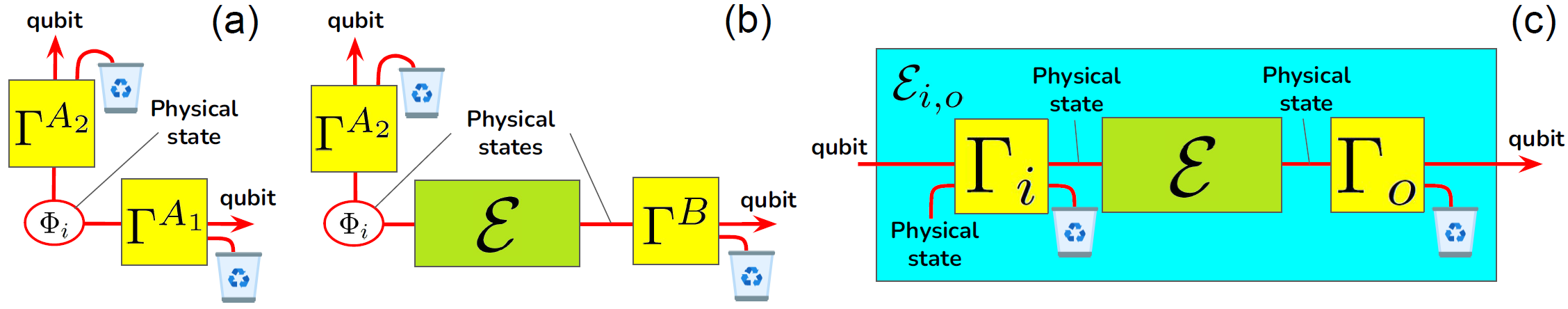}
    \caption{Schematic representation of isometries' actions on (a) the input state $\Phi_i$, (b) the output state $\Phi_o$, and (c) the quantum channel $\mathcal{E}$. All isometries, except $\Gamma_i$, extract a qubit state from a physical system. Only the qubit state remains as the other degrees of freedom are discarded. The isometry $\Gamma_i$ encodes a qubit state onto a physical state that can be fed into the quantum channel $\mathcal{E}$. $\Gamma_i$ schematically performs the inverse operation than $\Gamma^{\mathcal{A}_1}$. Together, $\Gamma_i$ and $\Gamma_o$ extract a qubit-to-qubit channel from a physical channel.}
    \label{fig:isometries}
\end{figure}

The isometries mentioned in the proof are fundamental in a device independent study, in order to extract ideal qubit spaces to real-world infinite-dimension physical Hilbert spaces.  $\Gamma^{\mathcal{A}_1}$, $\Gamma^{\mathcal{A}_2}$ and $\Gamma^{\mathcal{B}}$ are the same type of isometries as in all standard self-testing results \cite{supic2020}, and they extract a qubit state from the full state of a physical system, which encompasses all other degrees of freedom. The unused degrees of freedom are then thrown away. The channel isometries $\Gamma_i$ and $\Gamma_o$ were introduced more recently \cite{Sekatski2018} and together extract a qubit channel from a physical channel acting on all degrees of freedom of a physical system. The decoding map $\Gamma_o$ performs the same operation as $\Gamma^{\mathcal{B}}$, extracting a qubit out of a physical system. The encoding map $\Gamma_i$ however, performs the inverse operation than the other isometries, encoding the qubit state into a physical system, such that it can be fed into the physical channel. We give a schematic view of these channels in Fig.~\ref{fig:isometries}. In Protocol 1, the input state $\Phi_i$ is assumed to be fully characterized, so we can ignore the encoding map and $\Gamma_i = \Gamma^{\mathcal{A}_1} = \mathds{I}$. Yet, we must include that isometry when building the fully device independent Protocol 2.\\

The result we just showed allows us to deduce the protocol's success probability, by evaluating the fidelities $F^i$ and $F^o$ to a Bell state, as well as the transmissivity $t(\bar{\mathcal{E}}\vert\Phi_i)$. The two following paragraphs are dedicated to evaluating $F^o$ and $t(\bar{\mathcal{E}}\vert\Phi_i)$, using data received by Alice and Bob only. In order to tie up the security of Protocol 2, we tackle the certification of $F^i$ in a later paragraph.

\subsection{Certifying the average Bell output state}

In order to certify the average output probe state $\bar{\Phi}_o = (\bar{\mathcal{E}}\otimes\mathds{I})[\Phi_i]/t(\bar{\mathcal{E}}\vert\Phi_i)$, we use self-testing results from previous works \cite{unnikrishnan} that consider steering-based certification of the Bell pair in a finite number of measurement rounds, without making the common IID assumption. In a non-IID scenario the channel may change its behaviour throughout the protocol, such that we define $\mathcal{E}_{k|[k-1]}$ the expression of the channel when Alice sends the $k-$th state. Then, we call the output state $\Phi_k = (\mathcal{E}_{k|[k-1]}\otimes\mathds{I})[\Phi_i]/t_k$ when Alice sends the state $\Phi_i$, with $t_k = t(\mathcal{E}_{k|[k-1]}\vert\Phi_i)$ being the transmissivity of the state $\Phi_i$ through the channel $\mathcal{E}_{k|[k-1]}$. Using this notation, we can define the following state:
\begin{equation}\label{eq:avgTransmittedState1}
    \bar{\Phi}_t = \bigl(\sum_{k=1}^{N+1}\mathcal{T}_k\Phi_k\bigr)/(K+1)
\end{equation}
where $\mathcal{T}_k = 1$ when a state is detected by Bob, and $\mathcal{T}_k = 0$ otherwise, such that $\sum_{k=1}^{N+1}\mathcal{T}_k = K+1$. We take $\mathcal{T}_r = 1$, in order to include the state $\Phi_r = (\mathcal{E}_{r|[r-1]}\otimes\mathds{I})[\Phi_i]/t_r$  in the sum. $\bar{\Phi}_t$ is the average output state of the protocol, in the particular case $\rho_i = \Phi_i$ and when the protocol did not abort. Therefore, we expect $\bar{\Phi}_t$ to be a good approximation for $\bar{\Phi}_o$, the output state when sending $\Phi_i$ through the average channel $\bar{\mathcal{E}}_{i,o}$. However, we leave that consideration for the next subsection, and now show certification results for $\bar{\Phi}_t$ in place of $\bar{\Phi}_o$.\\

When $\rho_i = \Phi_i$, we can see our protocol as an attempt to authenticate an unmeasured Bell pair, emerging from an untrusted source. The latter is made of Alice's trusted source, sending copies of $\Phi_i$ in the untrusted quantum channel. The state emerging from the $\mathcal{E}_r$ is the unmeasured pair, and the $K$ other output states are measured by Alice and Bob in order to perform a Bell test. In that case, our protocol corresponds to that described in \cite{unnikrishnan,Anu2019}, such that we can apply the self-testing-based security results from that work, in a non-IID and 1sDI setting, to our protocol:

\begin{prop}\label{thm:AnuTheorem}
Let us consider our protocol where $\rho_i = \Phi_i$, Alice and Bob measure $K$ states and witness an average violation of either steering inequality of $2-\epsilon$. We can bound the fidelity of the average state $\bar{\Phi}_t$ to a maximally-entangled state $\Phi_+$, up to isometry. More precisely, there exist isometries $\Gamma^{\mathcal{A}_2}$ and $\Gamma^{\mathcal{B}}$ acting respectively on $L(\mathcal{H}_{\mathcal{A}_2})$ and $L(\mathcal{H}_{\mathcal{B}})$, such that by defining the local maps $\Lambda^{\mathcal{A}_2}[\cdot]=\emph{\tr}_{\mathcal{A}_2}(\Gamma^{\mathcal{A}_2}[\cdot])$ and $\Lambda^{\mathcal{B}}[\cdot]=\emph{\tr}_{\mathcal{B}}(\Gamma^{\mathcal{B}}[\cdot])$, for any $x>0$ we have with probability at least $(1-e^{-x})$:
\begin{equation}\label{AnuResultSteering}
F((\Lambda^{\mathcal{B}}\otimes\Lambda^{\mathcal{A}_2})[\bar{\Phi}_t],\Phi_+) \geq 1 - \alpha \cdot f_x(\epsilon,K) \underset{K\to+\infty}{\longrightarrow} 1-\alpha\epsilon
\end{equation}
with $\alpha$ a constant and $f$ a function which both depend on the inequality used:
\begin{equation}
f_x(\epsilon,K) = 8\sqrt{\dfrac{x}{K}} + \dfrac{\epsilon}{2} + \dfrac{\epsilon + 8/K}{2 + 1/K} 
\end{equation}\label{AnuResultCHSH}
and $\alpha = 1.26$
\end{prop}

It is worth noting that as the $r$-th state is left unmeasured in this protocol, and we assume the channel's operator has no way of guessing $r$, then the measurements performed on the test EPR pairs follow the same statistics in the general case than in the special case $\rho_i = \Phi_i$. We can therefore use the correlations witnessed in our protocol in Proposition~\ref{thm:AnuTheorem}, even when sending any quantum message $\rho_i$ in $r$-th position, in order to certify the hypothetical state $\bar{\Phi}_t$ up to isometry.\\

\begin{figure}[htbp]
  \centering
  \includegraphics[height=.3\linewidth]{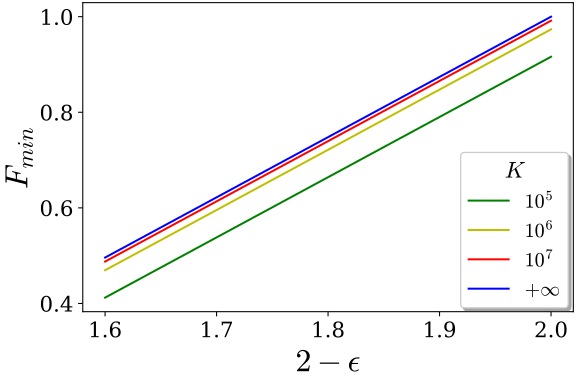}  
\caption{Minimum fidelity of the average output state to a Bell state, up to isometries, as a function of the deviation to maximum violation. As we make no IID assumption, we give the evolution for different numbers $K$ of states measured. We set a confidence level $1-e^{-x} \approx 0.999$.}
\label{fig:STAnuSteering}
\end{figure}

Finally, we give some insight on the behaviour of those bounds with the parameters of the problem. First, we can take $x=7$ in order to get a bound with almost absolute certainty, as $(1-e^{-x}) \approx 0.999$. The corresponding term in $\sqrt{x/K}$ can be made arbitrarily small by measuring a large number $K$ of states. Similarly, when measuring a reasonable amount of states $K > 10^8$, we reach the asymptotic regime where the fidelity is simply bounded by $1-\alpha\epsilon$. These results are presented in Fig.~\ref{fig:STAnuSteering}.

\vspace{0.5cm}


\subsection{Errors due to Post-Selection and Finite Statistics}\label{StatError}

We now show the validity of approximating the state $\bar{\Phi}_o$ (\ref{eq:avgBell}) with $\bar{\Phi}_t$ (\ref{eq:avgTransmittedState1}), as well as the following approximation:
\begin{equation}\label{eq:evaluateAvgPassProba}
    t(\bar{\mathcal{E}}\vert\Phi_i) \approx R = \dfrac{K+1}{N+1}
\end{equation}
where $K+1=\vert \mathbb{S}_P \vert$ is the number of states that Bob is able to measure after they are sent through the channel. Alice and Bob have direct access to the value $R$ in the end of the protocol, as the fraction of states that successfully pass through the channel, which we identify as the heralding efficiency $\eta_s$. Therefore, they can easily evaluate $t(\bar{\mathcal{E}}\vert\Phi_i)$ by using (\ref{eq:evaluateAvgPassProba}).

\begin{prop}\label{prop:staterror}
In our protocol, provided that Bob measured a large enough number $K+1$ of states, the transmissivity $t(\bar{\mathcal{E}}\vert\Phi_i)$ of $\Phi_i$ through the average channel $\bar{\mathcal{E}}$ can be approximated by the proportion $R$ of states which were successfully detected by Bob, and the state $\bar{\Phi}_o$ can be approximated by $\bar{\Phi}_t$. More precisely, for any $x>0$ we have with probability at least $(1-2e^{-x})^2$:
\begin{align}
\arccos\sqrt{F(\bar{\Phi}_t,\bar{\Phi}_o)} &\leq \Delta_x(R,K) \label{eq:ApproxState}\\
t(\bar{\mathcal{E}}\vert\Phi_i) &\geq \tau_x(R,K),\label{eq:ApproxTransm}
\end{align}
where  
\begin{align}
\Delta_x(R,K) &= \arccos{\tfrac{1-3\: \delta_x(R,K)}{1-\delta_x(R,K)}}\\
\tau_x(R,K) &= R\:(1-\delta_x(R,K))\\
\delta_x(R,K) &= \tfrac{1}{K+1} + \sqrt{\tfrac{2x}{R(K+1)}}
\end{align}
\end{prop}

In particular, this proposition gives the error terms mentioned given in Eqs.~(\ref{eq:RelativeDifferencePassProbaMain}) and (\ref{ineq:FinalResultMain}) from the Methods.\\ \

\textit{\textbf{Proof.}} We prove this proposition in two main steps, first showing bound (\ref{eq:ApproxTransm}) on the transmissivity $t(\bar{\mathcal{E}}\vert\Phi_i)$ with a certain probability, and secondly assuming (\ref{eq:ApproxTransm}) in order to derive bound (\ref{eq:ApproxState}) on the trace distance $D(\bar{\Phi}_t,\bar{\Phi}_o)$ with another probability. In each step, we define a random variable which, without assuming IID statistics, is identified as a martingale. The bounds are therefore derived from the Azuma-Hoeffding inequality.\\

First let us rewrite the transmissivity using the notation from the last paragraph:
\begin{equation}\label{eq:avgPassProbaRewrite}
    t(\bar{\mathcal{E}}\vert\Phi_i) = \tr\biggl(\dfrac{1}{N+1}\sum_{k=1}^{N+1}\mathcal{E}_k[\Phi_i]\biggr)= \dfrac{1}{N+1}\sum_{k=1}^{N+1}t_k
\end{equation}
Alice and Bob do not have direct access to that quantity, as they cannot measure $t_k$ individually. However, they have access to the random variables $\{\mathcal{T}_k\}_{1\leq k\leq N+1}$ defined in the previous subsection, the sum of which gives the number of states that were measured by Bob during the protocol:
\begin{equation}\label{eq:sumTransmVar}
    K+1 = \vert \mathbb{S}_P \vert =  \sum_{k=1}^{N+1}\mathcal{T}_k 
\end{equation}
As no IID assumption is made, the variables $\mathcal{T}_k$ may differ from one another and depend on the experiment's history. Taking the difference with transmissivities, we define a new random variable, for $j\neq k$ :
\begin{equation}\label{eq:DiffVarDef}
    \mathcal{D}_j = \sum_{\substack{k=1\\k\neq r}}^{j}(\mathcal{T}_k-\mathbb{E}[\mathcal{T}_k]) =  \sum_{\substack{k=1\\k\neq r}}^{j}(\mathcal{T}_k-t_k) 
\end{equation}
and $\mathcal{D}_r = \mathcal{D}_{r-1}$. The expectation value of $\mathcal{D}_j$ is finite for any $j$, as it is zero, and we have $\mathbb{E}[\mathcal{D}_{j+1}\vert H_j] = \mathcal{D}_j$, where $H_j$ is the history of the experiment after the $j$-th state is sent through the channel. This makes $\mathcal{D}_j$ a martingale. We also note that $\vert \mathcal{D}_{j+1} - \mathcal{D}_j\vert \leq 1$ for any $j$, such that we can apply the Azuma-Hoeffding inequality, giving:
\begin{equation}\label{eq:AzumaPass1}
    \Pr (\vert\mathcal{D}_j\vert \geq \gamma) \leq 2\exp\biggl(-\dfrac{\gamma^2}{2j}\biggr)
\end{equation}

Now we note that $\mathcal{D}_{N+1} = (N+1)\cdot(R  -  t(\bar{\mathcal{E}}\vert\Phi_i) )- 1 +t_r$, such that by taking $j = N+1$ we get:

\begin{equation}\label{eq:AzumaPass2}
    \Pr \bigl(\tfrac{-\gamma+1-t_r}{N+1}\leq  R- t(\bar{\mathcal{E}}\vert\Phi_i) \leq \tfrac{\gamma+1-t_r}{N+1}\bigr) \geq 1-2\exp\biggl(-\dfrac{\gamma^2}{2(N+1)}\biggr)
\end{equation}
Now considering $0\leq 1-t_r \leq 1$, and taking the relative difference we get:
\begin{equation}\label{eq:AzumaPass3}
        \Pr \bigl( \tfrac{\vert R - t(\bar{\mathcal{E}}\vert\Phi_i) \vert}{R}\leq \tfrac{\gamma+1}{K+1}\bigr)\geq 1-2\exp\biggl(-\dfrac{\gamma^2}{2(N+1)}\biggr)
\end{equation}
such that by taking $x = \tfrac{\gamma^2}{2(N+1)} > 0$ we get the following bound with probability at least $(1-2e^{-x})$:
\begin{equation}\label{eq:RelativeDifferencePassProba2}
    \vert \Delta_1 \vert = \dfrac{\vert R- t(\bar{\mathcal{E}}\vert\Phi_i) \vert}{R} \leq \delta_x(R,K)
\end{equation}
where $\delta_x(R,K) = \tfrac{1}{K+1} + \sqrt{\tfrac{2x}{R(K+1)}}$. This straightforwardly gives the inequality in (\ref{eq:ApproxTransm}): 
\begin{equation}\label{eq:ApproxTransm2}
    t(\bar{\mathcal{E}}\vert\Phi_i) \geq \tau_x(R,K)
\end{equation}
where $\tau_x(R,K) = R\:(1-\delta_x(R,K))$. Note that as the value of $x$ can be chosen arbitrarily, we can take the same value as in Proposition~\ref{thm:AnuTheorem}, which will simplify the notation.\\

To show the bound (\ref{eq:ApproxState}), we now assume (\ref{eq:RelativeDifferencePassProba2}) such that $\vert \Delta_1\vert\leq \delta_x(R,K)$. We note that one can re-write $\bar{\Phi}_o$ using the states $\Phi_k$ and transmissivities $t_k$:
\begin{equation}\label{eq:rewritePhio}
    \begin{aligned}
    \bar{\Phi}_o \: &= (\bar{\mathcal{E}}\otimes\mathds{I})[\Phi_i]/t(\bar{\mathcal{E}}\vert\Phi_i) \\&= \bigl(\dfrac{1}{N+1}\sum_{k=1}^{N+1}(\mathcal{E}_k\otimes\mathds{I})[\Phi_i]\bigr)/ t(\bar{\mathcal{E}}\vert\Phi_i)\\
    &= \bigl(\dfrac{1}{N+1}\sum_{k=1}^{N+1}t_k \Phi_k\bigr)/ t(\bar{\mathcal{E}}\vert\Phi_i)
    \end{aligned}
\end{equation}
We pick a projector $P$ that allows to express the trace distance between $\bar{\Phi}_o$ and $\bar{\Phi}_t$:
\begin{equation}\label{eq:distanceAvgPS}
    \begin{aligned}
    D(\bar{\Phi}_t,\bar{\Phi}_o) &= \tr(P(\bar{\Phi}_t-\bar{\Phi}_o))\\
    &= \sum_{k=1}^{N+1}\bigl(\tfrac{\mathcal{T}_k}{K+1}-\tfrac{t_k}{(N+1)t(\bar{\mathcal{E}}\vert\Phi_i)}\bigr) \tr(P\Phi_k)\\
    &\leq \biggl(\biggl\vert\sum_{k=1}^{N+1} \bigl(\tfrac{t(\bar{\mathcal{E}}\vert\Phi_i)}{K+1} - \tfrac{1}{N+1}\bigr) \mathcal{T}_k\tr(P\Phi_k) \biggr\vert + \biggl\vert\sum_{k=1}^{N+1}\tfrac{\mathcal{T}_k - t_k}{N+1}\tr(P\Phi_k) \biggr\vert\biggr)/t(\bar{\mathcal{E}}\vert\Phi_i)
\end{aligned}
\end{equation}
Let us call the second term in parenthesis $\vert \Delta_2 \vert$ and bound the first term:
\begin{equation}\label{eq:distanceAvgPS2}
    \begin{aligned}
    \biggl\vert\sum_{k=1}^{N+1} \biggl(\tfrac{t(\bar{\mathcal{E}}\vert\Phi_i)}{K+1} - \tfrac{1}{N+1}\biggr) \mathcal{T}_k\tr(P\Phi_k) \biggr\vert &= \sum_{k=1}^{N+1} \mathcal{T}_k\tr(P\Phi_k)\biggl\vert \tfrac{t(\bar{\mathcal{E}}\vert\Phi_i)}{K+1}-\tfrac{1}{N+1} \biggr\vert \\
    &\leq (K+1) \biggl\vert \tfrac{t(\bar{\mathcal{E}}\vert\Phi_i)}{K+1}-\tfrac{1}{N+1} \biggr\vert \\ 
    & = \biggl\vert t(\bar{\mathcal{E}}\vert\Phi_i)- R\biggr\vert\\
    &\leq R\: \delta_x(R,K)
\end{aligned}
\end{equation}

In order to bound $\vert\Delta_2\vert$, we make the exact same proof as for $\vert\Delta_1\vert$, taking $\tr(P\Phi_k)\cdot\mathcal{T}_k$ in place of $\mathcal{T}_k$ and $\tr(P\Phi_k)\cdot t_k$ in place of $t_k$, when defining $\mathcal{D}_j$ in equation (\ref{eq:DiffVarDef}). This new sum of variables $\Tilde{\mathcal{D}}_j$ is still a martingale such that $\vert \Tilde{\mathcal{D}}_{j+1} - \Tilde{\mathcal{D}}_j\vert \leq 1$. Therefore it still verifies equation (\ref{eq:AzumaPass1}), and $\Tilde{\mathcal{D}}_{N+1} = (N+1)\Delta_2   - \tr(P\Phi_r)(1 - t_r) $ such that:

\begin{equation}\label{eq:AzumaPass4}
\begin{aligned}
    &\Pr \bigl(\tfrac{-\gamma+\tr(P\Phi_r)(1-t_r)}{N+1}\leq  \Delta_2 \leq \tfrac{\gamma+\tr(P\Phi_r)(1-t_r)}{N+1}\bigr) \\&\hspace{2cm}\geq 1-2\exp\biggl(-\dfrac{\Tilde{\gamma}^2}{2(N+1)}\biggr)
\end{aligned}
\end{equation}
As $0\leq \tr(P\Phi_r)(1-t_r)\leq 1 $ we can simplify:

\begin{equation}\label{eq:AzumaPass5}
    \Pr \biggl(\vert\Delta_2\vert\leq \tfrac{\Tilde{\gamma}+1}{N+1}\biggr) \geq 1-2\exp\biggl(-\dfrac{\Tilde{\gamma}^2}{2(N+1)}\biggr)
\end{equation}
such that by taking $\tfrac{\Tilde{\gamma}^2}{2(N+1)} = x$ we get the following bound with probability at least
$(1-2e^{-x})$:
\begin{equation}\label{ineq:boundDelta2}
    \vert\Delta_2\vert \leq R\: \delta_x(R,K)
\end{equation}
This way, coming back to (\ref{eq:distanceAvgPS2}) we get:
\begin{equation}
     D(\bar{\Phi}_t,\bar{\Phi}_o)  \leq \dfrac{2R\: \delta_x(R,K)}{t(\bar{\mathcal{E}}\vert\Phi_i)} \leq \dfrac{2\: \delta_x(R,K)}{1-\delta_x(R,K)}
\end{equation}
Now we use a comparison between fidelity and trace distance $1-\sqrt{F} \leq D$ in order to bound the Bures' angle distance between $\bar{\Phi}_t$ and $\bar{\Phi}_o$:
\begin{equation}\label{eq:distanceAvgPS4}
\begin{aligned}
    A(\bar{\Phi}_t,\bar{\Phi}_o) &= \arccos{\sqrt{F(\bar{\Phi}_t,\bar{\Phi}_o)}}\leq \Delta_x(R,K)\\
    \textrm{where } \Delta_x(&R,K) = \arccos{\tfrac{1-3\: \delta_x(R,K)}{1-\delta_x(R,K)}}
\end{aligned}
\end{equation}

Finally, we point out that this bound is true with probability $(1-2e^{-x})$ and at the condition that bound (\ref{eq:RelativeDifferencePassProba2}) holds, which also happens with probability $(1-2e^{-x})$, such that both bounds hold with probability $(1-2e^{-x})^2$. This ties up the proof of Proposition~\ref{prop:staterror}. $\blacksquare$\\

This proposition highlights the purely statistics-induced error on states and transmissivities. It is mostly due to the fact that Alice and Bob only have access to a finite number of states, in a non-IID setting. Most importantly, as the channel is allowed to be lossy, these states only give information on a sample of the different expressions $\mathcal{E}_{k|[k-1]}$ that it might take during the protocol, causing more uncertainty than when certifying a source of state without channel. This error must be included in the bounds in order to derive the protocol's security. Also note that we can use this theorem when applying the injection map $\Lambda^{\mathcal{B}}\otimes\Lambda^{\mathcal{A}_2}$ defined in the previous subsection to both states, as we always have:
\begin{equation}
   F\bigl((\Lambda^{\mathcal{B}}\otimes\Lambda^{\mathcal{A}_2})[\bar{\Phi}_t],(\Lambda^{\mathcal{B}}\otimes\Lambda^{\mathcal{A}_2})[\bar{\Phi}_o]\bigr) \geq  F(\bar{\Phi}_t,\bar{\Phi}_o) 
\end{equation}

\begin{figure}[htbp]
\begin{subfigure}{.46\textwidth}
  \centering
  \includegraphics[height=.65\linewidth]{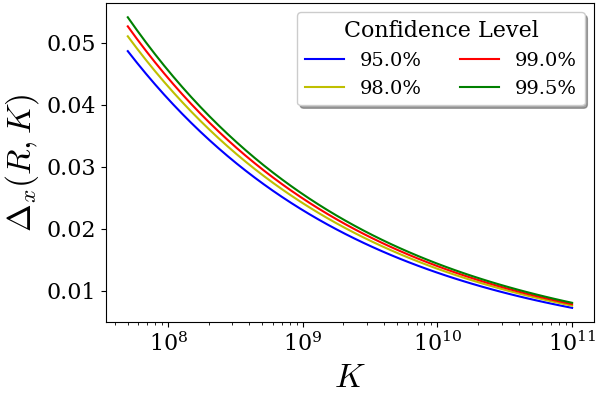}  
  \caption{For different minimum confidence levels, with a transmission ratio $R=50\%$.}
  \label{fig:CconfErr}
\end{subfigure}
\hspace{0.05\textwidth}
\begin{subfigure}{.46\textwidth}
  \centering
  \includegraphics[height=.65\linewidth]{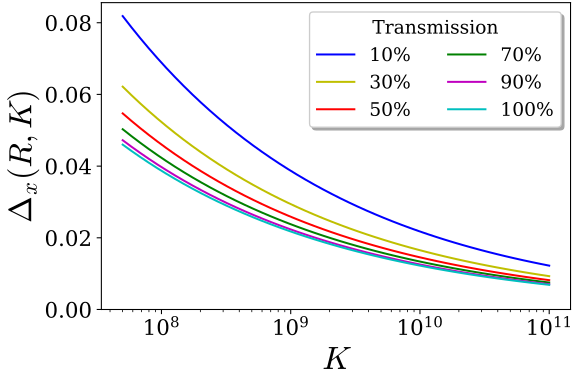}  
  \caption{With a minimum confidence level $99.5\%$, and for different transmission ratios $R$.}
  \label{fig:CtransErr}
\end{subfigure}
\caption{Minimum statistics-induced error $\Delta_x(R,K)$, as a function of the number of states measured $K$.}
\label{fig:error}
\end{figure}

This is fundamental to derive the final security bound for our protocol. Finally, we give some insight on the dependence of this error on the different parameters of the problem. First we notice that this error can be made arbitrarily small by measuring a large enough number $K$ of states, which still needs to be limited for practical applications. The error tends to increase with the confidence level, such that we need more states $K$ in order to ensure a smaller error with reasonable certainty. Similarly, the more lossy the channel is, \textit{i.e.} the smaller $R$, the bigger the error. Therefore having a lossy channel also imposes to measure more states in order to accurately certify the protocol. We give an idea of the evolution of that error in Fig.~\ref{fig:error}, for different confidence levels and different channel transmission ratios $R$. We see that with a transmission ratio $R=50\%$, corresponding to telecom light propagating in a 15km-long optical fiber or ideal quantum teleportation, we can ensure an error $\Delta_x(R,K)\leq 0.015$ with a confidence level of $99.5\%$, by measuring a reachable number of states $K\approx 10^{10}$.

\vspace{2cm}

\newpage 

\subsection{Certifying the transmitted quantum message}\label{sec:CertifAvg}

Combining the last three subsections allows us to extract a bound for the fidelity of the expected output state $\bar{\rho}_o$ to the quantum message $\rho_i$ up to isometry. We assume that Alice prepared $N$ states with fidelity $F^i$ to a Bell state, that Bob received $K$ of those states during the protocol, and that they measured an $\epsilon$-close to maximum violation of the steering inequality. First, they can use Lemma~\ref{thm:channelCertif}, implying that there exist isometries $\Gamma_i$, $\Gamma_o$, $\Gamma^{\mathcal{A}_1}$, $\Gamma^{\mathcal{A}_2}$, and $\Gamma^{\mathcal{B}}$, giving the result from (\ref{ineq:MainResult3}):
\begin{equation}\label{eq:OutputStateFidFinale1}
\begin{aligned}
    \sqrt{1-F\bigl((\Lambda^{\mathcal{B}}\otimes\Lambda^{\mathcal{A}_2})[\bar{\rho}_o],\rho_i\bigr)} &\leq  \sqrt{1-\mathcal{F}_\diamond(\bar{\mathcal{E}}_{i,o},\mathcal{E}_0)} \\&= \mathcal{C}_\diamond(\bar{\mathcal{E}}_{i,o},\mathcal{E}_0)\\ &\leq 2\: \mathcal{C}_J(\bar{\mathcal{E}}_{i,o},\mathcal{E}_0)\\ &\leq 2\: \sin\bigl( \arcsin\bigl(C^i/t(\bar{\mathcal{E}}\vert\Phi_i)\bigr) + \arcsin(C^o)\bigr)
\end{aligned}
\end{equation}
Now we fix $x>0$ in order to apply Proposition~\ref{prop:staterror}, such that we have both:
\begin{align}
    t(\bar{\mathcal{E}}\vert\Phi_i) &\geq \tau_x(R,K)\\
\label{ineq:statSineStates}
        \arccos{\sqrt{F}\bigl((\Lambda^{\mathcal{B}}\otimes\Lambda^{\mathcal{A}_2})[\bar{\Phi}_t],(\Lambda^{\mathcal{B}}\otimes\Lambda^{\mathcal{A}_2})[\bar{\Phi}_o]\bigr)} &\leq  \arccos{\sqrt{F}(\bar{\Phi}_t,\bar{\Phi}_o)} \\ &  \leq \Delta_x(R,K)
 \end{align}
with probability at least $(1-2e^{-x})^2$, where $\tau_x$ and $\Delta_x$ are functions detailed in paragraph~\ref{StatError}. In that case, we can apply the triangular inequality to $\arcsin(C^o)$:
\begin{equation}\label{ineq:TriangCoutAnnex}
    \arcsin(C^o)\leq \arcsin C\bigl((\Lambda^{\mathcal{B}}\otimes\Lambda^{\mathcal{A}_2})[\bar{\Phi}_t],\Phi_+\bigr) + \Delta_x(R,K) 
\end{equation}
and bound $t(\bar{\mathcal{E}}\vert\Phi_i)$ in order to get:
\begin{equation}\label{ineq:BoundInputNiid}
\arcsin\bigl(C^i/t(\bar{\mathcal{E}}\vert\Phi_i)\bigr) \leq \arcsin\bigl(C^i/ \tau_x(R,K)\bigr)
\end{equation}

We can then bound $C((\Lambda^{\mathcal{B}}\otimes\Lambda^{\mathcal{A}_2})[\bar{\Phi}_t],\Phi_+)$ using Proposition~\ref{thm:AnuTheorem}, with probability $(1-e^{-x})$:
\begin{equation}\label{ineq:AnuBoundFinaleDemo}
    \arcsin\bigl( C((\Lambda^{\mathcal{B}}\otimes\Lambda^{\mathcal{A}_2})[\bar{\Phi}_t],\Phi_+)\bigr) \leq \arcsin\sqrt{\alpha f_x(\epsilon,K)}
\end{equation}

Combining (\ref{eq:OutputStateFidFinale1}), (\ref{ineq:TriangCoutAnnex}), (\ref{ineq:BoundInputNiid}), and (\ref{ineq:AnuBoundFinaleDemo}) we can bound the input-output fidelity up to isometries:
\begin{equation}\label{eq:OutputStateFidFinale3}
    \sqrt{1-F(\bar{\rho}_o,\rho_i)} \leq 2\cdot\sin\biggl( \arcsin\bigl(C^i/\tau_x(R,K)\bigr) + \arcsin\sqrt{\alpha f_x(\epsilon,K)} + \Delta_x(R,K)\biggr)
\end{equation}
where $\alpha$ and $f$ are given in Proposition~\ref{thm:AnuTheorem}. This way, for any $x>0$ we can bound the output state fidelity to the input quantum message with probability at least $(1-e^{-x})\cdot (1-2e^{-x})^2$:
\begin{equation}\label{ineq:FinalResultAnnex}
    F(\bar{\rho}_o,\rho_i) \geq 1 - 4\cdot\sin^2\biggl( \arcsin\bigl(C^i/\tau_x\bigr) + \arcsin\sqrt{\alpha f_x(\epsilon,K)} + \Delta_x \biggr)
\end{equation}

\subsection{Input probe state certification and full device independence}

In protocol 2 Alice does not trust her measurement setup anymore, nor the source of input probe state $\Phi_i$. However we still make the IID assumption on that source. In that case we deduce the following theorem from a previous work \cite{unnikrishnan}:

\begin{prop}\label{thm:AnuTheoremDI}
When Alice measures an average violation of Bell inequality $2\sqrt{2}-\eta$ on $M$ identical copies of $\Phi_i$ with untrusted measurement apparatus, then for any $x>0$ we can bound the fidelity of $\Phi_i$ to $\Phi_+$ up to isometries, with probability $(1-e^{-x})$, meaning that there exists two isometries $\Gamma^{\mathcal{A}_1}$ and $\Gamma^{\mathcal{A}_2}$ on $\mathcal{L}(\mathcal{H}_{\mathcal{A}_1})$ and $\mathcal{L}(\mathcal{H}_{\mathcal{A}_2})$ such that: 
\begin{equation}\label{ineq:DIIIDfidelity}
      F\bigl((\Lambda^{\mathcal{A}_1}\otimes\Lambda^{\mathcal{A}_2})[\Phi_i],\Phi_+\bigr)  \geq 1-\alpha\cdot g_x(\eta,M) \underset{M\to+\infty}{\longrightarrow} 1-\alpha\cdot\eta
  \end{equation}
with $\Lambda^{\mathcal{A}_1}[\cdot]=\tr_{\mathcal{A}_1}(\Gamma^{\mathcal{A}_1}[\cdot])$, $\Lambda^{\mathcal{A}_2}[\cdot]=\tr_{\mathcal{A}_1}(\Gamma^{\mathcal{A}_2}[\cdot])$, $\alpha = 1.19$, and $g_x(\eta,M) = 8\sqrt{2x/M}+\eta$.\\

Then, if Alice and Bob measure $K$ states at the output of the channel with untrusted measurement apparatus, and witness an average violation of CHSH inequality of $2\sqrt{2}-\epsilon$, we can bound the fidelity of the average state $\bar{\Phi}_t$ to a maximally entangled state $\Phi_+$, up to isometries, with probability at least $(1-e^{-x})$, meaning that there exist isometries $\Gamma^{\mathcal{A}_2}$ and $\Gamma^{\mathcal{B}}$ on $L(\mathcal{H}_{\mathcal{A}_2})$ and $L(\mathcal{H}_{\mathcal{B}})$, such that: 
\begin{equation}\label{AnuResultOutputDI}
        F((\Lambda^{\mathcal{B}}\otimes\Lambda^{\mathcal{A}_2})[\bar{\Phi}_t],\Phi_+)  \geq 1 - \alpha \cdot f_x(\epsilon,K) \underset{K\to+\infty}{\longrightarrow} 1-\alpha\cdot\epsilon
\end{equation}
with $\Lambda^{\mathcal{A}_2}[\cdot]=\tr_{\mathcal{A}_2}(\Gamma^{\mathcal{A}_2}[\cdot])$, $\Lambda^{\mathcal{B}}[\cdot]=\tr_{\mathcal{B}}(\Gamma^{\mathcal{B}}[\cdot])$, $\alpha = 1.19$ and $f_x(\epsilon,K)  = 16\sqrt{\tfrac{2x}{K}} + \tfrac{3\epsilon}{4} + \tfrac{\epsilon + (4+2\sqrt{2})/K}{4 + 4/K}$.
\end{prop}

\begin{figure*}[htbp]
\begin{subfigure}{.32\textwidth}
  \centering
  \includegraphics[height=0.8\linewidth]{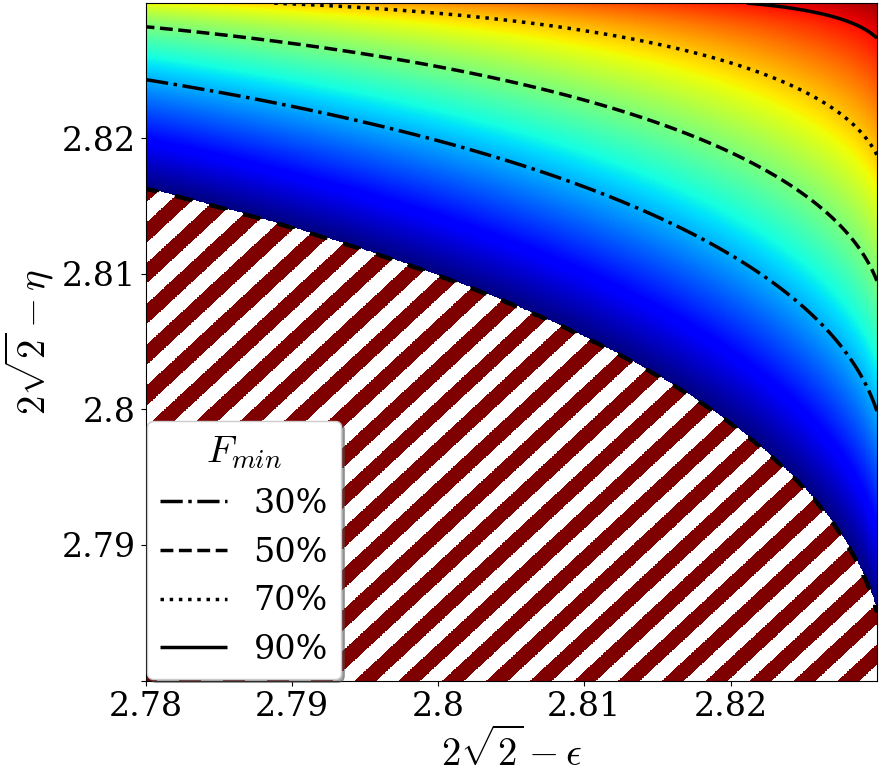}  
  \caption{R = 50\%}
  \label{fig:FullDI50}
\end{subfigure}
\begin{subfigure}{.32\textwidth}
  \centering
  \includegraphics[height=0.8\linewidth]{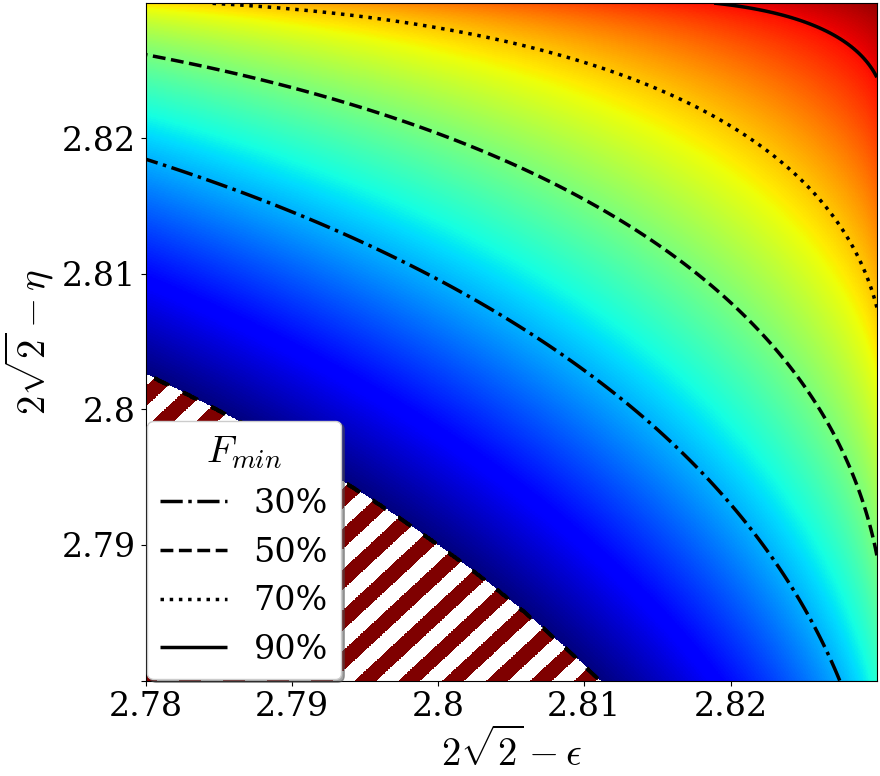}  
  \caption{R = 70\%}
  \label{fig:FullDI70}
\end{subfigure}
\begin{subfigure}{.32\textwidth}
  \centering
  \includegraphics[height=0.8\linewidth]{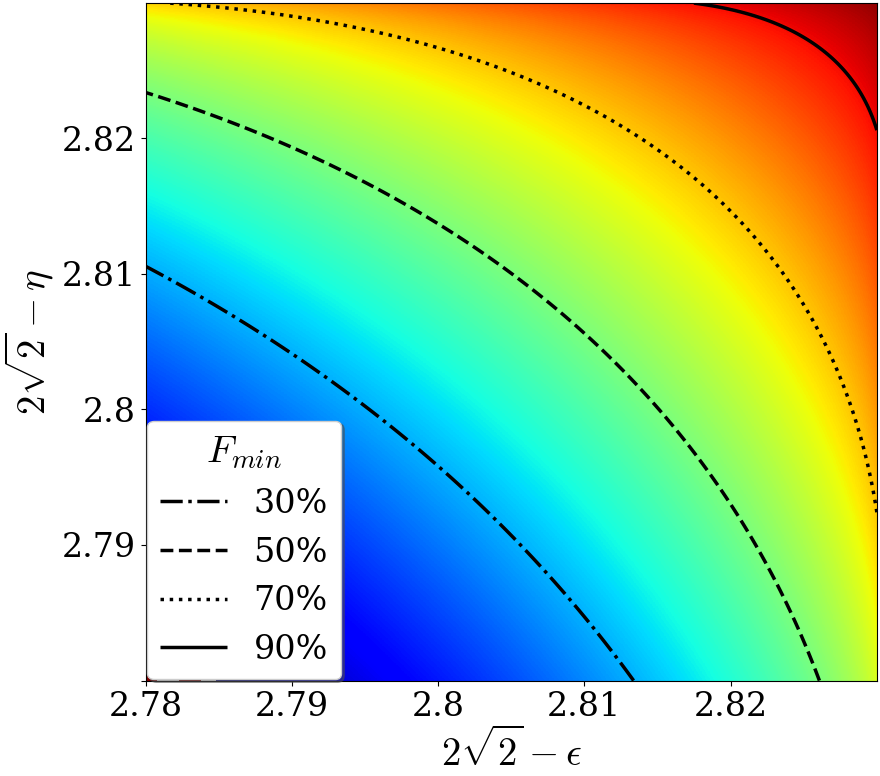}  
  \caption{R = 90\%}
  \label{fig:FullDI90}
\end{subfigure}
\caption{Minimum certified fidelity of the output state of Protocol 2, to the state sent through the channel, as a function of the deviations $\eta, \epsilon$ from maximum violation of CHSH inequality. We set $x=7$ for a confidence level $> 99.4\%$, $M = K = 10^{10}$, and different ratios of transmission $R=K/N$.}
\label{fig:FullDI}
\end{figure*}

Thanks to the IID assumption made on the probe-state source, we still consider all input probe states to be equal to $\Phi_i$, so the first part of Proposition \ref{thm:AnuTheoremDI} enables Alice and Bob to certify the quantity $F^i$ once, for the whole protocol. This way, compared to Proposition \ref{thm:AnuTheorem} for protocol 1, we bound $C^i \leq \sqrt{\alpha g_x(\eta,M)}$, and replace the expression of $f_x$ and $\alpha$. We also multiply the confidence level by $(1-e^{-x})$ to account for the confidence on the input bound, due to the finite number $M$ of input state tested. This straightly gives the bound:
\begin{equation}\label{ineq:FinalResultDI2}
    F(\bar{\rho}_o,\rho_i) \geq 1 - 4\cdot\sin^2\biggl( \arcsin\bigl(\sqrt{\alpha g_x(\eta,M)}/\tau_x\bigr) + \arcsin\sqrt{\alpha f_x(\epsilon,K)}+\Delta_x\biggr)
\end{equation}
with confidence level at least $(1-e^{-x})^2\cdot (1-2e^{-x})^2$ for any $x>0$, therefore showing the security bound for protocol 2. We show the corresponding certified fidelity with examples of experimental parameters in Fig.~\ref{fig:FullDI}.
\vspace{1cm}

\section{Details on the Experimental Protocol}\label{app:ExpProt}

\subsection{Probe State Source}\label{sourceDetails}
\begin{figure}[htbp]
\begin{subfigure}{.48\textwidth}
 \centering
    \includegraphics[height=6cm]{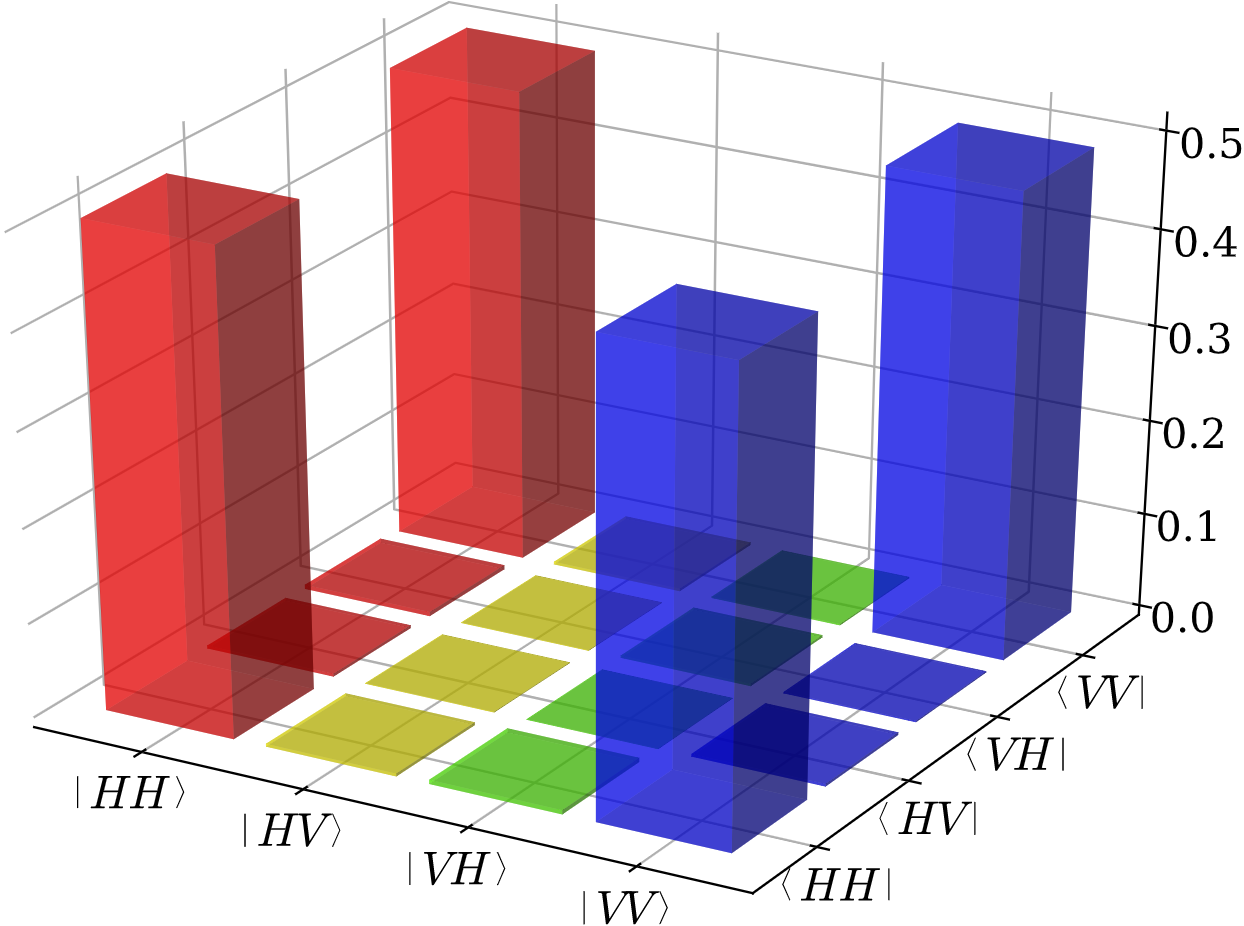}
    \caption{Real part.}
    \label{fig:stateReal}
\end{subfigure}
\hfill
\begin{subfigure}{.48\textwidth}
 \centering
    \includegraphics[height=6cm]{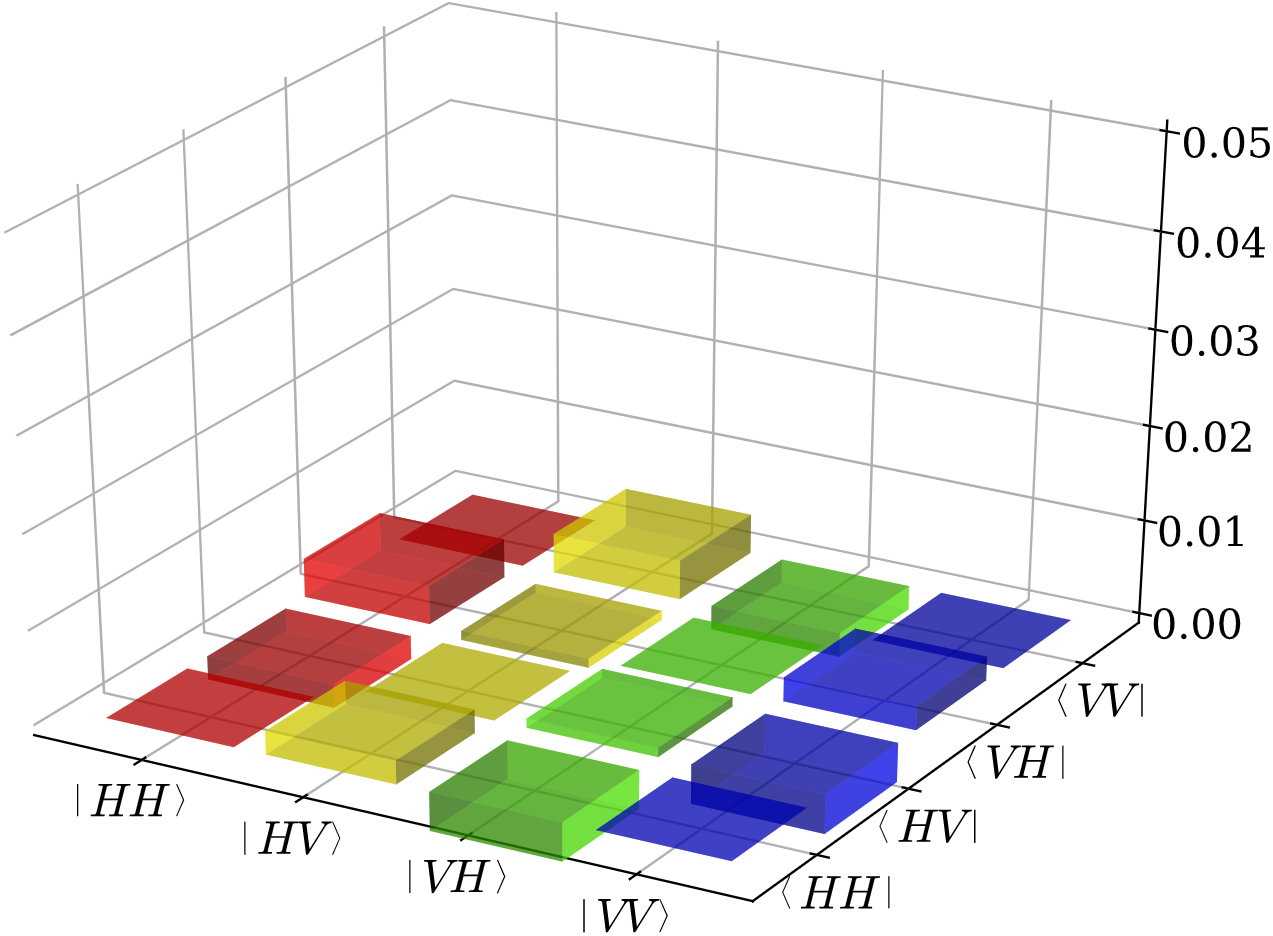}
    \caption{Imaginary part.}
    \label{fig:stateImag}
\end{subfigure}
\caption{Density matrix reconstructed by tomography of the probe quantum state emitted by the Sagnac source, in one iteration of the protocol. Real and imaginary parts are not at the same scale.}
\label{fig:state}
\end{figure}

Here we give a few details on the probe state source. We first provide an example of the polarization state of photon pairs, reconstructed via quantum state tomography. This state $\Phi_i$ was measured for the protocol implementation with heralding efficiency $\eta_s = 0.444$, and shows a fidelity $F(\Phi_i,\Phi_+) = 99.43\% \pm 0.05\%$ to the maximally-entangled state $\ket{\Phi_+} = \tfrac{\ket{HH}+\ket{VV}}{\sqrt{2}}$. As the imaginary part of the density matrix is negligible, we display the real part only, on Fig.\ref{fig:state}.\\

\begin{figure}[htbp]
\centering
  \includegraphics[height=10cm]{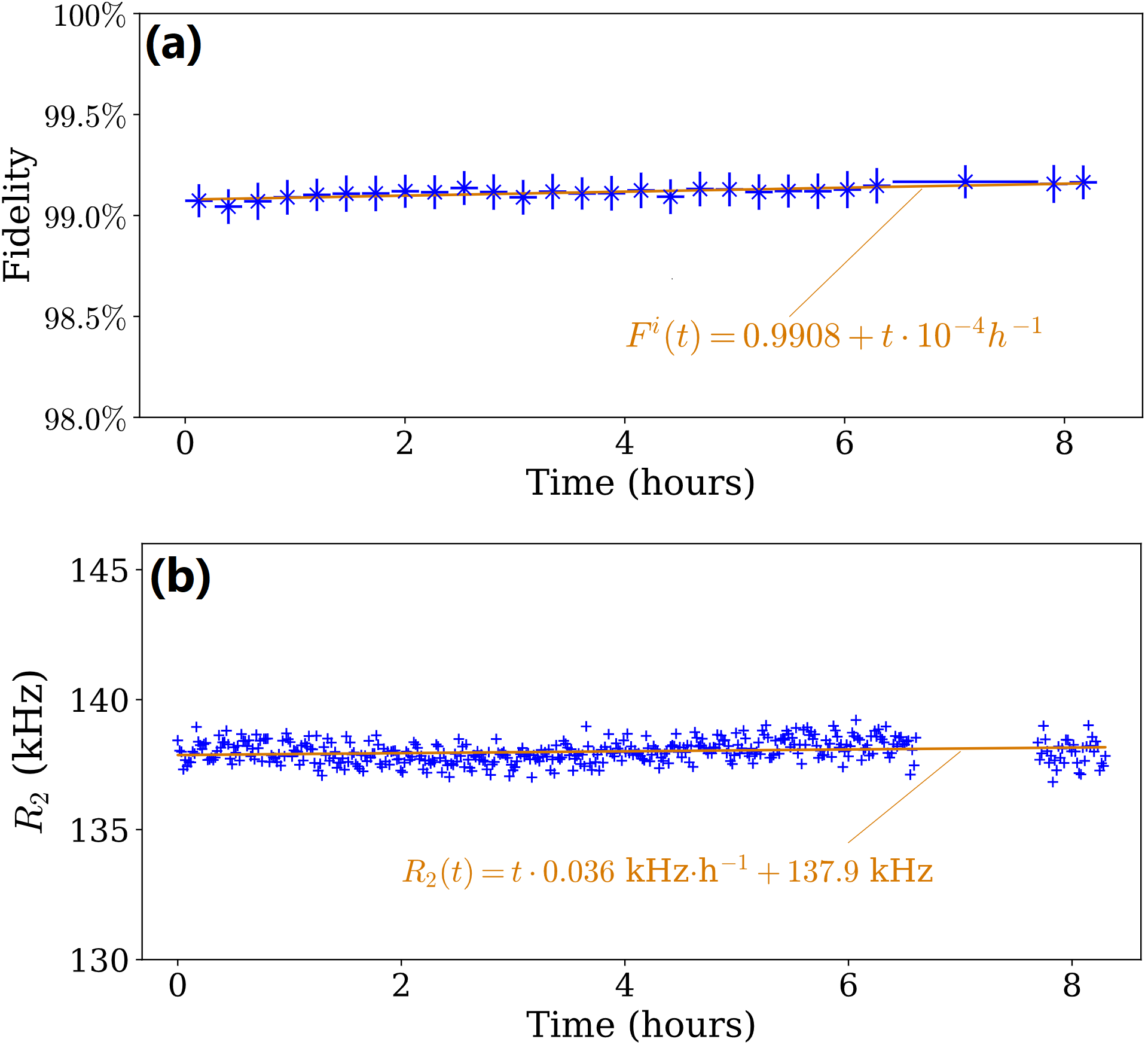}  
\caption{Features of the source measured over an 8-hours time-span. The 1-hour gap at the end of the data series is due to a cooling cycle of the detectors. \textbf{(a)} Biphoton detection rate $R_2$. \textbf{(b)} Fidelity of the source's state to a Bell state.}
\label{fig:stability}
\end{figure}

We also performed a continuous measurement of the quantum state via quantum state tomography, over an $8-$hours time span, in order to evaluate the stability of the quantum state during a protocol run. As we show in Fig.~\ref{fig:stability}, the low standard deviation and drift in the fidelity to $\Phi_+$, as well as in the photon detection rate, motivate the IID assumption we make on the probe state during the protocol. 

\newpage

\vspace{1cm}
\vspace{1cm}

\newpage

\subsection{Detailed Protocol Results}
The main results for the certification of lossy honest quantum channels are displayed in Fig.~\ref{fig:HonestResults} in the main text, but we do not detail the different measurements which were performed during this protocol. In the following Fig.~\ref{fig:HonestFidViolation}, we display the results for the two main experimental measurements, namely the probe state's fidelity to a Bell state and the steering inequality violation, that we feed in our main result~(\ref{ineq:MainResult3}) in order to bound the transmission fidelity. In particular, note that the correlations we measure are always more than $\epsilon$-close to maximum violation of steering inequality with $\epsilon=0.015$. Yet we witness a drop in the probe's state fidelity to a maximally entangled state, for the second point on the graph. This causes the corresponding points in Fig.~\ref{fig:HonestResults} to deviate from the average curves, and is purely due to experimental mishandling, causing some misalignment during one iteration of the protocol.

\begin{figure}[htbp]
\begin{minipage}[c]{0.68\linewidth}
    \centering
    \includegraphics[width=120mm]{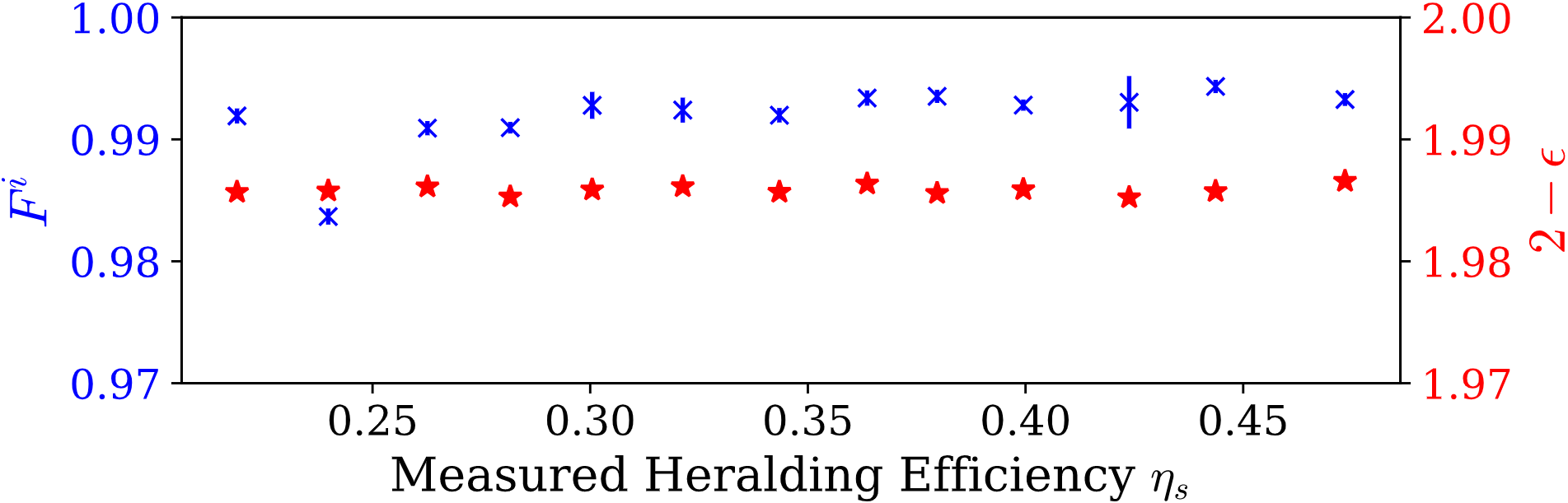}
    \caption{Measured probe-state fidelity $F^i$ to a maximally-entangled state, and close-to-maximum violation of steering inequality $2-\epsilon$.}
    \label{fig:HonestFidViolation}
\end{minipage}
\end{figure}

\subsection{Detectors Model in Experiment}\label{FairSampling}
\begin{figure}[htbp]
\begin{minipage}[c]{0.65\linewidth}
    \centering
    \includegraphics[width=100mm]{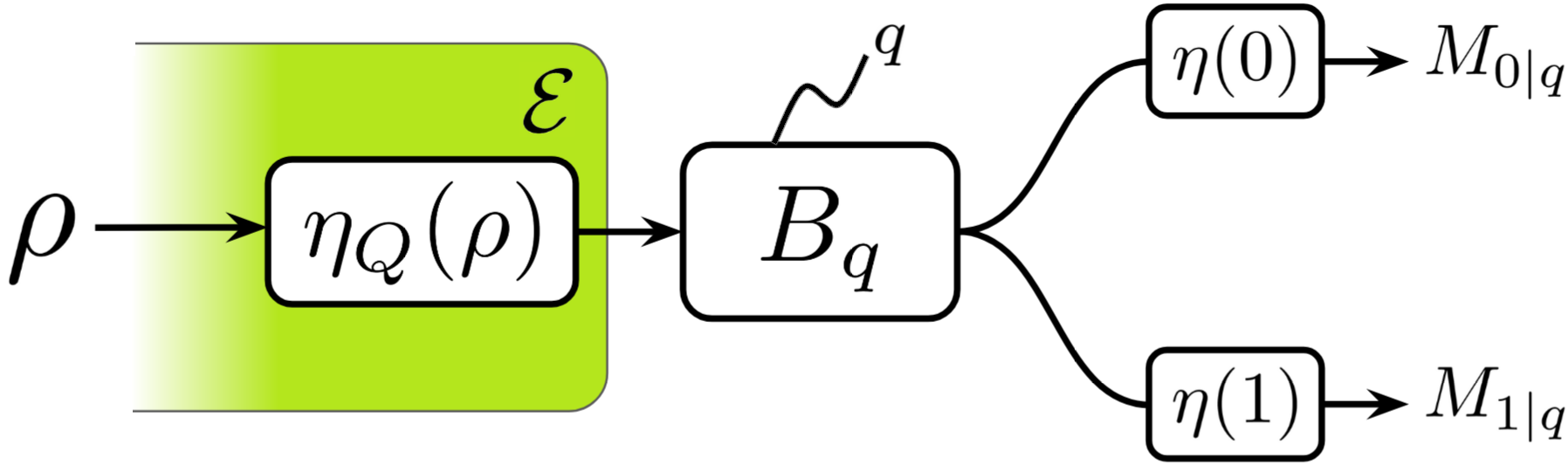}
    \caption{Schematic representation of Bob's measurement apparatus, taking our assumptions into account. The apparatus first displays some state-dependent transmissivity $\eta_Q$, that we can include inside the channel $\mathcal{E}$. Bob then measures the observable $B_q$, the result $l\in\{0,1\}$ of which is filtered with efficiency $\eta(l)$.}
    \label{fig:fairSampling}
\end{minipage}
\end{figure}

We now detail the assumptions taken on the players' detection systems, in order to perform our proof-of-principle experimental protocol, as well as the consequences on the protocol's results. We focus on the detectors used in order to certify the output probe state in the one-sided device independent protocol, and therefore omit the system that Alice uses in order to certify the input state $\Phi_i$. Both Alice and Bob each possess a local measurement apparatus, ideally made of 2-outcome POVMs $\{M^{\mathcal{A}_2}_{l|q}\}_{l=0,1}$ and $\{M^{\mathcal{B}}_{l|q}\}_{l=0,1}$, for $q=0,1$. In reality, these detectors have non-unit efficiency, meaning they only return a result with a certain probability which may depend on the parameter $q$, the outcome $l$, or even the quantum state $\rho$. This way we adopt a similar description as that of \cite{orsucci2020}, such that we get the probabilities of returning outcome $l$, when measuring $\rho$ with measurement parameter $q$:
\begin{align}
    &\mathbb{P}_{A}(l| q,\rho) = \tra(\rho M^{\mathcal{A}_2}_{l|q})\cdot \eta^{A}(l,q,\rho)\\
    &\mathbb{P}_{\mathcal{B}}(l| q,\rho) = \tra(\rho M^{\mathcal{B}}_{l|q})\cdot \eta^{\mathcal{B}}(l,q,\rho)
\end{align}
where $\eta^{A}$ and $\eta^{\mathcal{B}}$ are the efficiencies. For a bipartite state, the probability of getting outcomes $(l_{\mathcal{A}},l_{\mathcal{B}})$ with parameters $(q_{\mathcal{A}},q_{\mathcal{B}})$ becomes: 
\begin{equation}
    \mathbb{P}(l_{A},l_{\mathcal{B}}| q_{\mathcal{A}},q_{\mathcal{B}},\rho) = \tra\bigl(\rho\cdot M^{\mathcal{A}_2}_{l_{\mathcal{A}}|q_{\mathcal{A}}}\otimes M^{\mathcal{B}}_{l_{\mathcal{B}}|q_{\mathcal{B}}}\bigr)\cdot \eta^{A}(l_{\mathcal{A}},q_{\mathcal{A}},\rho_{\mathcal{A}}) \cdot \eta^{\mathcal{B}}(l_{\mathcal{B}},q_{\mathcal{B}},\rho_{\mathcal{B}})
\end{equation}
where $\rho_{\mathcal{A}} = \tra_{\mathcal{B}}(\rho)$ and $\rho_{\mathcal{B}} = \tra_{\mathcal{A}}(\rho)$ are the local states, such that the efficiencies are local. In the following we focus on the assumptions made on these efficiencies in our protocol, and the consequences on the results. First, in a one-sided device independent scenario, we assume that Alice fully characterizes her measurement apparatus, and proves her efficiency to be independent of the state $\rho$ and the measurement parameter $q$, such that: 
\begin{equation}
    \eta^{\mathcal{A}_2}(l,q,\rho) = \eta^{\mathcal{A}_2}(l)
\end{equation}
The values of $\eta^{\mathcal{A}_2}(l)$ are accessible to Alice, as part of her detectors' characterization. This way, for $l_+$ and $l_-$ such that $\eta^{\mathcal{A}_2}(l_+)>\eta^{\mathcal{A}_2}(l_-)$, Alice can ignore the outcomes $l_+$ with probability $1-\eta^{\mathcal{A}_2}(l_-)/\eta^{\mathcal{A}_2}(l_+)$ in order to effectively equalize the efficiencies of the two outcomes. In that case the efficiency on Alice's side is a constant $\eta^{\mathcal{A}_2}$, such that
\begin{equation}
    \eta^{\mathcal{A}_2}(l,q,\rho) =  \eta^{\mathcal{A}_2}
\end{equation}
On Bob's side, we first make the weak fair sampling assumption \cite{orsucci2020}, stating that we can factorize the efficiencies due to classical parameters from those due to quantum states:
\begin{equation}
    \eta^{\mathcal{B}}(l,q,\rho) = \eta^{\mathcal{B}}_C(l,q)\cdot \eta^{\mathcal{B}}_Q(\rho)  
\end{equation}
We then make a form of strong fair-sampling assumption, stating the efficiency does not depend on $q$, such that:
\begin{equation}
    \eta^{\mathcal{B}}(l,q,\rho) = \eta^{\mathcal{B}}_C(l)\cdot \eta^{\mathcal{B}}_Q(\rho)  
\end{equation}
Now we could assume the state-dependent efficiency to be unit, which leads to an unbalanced-outcomes homogeneous fair-sampling assumption, and leaves the protocol more vulnerable to attacks. Another solution is to consider $\eta^{\mathcal{B}}_Q(\rho)$ as a part of the quantum channel being tested, as shown in Fig.~\ref{fig:fairSampling}. In that case our protocol is more secure but certifies a different channel, the output of which is necessarily measured by Bob measurement apparatus. This would require further investigation if the quantum communication is followed by another protocol which does not involve Bob's measurement apparatus. In both cases, we can neglect the state-dependent efficiency, such that
\begin{equation}
    \eta^{\mathcal{B}}(l,q,\rho) = \eta^{\mathcal{B}}_C(l)  
\end{equation}
is an efficiency which \emph{a priori} depends on the result $l$. The detection probability then becomes 
\begin{equation}
    \mathbb{P}_{\mathcal{B}}(l|q,\rho) = \tra(\rho M^{\mathcal{B}}_{l|q})\cdot \eta^{\mathcal{B}}(l)
\end{equation}

Similarly to \cite{orsucci2020}, we now show that even though the efficiency $\eta^{\mathcal{B}}$ slightly varies with the outcome $l$, we can still use the measured outcome without any correction on Bob's side, and still get a good evaluation of $\beta = \vert\langle A_0B_0 \rangle + \langle A_1B_1 \rangle  \vert$. By definition we have:
\begin{equation}
    \langle A_qB_q\rangle = \langle M^{\mathcal{A}_2}_{0|q}  M^{\mathcal{B}}_{0|q}\rangle + \langle M^{\mathcal{A}_2}_{1|q}  M^{\mathcal{B}}_{1|q}\rangle - \langle M^{\mathcal{A}_2}_{0|q}  M^{\mathcal{B}}_{1|q}\rangle + \langle M^{\mathcal{A}_2}_{1|q}  M^{\mathcal{B}}_{0|q} \rangle
\end{equation}
With their imperfect detectors, Alice and Bob approximate that quantity by measuring the following:
\begin{equation}
     \overline{A_qB_q} =  \dfrac{n_{0,0\vert q}+n_{1,1\vert q}-n_{0,1\vert q}-n_{1,0\vert q}}{n_{0,0\vert q}+n_{1,1\vert q}+n_{0,1\vert q}+n_{1,0\vert q}}
\end{equation}
where $n_{l_{\mathcal{A}},l_{\mathcal{B}}\vert q}$ is the number of times the measurement of a pair gave the outcome $(l_{\mathcal{A}},l_{\mathcal{B}})$, when Alice and Bob both measured with parameter $q$. When measuring a big number of state $\mathcal{N}$ we approximate 
\begin{equation}
    n_{l_{\mathcal{A}},l_{\mathcal{B}}\vert q} = \mathcal{N} \cdot \mathbb{P}(l_{A},l_{\mathcal{B}}| q_{\mathcal{A}},q_{\mathcal{B}},\rho) =\mathcal{N} \cdot \tra(\rho \cdot M^{\mathcal{A}_2}_{l_{\mathcal{A}}|q}\otimes M^{\mathcal{B}}_{l_{\mathcal{B}}|q})\cdot \eta^{\mathcal{A}}\cdot\eta^{\mathcal{B}}(l_{\mathcal{B}})
\end{equation}
so we can rewrite the evaluation of $\langle A_qB_q\rangle$, symplifying the constant terms $\mathcal{N}$ and $\eta_{\mathcal{A}}$:
\begin{equation}
\begin{aligned}
     \overline{A_qB_q} &=  \dfrac{\tra\bigl[\rho \cdot (M^{\mathcal{A}_2}_{0|q}\otimes M^{\mathcal{B}}_{0|q} - M^{\mathcal{A}_2}_{1|q}\otimes M^{\mathcal{B}}_{0|q})\bigr]\cdot\eta^{\mathcal{B}}(0)+\tra\bigl[\rho \cdot (M^{\mathcal{A}_2}_{1|q}\otimes M^{\mathcal{B}}_{1|q} - M^{\mathcal{A}_2}_{0|q}\otimes M^{\mathcal{B}}_{1|q})\bigr]\cdot \eta^{\mathcal{B}}(1)}{\tra\bigl[\rho \cdot (M^{\mathcal{A}_2}_{0|q}\otimes M^{\mathcal{B}}_{0|q} + M^{\mathcal{A}_2}_{1|q}\otimes M^{\mathcal{B}}_{0|q})\bigr]\cdot\eta^{\mathcal{B}}(0)+\tra\bigl[\rho \cdot (M^{\mathcal{A}_2}_{1|q}\otimes M^{\mathcal{B}}_{1|q} + M^{\mathcal{A}_2}_{0|q}\otimes M^{\mathcal{B}}_{1|q})\bigr]\cdot \eta^{\mathcal{B}}(1)}\\
     &= \dfrac{\tra\Bigl[\rho \cdot A_q\otimes \bigl(M^{\mathcal{B}}_{0|q}\cdot\eta^{\mathcal{B}}(0)-M^{\mathcal{B}}_{1|q}\cdot\eta^{\mathcal{B}}(1)\bigr)\Bigr]}{\tra\Bigl[\rho \cdot \bigl(M^{\mathcal{B}}_{0|q}\cdot\eta^{\mathcal{B}}(0)+ M^{\mathcal{B}}_{1|q}\cdot \eta^{\mathcal{B}}(1)\bigr)\Bigr]}
\end{aligned}
\end{equation}
Then we take $\xi$ such that $\eta^{\mathcal{B}}(1)/\eta^{\mathcal{B}}(0) = 1 + \xi$, and we get
\begin{equation}
\begin{aligned}
     \overline{A_qB_q} &= \dfrac{\tra\Bigl[\rho \cdot A_q\otimes \bigl(M^{\mathcal{B}}_{0|q}-M^{\mathcal{B}}_{1|q}\cdot\eta^{\mathcal{B}}(1)/\eta^{\mathcal{B}}(0)\bigr)\Bigr]}{\tra\Bigl[\rho \cdot \bigl(M^{\mathcal{B}}_{0|q}+ M^{\mathcal{B}}_{1|q}\cdot \eta^{\mathcal{B}}(1)/\eta^{\mathcal{B}}(0)\bigr)\Bigr]}\\
     &= \dfrac{\langle A_qB_q\rangle-\tra\bigl[\rho \cdot A_q\otimes M^{\mathcal{B}}_{1|q}\bigr]\cdot\xi}{1+\tra\bigl[\rho \cdot M^{\mathcal{B}}_{1|q}\bigr]\cdot\xi}
\end{aligned}
\end{equation}
Considering $\eta^{\mathcal{B}}(1) \approx \eta^{\mathcal{B}}(0)$, such that $\vert\xi\vert \ll 1$, we can approximate the difference between the expected correlation $\langle A_qB_q\rangle$ and the measured correlation $\overline{A_qB_q}$, at first order:
\begin{equation}
\begin{aligned}
    \overline{A_qB_q}- \langle A_qB_q\rangle &\approx -\tra\bigl[\rho \cdot A_q\otimes M^{\mathcal{B}}_{1|q}\bigr]\cdot\xi-\langle A_qB_q\rangle\cdot\tra\bigl[\rho \cdot M^{\mathcal{B}}_{1|q}\bigr]\cdot\xi\\
    &= (1-\langle A_qB_q\rangle)\cdot\tra\bigl[\rho \cdot M^{\mathcal{A}}_{1\vert q}\otimes M^{\mathcal{B}}_{1|q}\bigr]\cdot\xi-(1+\langle A_qB_q\rangle)\cdot\tra\bigl[\rho \cdot M^{\mathcal{A}}_{0\vert q}\otimes M^{\mathcal{B}}_{1|q}\bigr]\cdot\xi
\end{aligned}
\end{equation}
Provided Alice and Bob witness a close-to-maximum violation of steering inequality, we also have $(1-\langle A_qB_q\rangle)\ll 1$ and $\tra\bigl[\rho \cdot M^{\mathcal{A}}_{0\vert q}\otimes M^{\mathcal{B}}_{1|q}\bigr]\ll 1$. This way, that difference is doubly negligible, such that even noticeable unbalance between the detectors efficiencies should not significantly deviate the measured correlation from the expected correlation. We therefore assume $\overline{A_qB_q}\approx \langle A_qB_q\rangle$, such that the value of $\beta$ can be accurately measured even without correction for the detectors efficiency. In our experiment, we measure the relative efficiency between Bob's detectors, for each protocol iteration. This way we get $\xi \lesssim 0.03$, while witnessing a close-to-maximum violation of steering inequality, legitimizing the approximation. We still compute the violation that would be measured if detectors were perfectly balanced, and $\eta^{\mathcal{B}}(1) = \eta^{\mathcal{B}}(0)$, by correcting the data with the relative efficiencies. The difference between the corrected and uncorrected data is included in the error bars displayed in Fig.~\ref{fig:HonestResults} in the main text.

\end{document}